\documentclass[fleqn,usenatbib,referee]{mnras}

\usepackage[T1]{fontenc}
\usepackage{ae,aecompl}
\usepackage{anyfontsize}

\usepackage{graphicx}	
\usepackage{amsmath}	
\usepackage{amssymb}	

\newcommand{\rlight}{r_{\rm L}}
\newcommand{\Rs}{R_{\rm s}}

\newcommand{\ex}{\mathbf{e}_{\rm x}}
\newcommand{\ey}{\mathbf{e}_{\rm y}}
\newcommand{\ez}{\mathbf{e}_{\rm z}}

\title[General-relativistic pulsar magnetospheric emission]{General-relativistic pulsar magnetospheric emission}

\author[J. P\'etri]{J.  P\'etri
\thanks{E-mail: jerome.petri@astro.unistra.fr} \\
  Universit\'e de Strasbourg, CNRS, Observatoire astronomique de Strasbourg, UMR 7550, F-67000 Strasbourg, France.
  }

\date{Accepted XXX. Received YYY; in original form ZZZ}

\pubyear{2017}

\begin{document}
\label{firstpage}
\pagerange{\pageref{firstpage}--\pageref{lastpage}}
\maketitle

\begin{abstract}
Most current pulsar emission models assume photon production and emission within the magnetosphere. Low frequency radiation is preferentially produced in the vicinity of the polar caps whereas the high-energy tail is shifted to regions closer but still inside the light-cylinder. We conducted a systematic study of the merit of several popular radiation sites like the polar cap, the outer gap and the slot gap. We computed sky maps emanating from each emission site according to a prescribed distribution function for the emitting particles made of an electron/positron mixture. Calculations are performed using a three dimensional integration of the plasma emissivity in the vacuum electromagnetic field of a rotating and centred general-relativistic dipole. We compare Newtonian electromagnetic fields to their general-relativistic counterpart. In the latter case, light bending is also taken into account. As a typical example, light-curves and sky maps are plotted for several power-law indices of the particle distribution function. The detailed pulse profiles strongly depend on the underlying assumption about the fluid motion subject to strong electromagnetic fields. This electromagnetic topology enforces the photon propagation direction directly, or indirectly, from aberration effects. We also discuss the implication of a net stellar electric charge on to sky maps. Taking into account the electric field strongly affects the light-curves originating close to the light-cylinder where the electric field strength becomes comparable to the magnetic field strength.
\end{abstract}




\section{Introduction}

Electromagnetic activity around neutron stars is indirectly evidenced by the broad band pulsed emission spectra detected on space and ground-based telescopes \citep{lyne_shape_1988}, \citep{abdo_second_2013}. More than 2000~pulsars are known today, each showing a unique distinctive fingerprint depicted by its pulse profile in radio, X-rays and gamma-rays. The multi-wavelength light-curve evolution offers a unique insight into the real nature of the emission mechanisms as well as on their location and spread within the magnetosphere. Among the most popular models are the polar cap \citep{ruderman_theory_1975}, the outer gap \cite{cheng_energetic_1986} and the slot gap \citep{arons_pair_1983, dyks_two-pole_2003} with possible extension to the striped wind \citep{kirk_pulsed_2002, petri_unified_2011-4}. Any recipe to compute such light-curves requires several ingredients: first a prescribed magnetic topology within the magnetosphere and the wind, second some ad-hoc particle acceleration and photon production sites (through curvature, synchrotron and/or inverse Compton radiation) and third particle distribution functions emerging from a balance between acceleration and radiation reaction. Obviously electromagnetic quantities, particle dynamics and photon productions are highly intertwined and in the best world should be computed self-consistently, taking into account bidirectional feedback between particle/radiation and particle/field. However, we are still far from such capabilities catching all the micro-physics of particle acceleration and radiation connected to the global electrodynamics of the magnetosphere although some modest attempts emerged recently \citep{cerutti_modelling_2016}.

An argument commonly used to solve for the particle trajectories claims that they follow magnetic field lines in the corotating frame dragged by the neutron star. Whereas this picture is tenable and well defined in for instance an ideal MHD context or in force-free flows, unfortunately, such approach reveals misleading and fallacious for vacuum fields where the concept of magnetic field line and its velocity is useless for particle motion \citep{newcomb_motion_1958}. The definition of a magnetic field line and of its velocity cannot be set unequivocally in regions where there exists a component of the electric field $\mathbf{E}$ parallel to the magnetic field $\mathbf{B}$, that is where $\mathbf{E} \cdot \mathbf{B} \neq 0$. But these places of non-vanishing $E_\parallel = \mathbf{E} \cdot \mathbf{B}/B$ are exactly where acceleration and therefore radiation occurs. Thus field lines and particle trajectories are not straightforwardly connected to each other. Nevertheless several authors attempted to describe the trajectory of emitting particles as a combination between motion along field lines and corotation enforced by the neutron star. Depending on assumptions about the precise path of these particles, the aberration formula for photons emitted along field lines follows the usual Lorentz boost \citep{dyks_two-pole_2003} or differs from it if the instantaneous corotation frame is taken into account \citep{bai_uncertainties_2010}. Actually the special relativistic aberration formula remains valid in the corotating frame as we remind in appendix~\ref{app:Aberration}. Indeed \cite{dyks_two-pole_2003} approach is physically as correct (or incorrect) as \cite{bai_uncertainties_2010} one. They differ by the assumption made about the particle motion and show large discrepancies between each other close to the light-cylinder. On one hand \cite{dyks_two-pole_2003} model breaks down at the light cylinder because the local inertial frame speed reaches the speed of light. On the other hand \cite{bai_uncertainties_2010} prescription has no physical solution at much larger distances, at several light-cylinder radii outside the light-cylinder and sometimes already at the light-cylinder too. Unfortunately therefore, none of these descriptions applies strictly to \cite{deutsch_electromagnetic_1955} vacuum field solution when distances become much greater than the light cylinder radius. Moreover the instantaneous corotation frame was used even for this vacuum field which is highly debatable. Indeed, as a starting point \cite{bai_uncertainties_2010} used in their primary assumption the ideal MHD or force-free case that differs significantly from vacuum and for which field lines can indeed be properly defined. Such treatment is more appropriate for force-free fields as done later by \cite{bai_modeling_2010}. Unfortunately, none of the aberration formula proposed in these works extend to the light-cylinder or beyond it, the situation going worse outside the magnetosphere where corotation would imply a speed larger than the speed of light and where the pulsar wind is launched. When corotation is assumed in vacuum, the corotation speed reaches the speed of light exactly at the light-cylinder by definition and all aberration formulas crash due to diverging Lorentz factors. This behaviour can be counterbalanced if magnetic field lines were sufficiently swept back in the sense that the poloidal component of the magnetic field decreases as fast as the toroidal component increases. This happens in the force-free limit where the magnetospheric currents generate a substantial toroidal field but not in vacuum as given by \citet{deutsch_electromagnetic_1955} solution. The side effect of these attempts using aberration transformations reflects in a high sensitivity of the emission maps on the cut-off radius where emission is supposed to stop. This radius needs to be arbitrarily set to values less than $\rlight=c/\Omega$, where $c$ is the speed of light and $\Omega$ the neutron star rotation rate, in order to avoid this divergence. The only exception that can handle arbitrary distances from the neutron star is the radiation reaction limit where the particle speed is directly deduced from the local properties of the electromagnetic field \citep{mestel_stellar_1999} and also called aristotelian electrodynamics by \cite{gruzinov_aristotelian_2013}. We will show that this ultra-relativistic radiation reaction limit is a special case of motion in a frame where electric and magnetic field are parallel and in which it moves at the speed of light. Therefore emission can be computed in whole space, no distinction is required between the notion of corotating magnetosphere and region outside the light-cylinder. Nevertheless complications arise because knowledge of the electric field is needed in addition to the magnetic field.

Nowadays more than 2000~pulsars are known as radio emitters. Although they have been observed since the early days of the discovery of pulsars fifty years ago, radio pulsars did not furnish severe constraints on the magnetosphere geometry and emission physics. This is largely due to the fact that energy produced in the radio waveband is negligible compared to the total spindown luminosity available. The situation drastically changed with the launch of Fermi/LAT in June~2008. Since then more than 250~pulsars are known to emit also gamma-rays. This number has doubled since the publication of the second Fermi pulsar catalogue \citep{abdo_second_2013}. Gamma-ray pulsars have sharpened our understanding of pulsar magnetospheres because contrary to radio pulsars, gamma-ray pulsars spend a substantial fraction of rotational kinetic energy into high energy radiation. We so to say indirectly see their magnetosphere as pulsed gamma-ray radiation. The flux remains significant even above several GeV severely constraining the emission sites to be well above the polar cap in order to avoid too strong magnetic absorption in magnetic field close to the critical value of $4.4\times10^9$~T \citep{daugherty_gamma-ray_1996}.

In the early ages of pulsar magnetospheric emission models, a vacuum electromagnetic field was used to predict the phase-resolved light-curves. The beauty of this approach was that an exact analytical solution exists and is known as \citet{deutsch_electromagnetic_1955} solution. As the numerical techniques to solve plasma problems improved to include the feedback of the flow onto the electromagnetic field according to the force-free prescription or even thanks to MHD simulations, a new trend naturally appeared to compare fluid (MHD/FFE) expectations to vacuum fields. Simulations first started with the aligned case looking for stationary solutions like the pioneer work of \cite{contopoulos_axisymmetric_1999}. \cite{timokhin_force-free_2006} relaxed the condition on the light-cylinder by moving the Y-point inside the magnetosphere as a free parameter. He found a set of FFE solutions with different energy loss rates arguing therefore that the time evolution of the spindown luminosity differs from the conventional magnetodipole formula. Followed then axisymmetric time-dependent simulations performed by several other authors using different finite volume \citep{komissarov_simulations_2006, mckinney_relativistic_2006} and spectral \citep{parfrey_introducing_2012, cao_spectral_2016} algorithms. However the most interesting case leading to pulsed emission concerns the oblique rotator. This problem was tackled by other authors beginning with \cite{spitkovsky_time-dependent_2006} and followed by \cite{kalapotharakos_extended_2012} and \cite{petri_pulsar_2012}. Although FFE models give more realistic electromagnetic field topologies and better results than vacuum fields (see the comparison made by \cite{bai_uncertainties_2010} and \cite{bai_modeling_2010}), the electric field being perpendicular to the magnetic field, no particle acceleration is allowed although this is compulsory to radiate high energy photons. Thus, little by little, magnetosphere started to include some dissipative processes.

To better stick to the wealth of observations of Fermi/LAT, dissipative effects have been included with several prescriptions like the one presented in \cite{li_resistive_2012}, in \cite{kalapotharakos_toward_2012} and in \cite{cao_oblique_2016} for magnetosphere models with conductivity. Light-curves a better fitted with dissipation according to recent work by \cite{kalapotharakos_gamma-ray_2012, kalapotharakos_gamma-ray_2014, brambilla_testing_2015}. In these models, curvature radiation was the main channel to produce high energy gamma-ray photons. The spatial inhomogeneity of the conductivity is controlled by the phase lag between radio and gamma-ray peak among others. \cite{kalapotharakos_fermi_2017} were even able to constrain the accelerating field and the conductivity according to the spindown losses. The polar cap was divided into an inner FFE region and an outer annulus with finite conductivity. Particles radiate curvature photons at the radiation reaction limit regime.

In some approximations, it is possible to set the particle velocity on hand of the electromagnetic field and known as aristotelian dynamics \citep{gruzinov_electrodynamics_2012}. Light-curves in this approximation are easily computed as shown by \cite{gruzinov_pulsar_2013} although the grid resolution could still be improved.

Phase-resolved radio polarization offers another useful insight into the magnetic field topology and emission sites. Rotational distortions of the static dipole have been extensively studied in the last decades to show that the polarization angle inflection point does not match with the maximum of the pulse profile. \cite{shitov_period_1983} found a deviation from a pure static dipole showing that the twisting angle is of the order
\begin{equation}
 \psi \approx 1.2 \, \left( \frac{r}{\rlight} \right)^3 \, \sin^2\chi
\end{equation}
where $\rlight$ is the light-cylinder radius, $r$ the spherical radial coordinate and $\chi$ the pulsar obliquity. This results was then used by \cite{shitov_pulsar_1985} to show the impact on the polarization angle profile with respect to the pulse profile. However this third order effect, was often neglected in subsequent analyses. Indeed \cite{blaskiewicz_relativistic_1991} only took into account aberration and retardation effects for the polarization angle. Some limitations of their approach have been underlined by \cite{craig_altitude_2012}. They found a much stronger lag of the order
\begin{equation}
 \Delta \psi = 4 \, \frac{r}{\rlight} .
\end{equation}
\cite{dyks_rotational_2004} even showed that the polarization angle can precede the pulse profile peak for very slow rotators because their so-called open volume subtended by the last open field lines scales as $\sqrt{\frac{r}{\rlight}}$ and is shifted in the counter-clockwise direction. All these results assume a rotating vacuum point dipole. Polar cap shapes are however very different in the vacuum and FFE approximations, see \cite{harding_gamma-ray_2016} for a review of vacuum, FFE and dissipative magnetosphere. However quoting the conclusion of \cite{harding_gamma-ray_2011} ``Although the force-free magnetosphere is a limiting case, these fits may indicate that the real pulsar magnetosphere solution is closer to the vacuum dipole in field geometry''. Thus a vacuum rotating dipole may still be useful to model pulsar light-curves from a geometrical point of view. Multi-wavelength polarization predictions including synchrotron and curvature radiation, extending from radio up to gamma-rays, will severely constraint the emission models and the geometry of rotation-powered pulsars \citep{harding_multiwavelength_2017}. Key properties and mysteries in neutron star and pulsar magnetospheres are reviewed in \cite{grenier_gamma-ray_2015} where the input of simulations in conjunction with gamma-ray modelling is emphasized.

Sharp features in the light curves are interpreted as caustic formation in the outer part of the magnetosphere due to the combined effect of aberration and retardation \citep{morini_inverse_1983, dyks_relativistic_2004}. Phase alignment between radio and gamma-ray pulses seen in some millisecond pulsars suggests that for these pulsars, radio and gamma-rays are produced at the same location, and according to \cite{venter_modeling_2012} corresponding to 30\% of the light cylinder radius. A comprehensive study of pulsar light-curve characterization was compiled by \cite{watters_atlas_2009} for the three main high energy models namely, polar cap, slot gap (two-pole caustic) and outer gap. See also \cite{venter_probing_2009} and later \cite{pierbattista_light-curve_2015, pierbattista_young_2016} for a similar investigation. Such atlas are useful to constrain the pulsar obliquity and the observer line of sight inclination as pulse profiles are very sensitive to these parameters. The assumptions about particle motion in the corotating (accelerated) frame and its transformation back to the observer (inertial) frame led to some discrepancies between several groups \citep{romani_constraining_2010}. However pointing out that some treatments are better than others is only a matter of point of view. In the end, what really counts is the particle velocity in the observer frame, independently on the assumed motion in the corotating frame, if this latter is required. Motion is usually claimed to be along magnetic field lines but this is too restrictive and even irrelevant for non ideal MHD flows departing from the $\mathbf{E} \cdot \mathbf{B} = 0$ condition.

Some refinements to the previous traditional views where proposed like the inner core and annular gaps by \cite{qiao_inner_2004} with some observational signatures shown by \cite{qiao_annular_2007}. Others used altitude-limited outer and slot gaps or low altitude slot gap models to better fit the light-curves especially for millisecond pulsars \citep{abdo_discovery_2010, venter_modeling_2012}.

So far most models relied on a centred dipole, which is a convenient and simple approximation of the magnetosphere because of its high degree of symmetry. Nevertheless, recently, interest has increased towards the consequences of an off-centred dipole that strongly affects the polar cap geometry and the light curves \citep{barnard_effect_2016, kundu_pulsed_2017}. Its polarization signature \citep{petri_polarized_2017} also differs from the rotating vector model \citep{radhakrishnan_magnetic_1969}. Such extensions of the standard picture should help in increasing the pair creation rate \citep{harding_pulsar_2011} and in adjusting the synchronisation between radio and high-energy light curves (explaining the time lag as seen by Fermi/LAT). This topology offers a natural explanation for the phase lag between thermal X-ray and radio emission as seen for instance in PSR~B1133+16 \citep{szary_xmm-newton_2017}.

In this paper, we study Newtonian as well as general-relativistic rotating dipoles and consider different prescriptions for particle motion in the vacuum electromagnetic field imposed by this dipole. We overcome several flaws from the usual aberration formula and corotating frame approach by considering a new frame where the electric field is parallel to the magnetic field and assuming that particles move along the common direction of $\mathbf{E}$ and $\mathbf{B}$ in that frame. It is well know that such frames always exist, whatever the electromagnetic field configuration, and that one solution has a speed strictly less than~$c$ \citep{gourgoulhon_relativite_2010} except for the null case where both electromagnetic field invariants vanish. This particular frame, which is not unique because any velocity component along the common direction of $\mathbf{E}$ and $\mathbf{B}$ can be added without modifying the electromagnetic field, does not require an intermediate corotating frame (let it be instantaneous or not) and therefore avoids divergent Lorentz factors. Note also that particles are not assumed to follow field lines as such lines are not easily or usefully defined in vacuum. Getting rid of the notion of corotation, we can use the traditional special relativistic aberration formula for light in whole space without trouble, going in principle to very large distances if needed. In order to highlight the differences induced by the different prescriptions for aberration which reflects in the particle motion, we show sky maps for the aberration used by \citet{dyks_two-pole_2003}, the instantaneous corotation frame technique by \citet{bai_uncertainties_2010} and our new assumption for particle motion valid in whole space (a generalization of \cite{mestel_stellar_1999} radiation drag view of particle motion). The plan of the paper is as follows. The magnetospheric emission models are exposed in Sec.~\ref{sec:Modele} detailing the gap models, the particle distribution functions, the aberration formulae and the general-relativistic electromagnetic field configuration. The shape of light-curves and the phase lag between radio and high-energy pulses are discussed in Sec.~\ref{sec:Results}. In Sec.~\ref{sec:Discussions} we propose a discussion about the influence of the total stellar electric charge on the sky maps. Conclusions and possible extensions are outlined in Sec.~\ref{sec:Conclusions}.

\section{Emission model}
\label{sec:Modele}

Whatever the underlying emission process truly operating in neutron star magnetospheres, there are some bottom lines that any model has to obey. Indeed, pulsar radiation models require three important but distinct and complementary ingredients, namely
\begin{itemize}
\item a geometrical description of the emission sites shaped by the prescribed magnetic field topology. We assume a rotating magnetic dipole evolving in vacuum for which an exact analytical solutions exist in flat spacetime and excellent numerical approximations have been computed for slowly rotating neutron star metrics. However, \cite{petri_multipolar_2017-1} has shown that frame-dragging is irrelevant and we neglect it in the remainder of this work by taking only the Schwarzschild part. The electric field is usually discarded or at best deduced from the infinite conductivity assumption. In vacuum, such relation does not hold and the full electric field solution for a rotating magnetic dipole is required.
\item a magnetospheric plasma configuration depicted by the dynamical properties of the radiating particles and their composition. In the simplest approach, the density follows the Goldreich-Julian expression \citep{goldreich_pulsar_1969}. A pair multiplicity factor~$\kappa$ could also be put into the picture. Moreover as a result of efficient acceleration, their kinetic energy repartition obeys a non-thermal distribution function relaxing to a power law. In the vacuum picture, we assume that the density is much less than the Goldreich-Julian value such that the current does not perturb significantly the electromagnetic field.
\item some radiation processes resulting from particle motion in the electromagnetic field. Synchrotron, curvature and inverse Compton mechanisms produce high and very high-energy photons up to the MeV/GeV range sometimes even to the TeV range. In this paper we do not specify the actual mechanisms producing the photons but the results are easily applied to any radiation fields although the spectra and pulse profile could slightly differ between them.
\end{itemize}
These three items are discussed in depth in the following paragraphs. Light bending by the stellar gravitational field is taken into account to produce sky maps. We then end this section by a discussion of the numerical algorithm used to produce pulsar light-curves.

\subsection{Electromagnetic topology}

The exact analytical expression for a magnetic dipole rotating in vacuum was given by \citet{deutsch_electromagnetic_1955}. We use his solution for the electromagnetic field topology outside the star, from the stellar surface up to the light-cylinder and beyond. A possible contribution from the wind, i.e. outside the light-cylinder could be considered but this is not touched in this work.

In order to describe properly the field, we fix the magnetic moment vector~$\bmu$ with respect to the rotation axis. The inclination angle between both axis is denoted by~$\chi$. Introducing a Cartesian coordinate system and the rotation rate as~$\Omega$ we have at any time~$t$
\begin{equation}
\bmu = \mu \, [ \sin \chi \, ( \cos (\Omega\,t) \, \ex + \sin (\Omega\,t) \, \ey ) + \cos \chi \, \ez ]
\end{equation}
where $(\ex, \ey, \ez)$ is the orthonormal basis of a Cartesian coordinate system. By convention at time $t=0$, the magnetic moment lies in the $(xOz)$ plane. The absolute value of the magnetic moment strength does not influence the emission process. For the remainder of the paper, we normalize it to $\mu=1$. The rotation rate can be conveniently normalized to the light-cylinder radius $\rlight=c/\Omega$ according to the ratio $R/\rlight$. It is always less than 0.1 corresponding to a 2~ms pulsar for standard neutron star parameters. A magnet rotating in vacuum does only produce radiation at its rotation frequency. To get a broad band spectrum, we still need to fill the magnetosphere with a relativistic plasma, although in a first step we neglect its back-reaction to the electromagnetic field.

The general-relativistic extension to the Deutsch solution has been given by a semi-analytical solution expanded into rational Chebyshev function \citep{petri_multipolar_2017-1} leading to generalized Hankel functions~$\mathcal{H}_\ell^{(1)}$ and is given for the aligned component of the magnetic field, with weight $\cos\chi$, by
\begin{subequations}
\label{eq:MagneticStatic}
\begin{align}
  \label{eq:MagneticStaticR}
  B^{\hat r}_\parallel & = - 6 \, B \, R^3 \, \left[ {\rm ln} \left( 1 - \frac{R_s}{r} \right) + \frac{R_s}{r} + \frac{R_s^2}{2\,r^2} \right] \, \frac{\cos\vartheta}{R_s^3} \\
  \label{eq:MagneticStaticT}
  B^{\hat \vartheta}_\parallel & = 3 \, B \, R^3 \, \left[ 2 \, \sqrt{ 1 - \frac{R_s}{r}} \, {\rm ln} \left( 1 - \frac{R_s}{r} \right) + \frac{R_s}{r} \, \frac{2\,r-R_s}{\sqrt{r\,(r-R_s)}} \right] \, \frac{\sin\vartheta}{R_s^3} \\
  B^{\hat \varphi}_\parallel & = 0
\end{align}
\end{subequations}
and for the perpendicular component, with weight $\sin\chi$, by
\begin{subequations}
\begin{align}
  \label{eq:DeutschEM}
  B^{\hat r}_\perp(\mathbf{r},t) & = \sqrt{\frac{3}{\pi}} \, \frac{f^{\rm B}_{1,1}(R)}{2\,r} \, \frac{\mathcal{H}^{(1)}_1(k\,r)}{\mathcal{H}^{(1)}_1(k\,R)} \, \sin \vartheta \, e^{i\,\psi} \\
  B^{\hat \vartheta}_\perp(\mathbf{r},t) & = \sqrt{\frac{3}{\pi}} \, \frac{f^{\rm B}_{1,1}(R)}{4} \, \left[ \frac{\alpha}{r} \, \frac{\frac{d}{dr} \left( r \, \mathcal{H}^{(1)}_1(k\,r) \right)}{\mathcal{H}^{(1)}_1(k\,R)} + \frac{\tilde\omega \, \tilde\omega_R \, R}{\alpha \, \alpha_R^2\,c^2} \, \frac{\mathcal{H}^{(1)}_2(k\,r)}{\frac{d}{dr} \left( r \, \mathcal{H}^{(1)}_2(k\,r) \right) |_{R}} \right] \,  \cos \vartheta \, e^{i\,\psi} \\
  B^{\hat \varphi}_\perp(\mathbf{r},t) & = \sqrt{\frac{3}{\pi}} \, \frac{f^{\rm B}_{1,1}(R)}{4} \, \left[ \frac{\alpha}{r} \, \frac{\frac{d}{dr} \left( r \, \mathcal{H}^{(1)}_1(k\,r) \right)}{\mathcal{H}^{(1)}_1(k\,R)} + \frac{\tilde\omega \, \tilde\omega_R \, R}{\alpha \, \alpha_R^2\,c^2} \, \frac{\mathcal{H}^{(1)}_2(k\,r)}{\frac{d}{dr} \left( r \, \mathcal{H}^{(1)}_2(k\,r) \right) |_{R}} \, \cos 2\vartheta \right] \, i \, \, e^{i\,\psi} .
\end{align}
\end{subequations}
When performing similar calculations for the electric field, we find for the aligned component, with weight $\cos\chi$,
\begin{subequations}
\begin{align}
  \label{eq:Drot1}
  D^{\hat r} & = - \varepsilon_0 \, \frac{B\,R^4}{R_s^3} \,
  \frac{\tilde{\omega}_R}{\alpha_R^2} \, C_1 \, C_2 \, \left[ \left( 3
      - 4\,\frac{r}{R_s} \right) \, \ln\alpha^2 + \frac{R_s^2}{6\,r^2}
    + \frac{R_s}{r} - 4 \right] \, ( 3\,\cos^2\vartheta - 1 ) \\
  D^{\hat \vartheta} & = 6 \, \varepsilon_0 \, \frac{B\,R^4}{R_s^3} \,
  \frac{\tilde{\omega}_R}{\alpha_R^2} \, \alpha \, C_1 \, C_2 \, \left[
    \left( 1 - 2\,\frac{r}{R_s} \right) \, \ln\alpha^2 - 2 -
    \frac{R_s^2}{6\,r^2\,\alpha^2} \right] \, \cos\vartheta \, \sin\vartheta \\
  D^{\hat \varphi} & = 0
\end{align}
\end{subequations}
where 
\begin{subequations}
\begin{align}
  \label{eq:C1}
  \alpha_R & = \sqrt{ 1 - \frac{R_s}{R} }\\
  \omega_R & = \frac{a\,R_s\,c}{R^3} \\
  \tilde{\omega}_R & = \Omega - \omega_R \\
  C_1 & = \ln\alpha_R^2 + \frac{R_s}{R} + \frac{R_s^2}{2\,R^2} \\
  C_2 & = \left[ \left( 1 - 2\,\frac{R}{R_s} \right) \, \ln\alpha_R^2 - 2 -
  \frac{R_s^2}{6\,R^2\,\alpha_R^2} \right]^{-1}
\end{align}
\end{subequations}
and for the perpendicular radiating part, with weight $\sin\chi$,
\begin{subequations}
\begin{align}
 D^{\hat r}_\perp & = \frac{3}{4} \, \sqrt{\frac{3}{\pi}} \, \varepsilon_0 \, f^{\rm B}_{1,1}(R) \, \frac{\tilde\omega_R\,\mathcal{H}^{(1)}_2(k\,r)}{\alpha_R^2 \, \frac{d}{dr} \left( r \, \mathcal{H}^{(1)}_2(k\,r) \right) |_R } \, \sin2\vartheta \, \, e^{i\,\psi} \\
 D^{\hat \vartheta}_\perp & = \sqrt{\frac{3}{\pi}} \, \frac{e^{i\,\psi}}{4} \, \varepsilon_0 \,f^{\rm B}_{1,1}(R) \, \left[ \frac{\alpha \, \tilde\omega_R \, R}
  {r \, \alpha^2_R} \, \frac{\frac{d}{dr} \left( r \, \mathcal{H}^{(1)}_2(k\,r)\right)}{\frac{d}{dr} \left( r \, \mathcal{H}^{(1)}_2(k\,r) \right) |_R } \, \cos2\vartheta - \frac{\tilde\omega}{\alpha} \, \frac{\mathcal{H}^{(1)}_1(k\,r)}{\mathcal{H}^{(1)}_1(k\,R)} \right] \\
 D^{\hat \varphi}_\perp & = \sqrt{\frac{3}{\pi}} \, \frac{i\,e^{i\,\psi}}{4} \, \varepsilon_0 \,f^{\rm B}_{1,1}(R) \, \left[ \frac{\alpha \, \tilde\omega_R \, R}
  {r \, \alpha^2_R} \, \frac{\frac{d}{dr} \left( r \, \mathcal{H}^{(1)}_2(k\,r)\right)}{\frac{d}{dr} \left( r \, \mathcal{H}^{(1)}_2(k\,r) \right) |_R } - \frac{\tilde\omega}{\alpha} \, \frac{\mathcal{H}^{(1)}_1(k\,r)}{\mathcal{H}^{(1)}_1(k\,R)} \right] \, \cos\vartheta .
\end{align}
\end{subequations}
See also \cite{rezzolla_general_2001} and  \cite{rezzolla_electromagnetic_2004} for similar formalism and expressions about general-relativistic rotating dipoles in vacuum. Expressions for multipolar rotating fields have also been computed up to $\ell=4$. For the aligned multipolar case, magnetic field lines satisfy the following equation 
\begin{equation}
 |r\,f_{\ell,0}^{\rm B} \, \sin\vartheta \, \partial_\vartheta Y_{\ell,0}| = cste
\end{equation}
which is the same as the one given by \cite{gonthier_general_1994} for $\ell=1$.

The light-cylinder radius in Schwarzschild spacetime~$\rlight^{\rm GR}$ is defined by the location where the corotation speed reaches the speed of light for a local observer with his own clock ticking with proper time $d\tau = \alpha \, dt$. There the speed of light is reached for $r\,\Omega=\alpha\,c$ leading to a cubic equation 
\begin{equation}
 \frac{r^2}{\rlight^2} + \frac{\Rs}{r} - 1 = 0.
\end{equation}
An approximate solution two second order in $\Rs/\rlight$ is given by
\begin{equation}
 \rlight^{\rm GR} \approx \rlight \, \left( 1 - \frac{1}{2} \, \frac{\Rs}{\rlight} - \frac{3}{8} \, \frac{\Rs^2}{\rlight^2} \right).
\end{equation}
We will use this expression for the light-cylinder radius in general relativity. Polar cap shapes and separatrix locations are computed according to this value.

\subsection{Particle distribution}

There is still no general and accepted consensus on the plasma chemical composition: pure electron-positron pairs, or adorn with a fraction of protons and/or ions. Nevertheless, heavy elements like nucleons radiate much less than light elements like leptons because emissivity strongly depends on the inverse of the mass of each species. It is therefore very likely to consider only electron-positron pairs that are accelerated in the electromagnetic field present around the polar caps and in the separatrix. We assume that they reach a stationary state where acceleration is perfectly compensated by radiation reaction. Their distribution function is therefore described by a power law in energy with a index~$p$ such that 
\begin{equation}
\label{eq:DistributionParticule}
f(\mathbf{r}, \gamma) = K(\mathbf{r}) \, \gamma^{-p}
\end{equation}
expressed in the rest frame of the fluid. $(\mathbf{r}$ represents the position vector and $\gamma$ the Lorentz factor in the rest frame of the fluid. This function is actually subject to a low and a high energy cut-off but we restrict ourselves to the global dependence on the power law index~$p$ only. $K(\mathbf{r})$ is related to the density of leptons in the magnetosphere. We could choose a variation following Goldreich-Julian density as a first guess with a possible pair multiplicity. But as a starting point, we consider more general density functional variations such that $K$ depends on distance as $r^{-q}$. Particles flow in the above prescribed electromagnetic field but how should they shine?

\subsection{Radiation properties}

There are several ways for particles to produce photons. We could consider any radiation mechanism easily implemented in the code. However, for our pure geometric consideration of light-curve production, studying a generic emission process is enough. Looking for broadband spectra and phase-resolved polarization properties would certainly require to consider a specific high-energy emission process. This quantitative study will not touched within the present work.

We assume an isotropic distribution of pitch angle in the comoving frame, therefore the emissivity reproduces the same isotropic pattern. In order to get the emissivity in the inertial frame, we need to perform a Lorentz boost from the rest frame (not necessarily the corotating frame) to the observer frame. This implies a Doppler factor
\begin{equation}
\label{eq:Doppler}
\mathcal{D} = \frac{1}{\Gamma\,(1-\bbeta\cdot\mathbf{n}_{\rm obs})}
\end{equation}
where $\bbeta=\mathbf{v}/c$ is the particle velocity in the inertial frame normalized to the speed of light, $\Gamma=(1-\beta^2)^{-1/2}$ the associated bulk Lorentz factor and
\begin{equation}
 \mathbf{n}_{\rm obs} = \sin\zeta \, \ex + \cos\zeta \, \ez
\end{equation}
the direction of the line of sight, underlying an angle $\zeta$ with respect to the rotation axis. In a Cartesian coordinate system, 
\begin{equation}
\bbeta\cdot\mathbf{n}_{\rm obs} = \beta_x \, \sin\zeta + \beta_z \, \cos\zeta 
\end{equation}
and in the most appropriate spherical coordinate system
\begin{subequations}
\begin{align}
\bbeta\cdot\mathbf{n}_{\rm obs} & = \beta_r \, f_r + \beta_\vartheta \, f_\vartheta + \beta_\varphi \, f_\varphi\\
f_r(\zeta, \vartheta,\varphi) & = \sin\zeta \, \sin\vartheta \, \cos\varphi + \cos\zeta\,\cos\vartheta\\
f_\vartheta(\zeta, \vartheta,\varphi) & = \sin\zeta \, \cos\vartheta \, \cos\varphi -\cos\zeta\,\sin\vartheta \\
f_\varphi(\zeta, \vartheta,\varphi) & = - \sin\zeta \, \sin\varphi .
\end{align}
\end{subequations}
The spectral shape depends on the radiation mechanism: curvature, synchrotron or inverse Compton. For concreteness we assume a formal dependence reminiscent of curvature radiation. In that case, with the particle distribution function given in the rest frame by eq.~(\ref{eq:DistributionParticule}), the emissivity perceived by an inertial observer becomes
\begin{equation}
\label{eq:EmissiviteCourbure}
j_{\rm cur}(\omega) \propto \mathcal{D}^{(p+5)/3} \omega^{-(p-1)/3} .
\end{equation}
The power law function is auto-similar in the sense that the beaming factor is independent of the frequency~$\omega$. Therefore we expect the same pulse profiles at different energies but with different relative intensities as a consequence of the factor $\omega^{-(p-1)/3}$. Pulse profiles are sensitive to relativistic beaming, impacted by the index~$p$. In a more realistic spectrum, the local slope and therefore the index~$p$ depends on the frequency~$\omega$. For typical Fermi/LAT pulsars in the gamma-ray range, the emissivity shows a sub-exponential cut-off \citep{abdo_second_2013} and is approximated locally by an increasing index~$p$. This means that beaming becomes more effective at higher energies and therefore pulses become thinner. This will be demonstrated later. Synchrotron emissivity leads to a slightly different functional dependence as
\begin{equation}
\label{eq:EmissiviteSynchrotron}
j_{\rm syn}(\omega) \propto \mathcal{D}^{(p+3)/2} \omega^{-(p-1)/2}
\end{equation}
but qualitatively with the same evolution with frequency~$\omega$. Inverse-Compton emission beamed to the observer frame shows another dependence on the Doppler factor. We should also distinguish between the Thomson and the Klein-Nishina regime leading to very different slopes for emissivity. Such studies are left for future work when an accurate broadband spectrum is required to fit data of particular pulsars.

\subsection{Emission sites}

Although our approach can deal with any shape of emission regions, we focus on the three standard sites: polar cap, slot gap and outer gap. We next explain how such regions are delimited in space to set the weights for emissivity.

\subsubsection{Polar cap}

\begin{figure}
\centering
\input{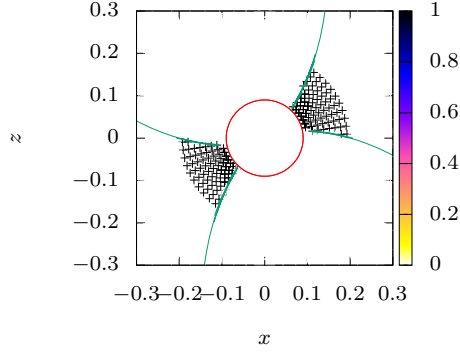}
\caption{Variation of emissivity following the weight defined for the polar caps. Black crosses show locations where photons are produced.}
\label{fig:poids_calotte}
\end{figure}

Polar caps are supposed to host the site of radio photon production. The altitude of emission, constrained by radio observations, seems to range from several stellar radii up to a substantial fraction of the light-cylinder, about 10\% \citep{mitra_comparing_2004, mitra_meterwavelength_2016}. In our prescription for emissivity, we distinguish between two cases. The first emission pattern forces photons to be emitted radially outwards and not embracing the electromagnetic field topology. This should mimic the thermal X-ray radiation from the hot spots on the surface by considering only the peak intensity for an isotropic pattern. The second emission rule forces photons to propagate tangentially to the particle motion in the observer frame at their launching position. This second option represents the traditional view about coherent radio emission from pulsars. In order to compute radio emission from the poles, integration must be done in a volume around the polar caps. Two options are envisaged: a sharp cut in the emissivity when moving away from field lines not rooting in the polar cap region or a smoother transition between maximal emission at the centre of the polar cap and very weak emission when crossing the last open field line. The first option simply switches from 100\% to 0\% whereas the second option decay from 100\% to 0\% within a transition layer that is set by the physics, here by the user. The latter is achieved by introducing two functions shaping the polar volumes according to
\begin{equation}
 f_{\rm pc}^\pm = 1 \pm (\cos\chi\,\cos\vartheta + \sin\chi\,\sin\vartheta\,\cos\varphi) .
\end{equation}
These functions are comprised in the interval $[0,1]$, vanishing exactly along the magnetic axis. The associated weight~$w_{\rm pc}$ for emission close to the polar cap is given by
\begin{equation}
\label{eq:poids_calotte}
 w_{\rm pc} = e^{-(x-1)^2 / \sigma_{\rm pc}^2} \, \left( e^{{f_{\rm pc}^+}^2/\Delta_{\rm pc}^2} + e^{{f_{\rm pc}^-}^2/\Delta_{\rm pc}^2} \right)
\end{equation}
where $x=r/R$. The smooth polar cap boundaries in eq.~(\ref{eq:poids_calotte}) do not closely follow the dichotomy between open and closed field lines. These are controlled by the thickness $\Delta_{\rm pc}$. This weight is maximal at the stellar surface $x=1$ when $f_{\rm pc}^+=0$ or $f_{\rm pc}^-=0$, which means at the north or south magnetic pole respectively. $\sigma_{\rm pc}\,R$ is the typical height of emission and $\Delta_{\rm pc}$ the transversal size of the transition from radio emission to extinction. Eventually, we already emphasize that the first option requires much less parameters to get at the end very similar sky maps. Thus we decided to only show results for the sharp boundary emission sites. Fig.~\ref{fig:poids_calotte} shows an example of weight in the $(x,z)$ plane used to simulate polar cap emission for an obliquity $\chi=60 \degr$.
For details about the possible dynamics of polar cap particle acceleration in general relativity we refer to \cite{zanotti_particle_2012} and to \cite{morozova_explaining_2014} for their relation with drifting subpulses.

\subsubsection{Slot gap}

\begin{figure}
\centering
\input{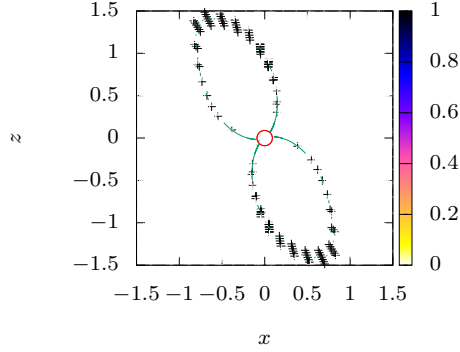}
\caption{Variation of emissivity following the weight defined for the slot gap model. Black crosses show locations where photons are produced.}
\label{fig:poids_slot_gap}
\end{figure}

High-energy emission must be put at higher altitude in order to circumvent the photon absorption process in a too strong magnetic field \citep{erber_high-energy_1966}. A commonly used acceleration gap where radiation leaves the star is the slot gap. It is a thin layer sticking on the last open field line surface, the so-called separatrix. Emission is maximal on this separatrix and decreases monotonically when moving out of this surface. Defining the distance to the separatrix~$h$ by the minimal distance between a field line and the light-cylinder and normalizing to the light-cylinder radius by $x_{\rm sg}=h/\rlight$, the slot gap weight is conveniently written as
\begin{equation}
\label{eq:poids_sg}
w_{\rm sg} = e^{- x_{\rm sg}^2 / \sigma_{\rm sg}^2} 
\end{equation}
where $\sigma_{\rm sg}\,\rlight$ is the typical thickness of the layer and $\sigma_{\rm r}\,\rlight$ the extinction depth when approaching the light-cylinder. A simplified version would consider weights to be 1 when $|h|<\sigma_{\rm r}\,\rlight$ and zero otherwise. Here also we opt for the sharp boundary region as both a qualitatively equivalent. An illustration of the weight is shown in Fig.~\ref{fig:poids_slot_gap} for $\chi=60 \degr$.

\subsubsection{Outer gap}

\begin{figure}
\centering
\input{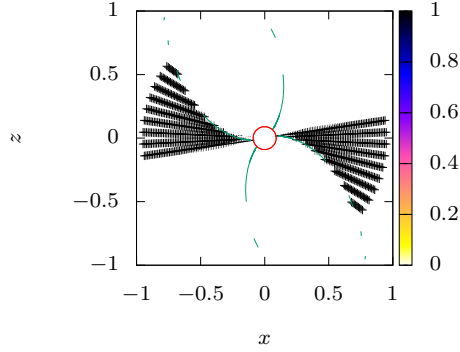}
\caption{Variation of emissivity following the weight defined for the outer gaps. Black crosses show locations where photons are produced.}
\label{fig:poids_outer_gap}
\end{figure}

Outer gaps are defined by the volume between the null surface where $\mathbf{\Omega} \cdot \mathbf{B}=0$ and the light-cylinder \citep{cheng_energetic_1986}. Smooth and sharp versions are considered here too. A smooth version could look like
\begin{equation}
\label{eq:poids_og}
w_{\rm og} = 
\begin{cases}
e^{- h_{\rm og}^2 / \sigma_{\rm og}^2} \textrm{ if } B_z>0 \\
0 \textrm{ if not}
\end{cases}
\end{equation}
where $h_{\rm og}=\rho_{\rm max}-\rlight$ is the maximal cylindrical distance between the light-cylinder and a given field line passing through the outer gap and $\sigma_{\rm og}\,\rlight$ is the typical thickness of the layer. But we restrict here again on the sharp boundary volume which is shown in Fig.~\ref{fig:poids_outer_gap} for an obliquity $\chi=60 \degr$ and easier to define.
Photons produces within these emission sites leave the magnetosphere following straight lines in the Newtonian view or curved trajectories when general relativity is included. The initial direction of propagation of light in the inertial frame depends on the assumptions about emissivity first defined in the corotating frame of the star or immediately in the inertial frame. The aberration formula for light differs in both cases. Starting with a description of photon production in the corotating frame leads to severe problems at the light-cylinder and beyond because no such frame exists at large distance. Staying in the inertial frame avoids such complications if emission is properly defined as we now show.

\subsection{Aberration}

By assumption, in some models, particles follow magnetic field lines in the corotating frame. Their distribution function is isotropic in the rest frame of the fluid. Following previous prescriptions by \cite{bai_uncertainties_2010}, we assume that their Lorentz factor~$\Gamma$ is constant in the observer frame such that the velocity which is a combination between propagation along field lines and corotation at speed 
\begin{equation}
\mathbf{V}_{\rm c} = \mathbf{\Omega} \wedge \mathbf{r}
\end{equation}
is
\begin{equation}
\label{eq:AberrationCorot}
\mathbf{v} = v_\parallel^{\rm c} \, \mathbf{t} + \mathbf{V}_{\rm c}
\end{equation}
where $\mathbf{t} = \pm \mathbf{B}/B$ is the outward pointing tangent to the field line. Solving for $v_\parallel^{\rm c}$ the only positive solution as long as $v> V_{\rm c}$ is
\begin{equation}
\label{eq:SolutionVparallel}
v_\parallel^{\rm c} = - \mathbf{t} \cdot \mathbf{V}_{\rm c} + \sqrt{(\mathbf{t} \cdot \mathbf{V}_{\rm c})^2 + v^2 - V_{\rm c}^2} .
\end{equation}
Knowing the velocity, we get the Doppler factor for radiation as explained in eq.~(\ref{eq:Doppler}). This velocity field assumes that the electric field vanishes in the corotating frame. But this requires a large amount of plasma to screen the electric field, in contradiction with the vacuum assumption we made here and also in \cite{bai_uncertainties_2010}. Therefore, the aberration formula eq.~(\ref{eq:AberrationCorot}) can only be an approximation in our case. Moreover, this approximation also fails at sufficiently large distance because the Deutsch field solution possesses a magnetic field structure for which the polo\"\i dal component does not decay fast enough with respect to the toroidal component. Real solutions to eq.(\ref{eq:SolutionVparallel}) do not exists at several $\rlight$ because the square root becomes negative. Indeed, taking an orthogonal rotator, it can be shown that in the equatorial plane the term in the square root of eq.~(\ref{eq:SolutionVparallel}) tends to $v^2-4\,c^2<0$ for $r\rightarrow+\infty$ on the spiral given by $\varphi + k\,r - \Omega\,t = \upi/2$. In the most favorable case for which $v=c$, it actually becomes negative already at the light-cylinder. Using the corotation frame does not help to go beyond the light-cylinder in vacuum.

Another aberration formula was originally used by \citet{dyks_two-pole_2003} to switch from the corotating frame to the observer frame, the usual textbook expression between two observers moving with constant relative velocity with respect to each other. It reads
\begin{equation}
\label{eq:AberrationLorentz}
 \mathbf{n} = \dfrac{1}{\mathcal{D}} \, \left[ \mathbf{n}' + \Gamma \, \left( \dfrac{\Gamma}{\Gamma+1} \, ( \bbeta \cdot \mathbf{n}' ) + 1 \right) \, \bbeta \right] .
\end{equation}
We will compare both expressions when computing phase-plots and field line projections. However, note that in the true aberration formula, the electric field even in the corotating frame must be taken into account.

The concept of magnetic field line in vacuum is misleading and specifying motion along a particular field line is not well defined in the general case. This requires some caution about the interpretation of the corotation speed $\mathbf{V}_{\rm c}$. Another way to follow the particle trajectory replaces this velocity by a special frame in which the electric field is parallel to the magnetic field. The velocity $\mathbf{\beta}_\parallel \, c$ required by the Lorentz transform to get this condition is
\begin{equation}
\frac{\mathbf{\beta}_\parallel}{1+\beta_\parallel^2} = \frac{c \, \mathbf{E} \wedge \mathbf{B}}{E^2 + c^2\, B^2}
\end{equation}
neglecting all other curvature, gradient and polarization drifts in the limit of vanishing Larmor radius which is correct in a super strong magnetic field. In that frame, motion is along the common direction of $\mathbf{E}'$ and $\mathbf{B}'$. To get the useful solution, we write $\mathbf{V}_\parallel = \alpha \, \mathbf{E} \wedge \mathbf{B}$. The constant $\alpha$ is the solution given by
\begin{equation}
 \alpha = \frac{E^2+c^2\,B^2 - \sqrt{\mathcal{I}_1^2 + 4\,c^2\,\mathcal{I}_2^2}}{2 \, (\mathbf{E} \wedge \mathbf{B})^2}
\end{equation}
where we introduced both relativistic electromagnetic invariants as $\mathcal{I}_1=E^2-c^2\,B^2$ and $\mathcal{I}_2 = \mathbf{E} \cdot \mathbf{B}$. The minus sign in front of the square root enforces a speed less than that of light. The electric and magnetic field in the frame moving at speed $\mathbf{V}_\parallel$ are found by a special-relativistic Lorentz boost of the electromagnetic field and gives
\begin{subequations}
\begin{align}
 \mathbf{E}' & = \Gamma \, [ ( 1 - \alpha \, B^2 ) \, \mathbf{E} + \alpha \, ( \mathbf{E} \cdot \mathbf{B}) \mathbf{B} ] \\
 \mathbf{B}' & = \Gamma \, [ ( 1 - \alpha \, E^2/c^2 ) \, \mathbf{B} + \alpha \, ( \mathbf{E} \cdot \mathbf{B}) \mathbf{E}/c^2 ] .
\end{align}
\end{subequations}
In this frame, particles move along the common direction of $\mathbf{E}'$ and $\mathbf{B}'$. Thus the local tangent to the trajectory becomes $\mathbf{t}'_\parallel = \pm \mathbf{E}'/E' = \pm \mathbf{B}'/B'$, the sign being chosen such that particles flow outwards. Therefore we replace $\bbeta$ by $\mathbf{V}_\parallel/c$ in eq.~(\ref{eq:AberrationLorentz}) to get a velocity field that should not be confused or seen as motion along field lines because this concept is usually ill defined for non-ideal plasmas when $\mathbf{E} \cdot \mathbf{B} \neq 0$. Our expression for the particle velocity resembles to the aristotelian expression given by \cite{gruzinov_aristotelian_2013}. Our velocity prescription is however more general because we do not assume that particles travel exactly at the speed of light. The speed along the common $\mathbf{E}$ and $\mathbf{B}$ direction is unconstrained and fixed by the ``user'' contrary to aristotelian electrodynamics. Indeed \cite{gruzinov_aristotelian_2013} introduced two new quantities $E_0>0$ and $B_0$ according to the following invariants
\begin{subequations}
\begin{align}
 \mathcal{I}_1 & = E^2 - c^2 \, B^2 = E_0^2 - c^2 \, B_0^2 \\
 \mathcal{I}_2 & = \mathbf{E} \cdot \mathbf{B} = E_0 \, B_0 .
\end{align}
\end{subequations}
Solving for the magnetic field~$B_0$, we get
\begin{equation}
 B_0^2 = \frac{-\mathcal{I}_1 \pm \sqrt{\mathcal{I}_1^2 + 4\,c^2\,\mathcal{I}_2^2}}{2\,c^2} .
\end{equation}
Only the solution with a positive sign~$+$ is real. In such a way, the quantity~$\alpha$ can be expressed as
\begin{equation}
 \alpha = \frac{c^2\,(B^2-B_0^2)}{(\mathbf E \wedge \mathbf B)^2}
\end{equation}
which is also usefully expressed as
\begin{equation}
 \alpha = \frac{c^2}{E_0^2 + c^2 \, B^2} = \frac{c^2}{E^2 + c^2 \, B_0^2} .
\end{equation}
Plugging this expression into the electromagnetic field we find
\begin{subequations}
\begin{align}
 \mathbf E' & = \alpha \, \Gamma \, E_0 \, \left[ \frac{E_0}{c^2} \, \mathbf E + B_0 \, \mathbf B \right] \\
 \mathbf B' & = \alpha \, \Gamma \, B_0 \, \left[ B_0 \, \mathbf B + \frac{E_0}{c^2} \, \mathbf E \right] .
\end{align}
\end{subequations}
They are therefore colinear because $E_0 \, \mathbf B' = B_0 \, \mathbf E'$. The frame velocity consequently simplifies into
\begin{equation}
 \mathbf V = \frac{\mathbf E \wedge \mathbf B}{E_0^2/c^2 + B^2} .
\end{equation}
If particles exactly move at the speed of light, in the comoving frame this velocity becomes $\mathbf v'= \pm \mathbf E'/E' = \pm \mathbf B'/B'$, the sign depending on the charge. Note also that $E'=E_0$ and $B'=B_0$. Doing the Lorentz transformation to the observer frame, noting that $\mathbf V$ and $\mathbf v'$ are orthogonal, this is nothing but aristotelian electrodynamics. Our treatment is more general because we do not enforce light speed in this frame.

\subsection{Ray tracing in Schwarzschild metric}

Generalization of Deutsch solution to curved space-time of slowly rotating neutron stars has been given by \cite{petri_multipolar_2017-1}. The electromagnetic topology used in this paper is extracted from these semi-analytical expressions. Thus in order to keep our investigation self-consistent, photons have to be subject to bending, time delay and gravitational redshift. In the present study, we only take into account light bending because time delay is negligible and we do not consider spectral properties thus neglect also photon reddening. Full consideration of the three effects will be investigated in detail in a forthcoming paper. Moreover, frame dragging does not impact neither on the electromagnetic field nor on the photon trajectories. We therefore decided to keep only the Schwarzschild metric as a representative geometry around neutron stars. This approximation improves for slowly rotating pulsar with period higher that several tenths of milliseconds. Ray tracing techniques around black holes have been developed by many authors \citep{vincent_gyoto:_2011, psaltis_ray-tracing_2012, chan_gray:_2013}. Basically two different approaches are used. The first one integrates the equations of motion starting from an initial position and with fixed constants of motion. This is usually easy to implement but becomes inaccurate for large distances and computationally expensive. The second approach integrates analytically the trajectories that are then given as integrals to be computed by any quadrature method. The latter is generally faster and more accurate for any distance but more involved for arbitrary motion \citep{rauch_optical_1994}. As we have to integrate millions of photon paths we prefer the second quadrature technique.

Photon path integration in Schwarzschild metric follows straightforwardly from the computation of 
\begin{equation}
 \Delta \varphi = \pm \int_{r_0}^r \frac{b \, dr}{r^2 \, \sqrt{1 - \left( 1 - \frac{R_{\rm s}}{r} \right) \, \frac{b^2}{r^2} }}
\end{equation}
where $r_0$ is the initial radius of the photon, $r$ its final radius and $\Delta \varphi$ the variation in position angle from $r_0$ to $r$. $b$ is the impact parameter defined by
\begin{equation}
 b = \frac{r_0 \, \sin \xi}{\sqrt{1-\Rs/r_0}} .
\end{equation}
and $\xi$ the emission angle with respect to the radial direction. The associated time of flight is then
\begin{equation}
 c\,\Delta t = \pm \int_{r_0}^r \frac{dr}{\left( 1 - \frac{R_{\rm s}}{r} \right) \, \sqrt{1 - \left( 1 - \frac{R_{\rm s}}{r} \right) \, \frac{b^2}{r^2} }} .
\end{equation}
The $\pm$ sign corresponds to a receding~$(-)$ or a distancing~$(+)$ photon.

\subsection{Numerical algorithm}

The time-dependent intensity received by a distant observer is computed by integration of the spatially and temporally varying emissivity. This three-dimensional spatial integration is performed via a Fourier-Chebyshev expansion of the integrand on a structured grid in spherical polar coordinates. The emissivity is computed in coefficient space in order to evaluate it at any time and any point in space, even between the azimuthal grid points through a very high order interpolation scheme. Series summation is greatly enhanced by the Clenshaw technique as explained in \citet{press_numerical_2007}. The domain of integration is a spherical shell with inner radius equal to the stellar radius~$R$ and an outer radius~$R_{\rm out}$ not necessarily less than the light-cylinder radius.

The distant (inertial) observer frame is set up with a Cartesian coordinate system ($x,y,z$). A point $M$ in space is represented by its spherical coordinates ($r,\vartheta,\varphi$) such that 
\begin{equation}
\mathbf{r} = \sin\vartheta\,\cos\varphi\,\ex + \sin\vartheta\,\sin\varphi\,\ey + \cos\vartheta\,\ez .
\end{equation}
The retarded time~$t_{\rm ret}$ for a distant observer $D\gg r$ is then expressed in Cartesian coordinates according to
\begin{equation}
 t_{\rm ret} \equiv t_{\rm obs} + \frac{\mathbf n_{\rm obs} \cdot \mathbf r}{c} = t_{\rm obs} + \frac{r}{c} \,f_r(\zeta, \vartheta,\varphi) 
\end{equation}
where $t_{\rm obs}$ is the time of emission as measured by the observer. Because the magnetosphere rotates at a constant angular speed~$\Omega$ we can evaluate the emissivity at any azimuth by shifting in phase according to $\varphi'=\varphi-\Omega\,t_{\rm ret}$. Calculations are done through a summation of the Fourier series in $\varphi'$, deduced from the Fourier series in $\varphi$. This technique is very versatile and can accommodate any particle distribution function, any electromagnetic topology and use any radiation mechanism in general relativity.

\section{Results}
\label{sec:Results}

We now discuss in depth the consequences of the emission regions and radiation properties as seen in the light curves extracted from several aberration formulae and particle distribution functions in flat and Schwarzschild space-time.

\subsection{Polar cap shape: flat vs curved spacetime}

The central region on which all magnetospheric emission models rely on is the polar cap. It is therefore crucial to determine accurately this surface on the neutron star. The polar caps are defined by the location of the feet of the last closed field lines on the stellar surface. It is an intersection, in the geometrical sense, between a curve and a surface (field line with a sphere). This shape must not be confused with the image given by photons escaping from the polar cap rims tangentially to field lines. We will see in the next section that both shapes do not agree, the first one being a geometric locus whereas the second one being an image of this geometric locus.

First, to check that our code gives the correct polar cap rims, we compare results from the special case of a static dipole with exact analytical expressions. It is indeed possible to get simple but exact analytical expressions for the polar cap shape for an aligned and a perpendicular rotator. It is well know that for the aligned case the polar cap is a perfect circle and the opening angle is
\begin{equation}
\label{eq:thetaPCaligned}
\vartheta_{\rm pc} = \arcsin \sqrt{\frac{R}{\rlight}} \approx \sqrt{\frac{R}{\rlight}}.
\end{equation}
This is the usual estimate used even for oblique and orthogonal rotators. But in general the polar cap is far from a circle. For the perpendicular rotator, the boundary of the polar cap is given in the magnetic axis coordinate system where $\bmu$ is along $\ez$ by
\begin{equation}
\label{eq:thetaPCperp}
\vartheta_{\rm pc} = 
\begin{cases}
\arcsin \left[ \sqrt{\frac{R}{\rlight} \, |\sin\varphi|} \right] & \text{for } \cos2\varphi\leqslant1/3 \\
\arcsin \left[ \frac{\sqrt{2}}{3^{3/4}\,|\cos\varphi|} \, \sqrt{\frac{R}{\rlight}}  \right] & \text{for }  \cos2\varphi \geqslant 1/3
\end{cases}
\end{equation}
where $\varphi$ is the longitudinal angle in the coordinate system attached to the magnetic axis. Analytical shapes of polar caps have been computed by many authors. For completeness we report again some shapes associated to a static and a rotating dipole in appendix~\ref{app:PolarCapShape} for several inclination angles of the dipole.

In this paper, we are interested in the polar cap variations induced by general-relativistic effects. These include
\begin{itemize}
\item the magnetic topology in curved space-time.
\item the radius of the light-cylinder due to gravitational time dilation.
\item the light bending effect as seen by a distant observer.
\end{itemize}
To keep the space-time metric simple but without renouncing to accuracy, we use the Schwarzschild metric in Boyer-Lindquist coordinates. In a previous work, we have shown that frame-dragging has no impact on the magnetic topology. Thus the only free parameter is the Schwarzschild radius or in other words the compactness of the neutron star.

\begin{figure}
\centering
\begin{tabular}{cc}
\resizebox{0.5\textwidth}{!}{\input{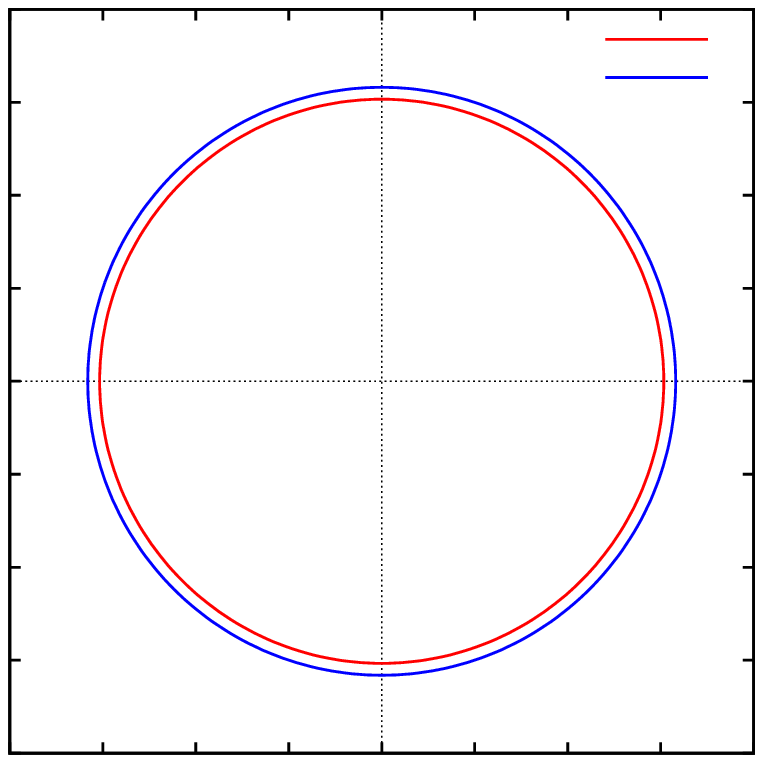}} &
\resizebox{0.5\textwidth}{!}{\input{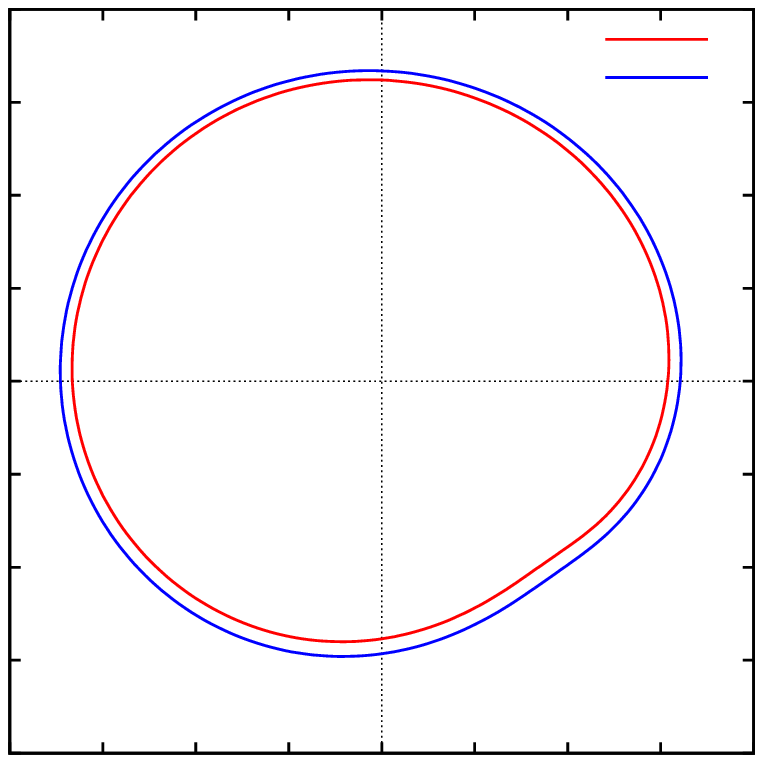}} \\
\resizebox{0.5\textwidth}{!}{\input{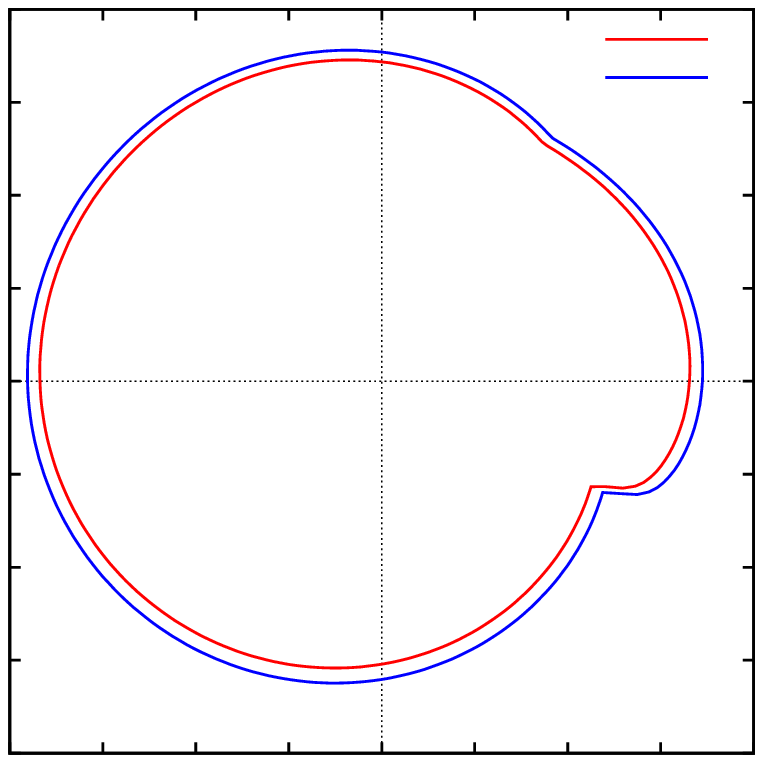}} &
\resizebox{0.5\textwidth}{!}{\input{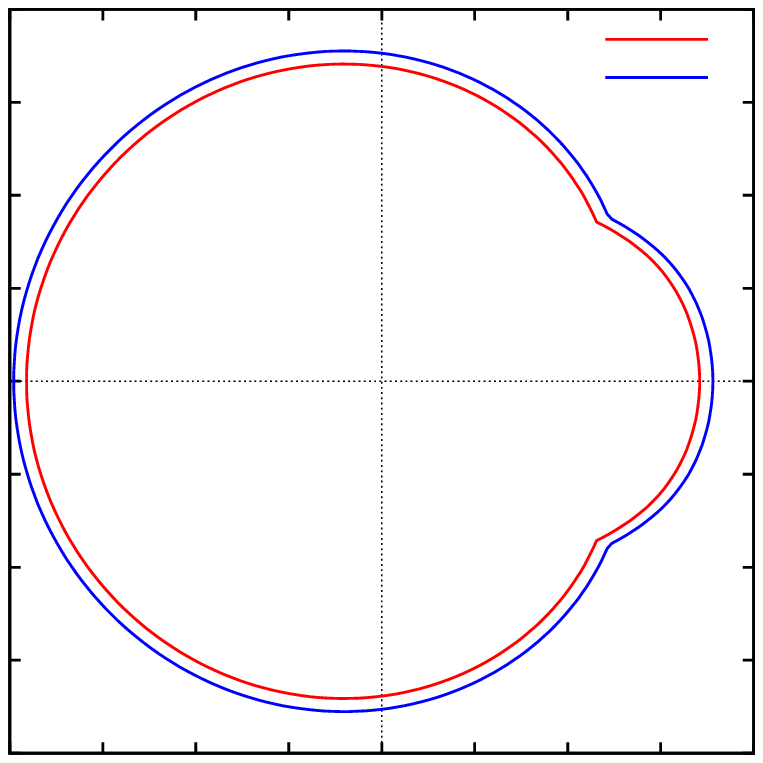}}
\end{tabular} 
\caption{Polar cap shape for a rotating dipole in GR compared to Newtonian space for four obliquities $\chi\in\{0,30,60,90\degr\}$.}
\label{fig:PolarCapDeutschGRvsN}
\end{figure}

As a starting point, to be as comprehensive as possible, we compare polar cap geometrical shapes in Newtonian and GR for a rotating dipole of arbitrary inclination angle~$\chi$. Fig.~\ref{fig:PolarCapDeutschGRvsN} shows the rim of one polar cap for $\chi\in\{0\degr, 30\degr, 60\degr, 90\degr\}$. GR always reduces the size and the surface of the polar cap.

\begin{figure}
\centering
\resizebox{0.5\textwidth}{!}{\input{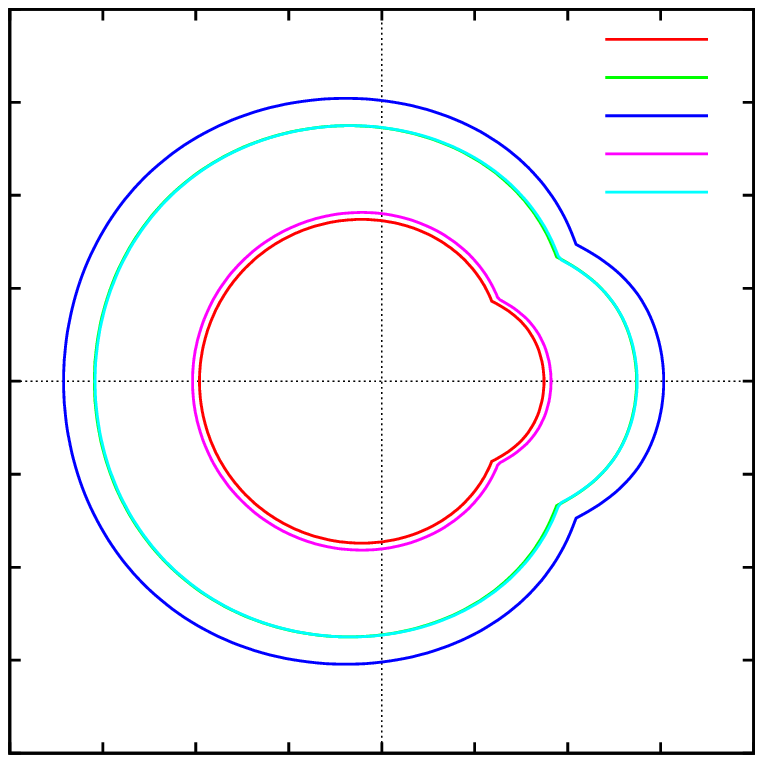}}
\caption{Polar cap shape for the orthogonal dipole in GR compared to Newtonian space. ``pc'' stand for the usual polar cap shape. ``gr pc'' stands for GR shape including magnetic topology and light-cylinder radius. ``ph'' and ``gr ph'' include propagation of photons tangentially to magnetic field lines in flat and curved spacetime respectively. ``gr ph b'' represents the bending of light in GR without any counterpart in Newtonian theory. The origin of the plot is centred on the magnetic axis.}
\label{fig:PolarCapCHI90DeutschGRvsN}
\end{figure}

Next we show the discrepancies between Newtonian and general-relativistic polar cap models in fig.\ref{fig:PolarCapCHI90DeutschGRvsN}. The magenta ``pc'' curve depicts the usual polar cap in flat (Newtonian) space-time. The red ``gr pc'' curve corresponds to the GR counterpart including magnetic topology and light-cylinder radius variations corresponding to the case $90\degr$ in fig.~\ref{fig:PolarCapDeutschGRvsN}. We note a decrease in the rim size due to GR. These rims are the geometric locus of the root of the last closed field lines. Photons are assumed to be emitted tangentially to these field lines. Thus the corresponding image of the polar caps are shown in fig.\ref{fig:PolarCapCHI90DeutschGRvsN} as ``ph'' in cyan for flat and ``gr ph'' in green for Schwarzschild space-time. Both images are almost identical. The change in the direction of the tangent to the field seems to compensate almost exactly for the variation in the magnetic field structure. The last blue curves denoted as ``gr ph b'' includes light bending effects that have no counterpart in Newtonian gravity. Bending opens up slightly more the final direction of propagation with respect to the direction of the local tangent. As a consequence, the polar cap image inflates. In these plots, for the GR case, we use a compactness of $\Rs/R=0.4$.

\begin{figure}
\centering
\resizebox{0.5\textwidth}{!}{\input{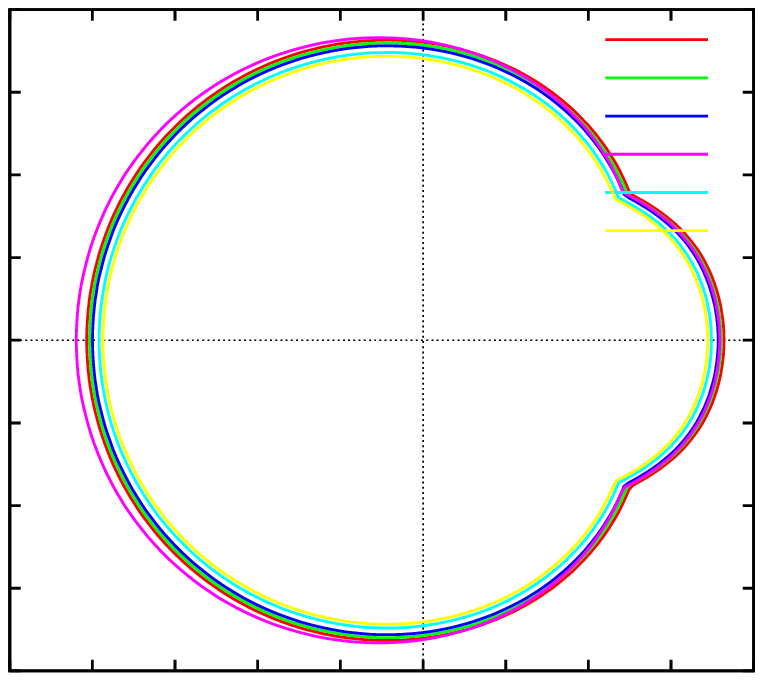}}
\caption{Polar cap shape deformation according to an increasing compacity of the neutron star from Newtonian gravity to very compact stars, $\Rs/R=\{0,0.1,0.2,0.3,0.4,0.5\}$. The origin of the plot is centred on the magnetic axis.}
\label{fig:PolarCapCHI90DeutschCompacite}
\end{figure}

Next we investigate the impact of the compactness on the polar cap size. Some illustrative examples are shown in fig.~\ref{fig:PolarCapCHI90DeutschCompacite} for typical compacity of $\Rs/R=\{0,0.1,0.2,0.3,0.4,0.5\}$. The cap size decreases with increasing compactness. However we have not seen a monotonic shrinking of the rim with respect to the compacity. For instance, the trailing and the leading side of the cap do not show the same behaviour with increasing compacity. Nevertheless, the variations are contained in a small range less than 10\%.

\begin{figure}
\centering
\resizebox{0.5\textwidth}{!}{\input{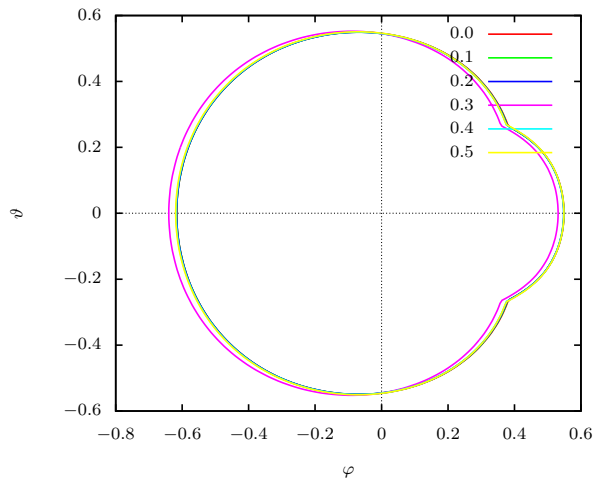}}
\caption{Tangent to the magnetic field line with increasing compacity of the neutron star $\Rs/R=\{0,0.1,0.2,0.3,0.4,0.5\}$. The origin of the plot is centred on the magnetic axis.}
\label{fig:TangentPolarCapCHI90DeutschCompacite}
\end{figure}

Photons are emitted tangentially to the local magnetic field lines. The influence of the compactness on the polar cap image when this change in the direction of propagation is including is shown in fig.~\ref{fig:TangentPolarCapCHI90DeutschCompacite} for the same compacity parameters as in the previous figure. The trend in the polar cap image with compactness is not clearly established. The only appreciable effect is a slight shift of the cap centre towards the leading side (in the direction of rotation).

\begin{figure}
\centering
\resizebox{0.5\textwidth}{!}{\input{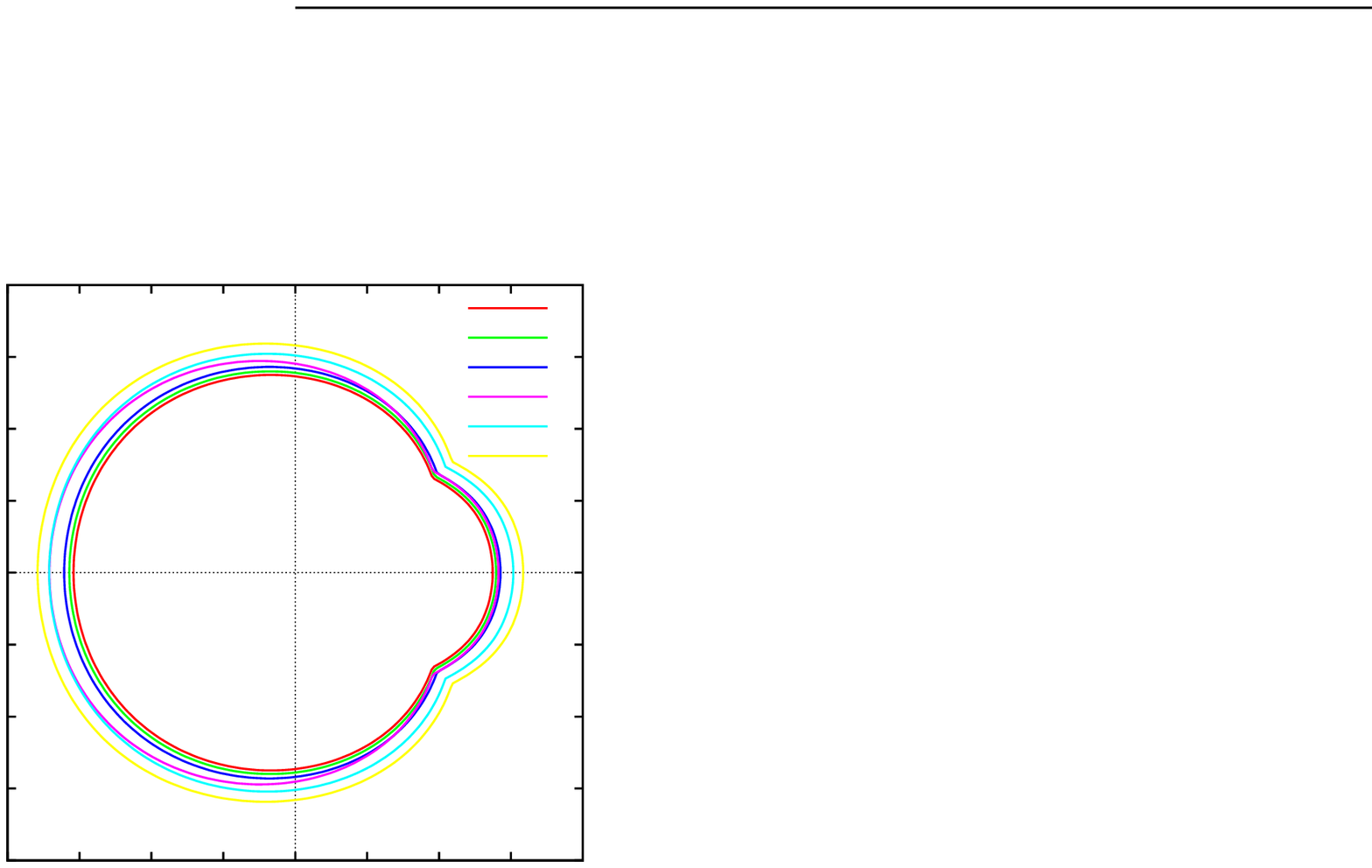}}
\caption{Photon propagation direction taking into account light bending with increasing compacity of the neutron star $\Rs/R=\{0,0.1,0.2,0.3,0.4,0.5\}$. The origin of the plot is centred on the magnetic axis.}
\label{fig:PhotonCHI90DeutschCompacite}
\end{figure}

In order to be fully consistent with our GR description, light bending must be taken into account. Photons firstly propagating tangentially to field lines are bend when moving to the observer located at very large distance. The compactness impact on the polar cap image is shown in fig.~\ref{fig:PhotonCHI90DeutschCompacite} for the same compacity parameters as in the previous figures. The polar cap image monotonically increases with increasing compactness. Stronger gravity bends more the photon paths towards direction opposite to the radial vector because of its attractive nature. The inflation reaches about 15\% in diameter for a compactness of $\Rs/R=0.5$.

This last plot corresponds to the combined effects of all GR corrections and represents the basic picture to understand the set of sky maps we now review in some details. Note that we assumed emission initially tangent to the local magnetic field line but this is generally not required. It is possible to compute light bending for any initial direction of emission. We next use this versatility to compute light-curve for any aberration formula.

\subsection{Sky maps}

A fixed pulsar geometry, meaning essentially obliquity~$\chi$ and gap models, leads to very different pulse profile depending on the inclination of the line of sight~$\zeta$. In order to capture the influence of $\zeta$, the full set of light curves is summarized in sky maps showing intensity maps with respect to pulsar rotation phase normalized to its period on the x-axis and inclination of line of sight on the y-axis. A relevant set of sky maps is given in the following figures including each six panels representing different aberration effects in Newtonian and GR approximations. The first formula corresponds to aberration in an inertial frame using the Lorentz transform (LBA for Lorentz Boost Aberration), the second takes into account aberration in the instantaneous corotating frame (CFA for Corotating Frame Aberration) and the last uses aberration arising from a velocity in the direction common to the electric and magnetic field in a special frame boosted to the observer frame (CDA for Common Direction Aberration).

Quantitative emission properties depend on the particle distribution power law index~$p$ and on their Lorentz factor $\gamma$. We use a constant spatial particle density number although any prescribed varying density could be used. For the sake of brevity, plots are given for an index $p=0$ and $p=2$ and for a Lorentz factor of $\gamma=10$.

\begin{figure*}
\centering
\resizebox{0.95\textwidth}{!}{\input{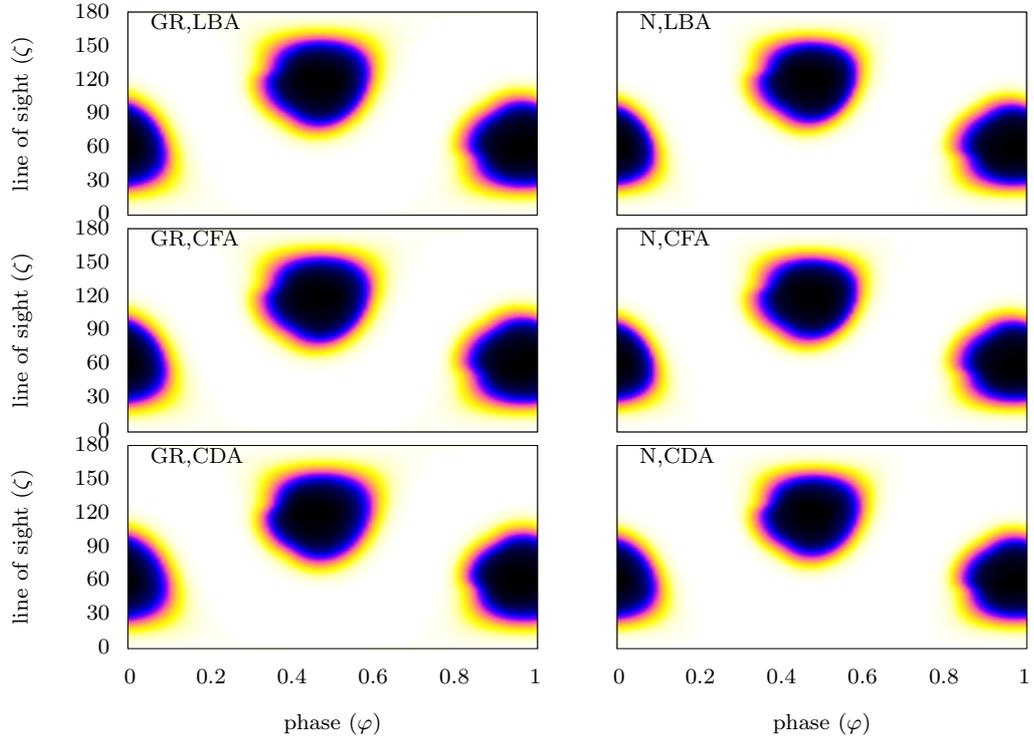}} 
\caption{Sky maps for the polar caps with obliquity~$\chi=60\degr$, power law index~$p=2$ and Lorentz factor~$\gamma=10$. In each panel N stands for Newtonian and GR for general-relativistic cases. LBA means Lorentz Boost Aberration, CFA means Corotating Frame Aberration and CDA means Common Direction Aberration.}
\label{fig:phase_plot_m1_c60_g10_p2_q0_qe0}
\end{figure*}

Fig.~\ref{fig:phase_plot_m1_c60_g10_p2_q0_qe0} shows sky maps for the polar cap model in the six regimes: with and without gravity correspondingly each with three aberration formulae. Emission is maximum around the centre of each polar cap as expected. Although our emission volume is sharply delimited by a step function, due to smearing arising from relativistic beaming, aberration and retardation effects, light curves appear smooth at any angle. This explains our choice not to use more complicated boundaries for polar cap. As we show later, the same smearing is used as a benefit for slot gap and outer gap models. The smoothing introduced by propagation effects shows that the emissivity to be integrated is a smooth function of space and time. It can therefore be integrated with Fourier-Chebyshev techniques without loss of accuracy and without Gibbs phenomenon provided the resolution is fine enough. From a practical point of view, we started with a low resolution and then increased it iteratively by a factor of two until the phase plots have converged. Starting with $N_{\rm r} \times N_\vartheta \times N_\varphi=33\times32\times64$, a resolution of $N_{\rm r} \times N_\vartheta \times N_\varphi=257\times256\times512$ gives good results for any index~$p$. However, high Lorentz factors require higher resolution because the features in the light-curves become sharper. Recall that relativistic beaming smears a profile with an angular opening of roughly $1/\gamma$.

Close to the stellar surface, all aberration prescriptions lead to the same sky maps. Discrepancies between them becomes apparent only when photons are predominantly produced close to the light-cylinder. However, GR sky maps result in larger spots compared to Newtonian approximations. This is reminiscent to light-bending effects as explained in the last plot of the previous section, fig.\ref{fig:phase_plot_m1_c60_g10_p2_q0_qe0}.

\begin{figure*}
\centering
\resizebox{0.95\textwidth}{!}{\input{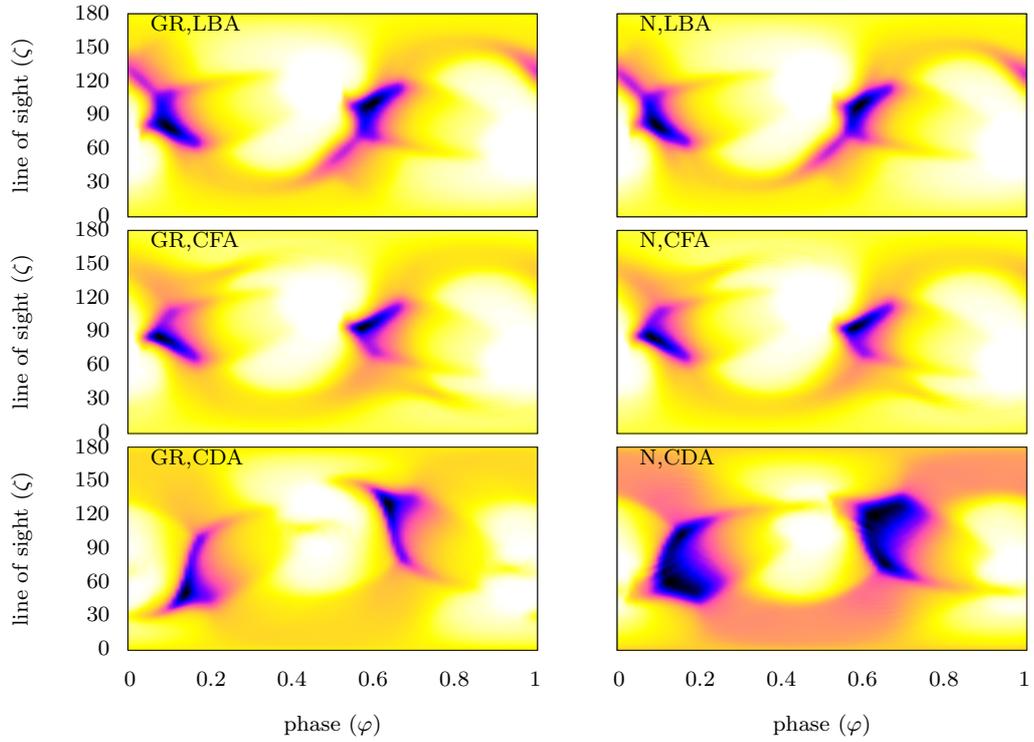}} 
\caption{Sky maps for the slot gap with the same parameters as in fig.~\ref{fig:phase_plot_m1_c60_g10_p2_q0_qe0}.}
\label{fig:phase_plot_m2_c60_g10_p2_q0_qe0}
\end{figure*}

Fig.~\ref{fig:phase_plot_m2_c60_g10_p2_q0_qe0} shows sky maps for the slot gap model in the very same six regimes. The diagrams share the same characteristic for all three aberration prescriptions. LBA and CFA show very similar trends with respect to the line of sight inclination angle. There is however a slight shift in the maximum of intensity. N-CDA forms an S-shape curve different from both previous diagrams. Close to the light-cylinder, the electric field becomes comparable to the magnetic field ($E \sim r/\rlight\,c\,B$) and therefore the aberration angle can significantly differ from one prescription to the other. In CDA, the electric field acts on the particle velocity and therefore also on the photon propagation direction. Therefore, light-curves can be significantly impacted by the total electric charge of the system. This is explored in more details in the discussion of Sec.~\ref{sec:Discussions}. GR tends to narrow the pulse profiles in CDA prescription as seen by comparing GR-CDA with N-CDA. Also some variations are observed between GR-LBA and N-LBA but the changes are less drastic. The same conclusion applies to GR-CFA with respect to N-CFA.

\begin{figure*}
\centering
\resizebox{0.95\textwidth}{!}{\input{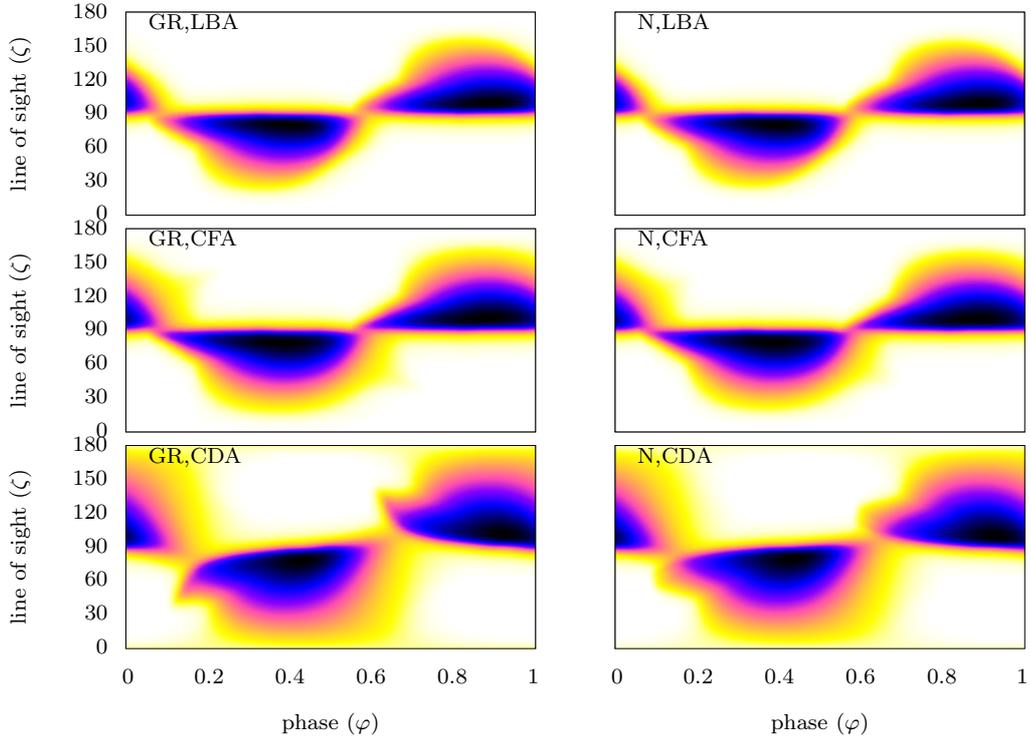}} 
\caption{Sky maps for the slot gap with the same parameters as in fig.~\ref{fig:phase_plot_m1_c60_g10_p2_q0_qe0}.}
\label{fig:phase_plot_m3_c60_g10_p2_q0_qe0}
\end{figure*}

Fig.~\ref{fig:phase_plot_m3_c60_g10_p2_q0_qe0} shows sky maps for the outer gap model again in the same six regimes. LBA shows an extension of relevant emission the smallest compared to CFA and CDA. The largest pulse profiles are seen in the CDA case. Not surprisingly, when photons are produced closer to the surface, the discrepancy between aberration formulas remains weak because the electric field is only of the order $E\sim R \, \Omega \,B$. But when most of emission comes from regions close to the light-cylinder as in outer gaps, the electric field is of the order $E\sim c\,B$ and deviations are clearly observed. This fact shows how sensitive light-curves can be according to the assumption for individual particle motion in the external electromagnetic field. We think that in vacuum, our CDA prescription is most appropriate for representing realistic particle trajectories. Note that emission is not sharply cut at $\zeta=90\degr$ as would be the case in previous works using single photon trajectories. Because of our finite Lorentz factor, and due to relativistic beaming, there is a spread in value of $\zeta$ around $90\degr$ where emission still occurs. This spread decreases when the Lorentz factor increases because of the $1/\gamma$ opening angle estimate. Moreover, in CDA, the line $\zeta=90\degr$ does not separate the emission from the two hemispheres sharply because $\mathbf{E}$ contributes significantly to the gaps only defined by the $\mathbf{B}$ field.

Whatever the emission process, the transition between faint radiation and maximum radiation occurs in a layer of thickness $1/\gamma$ due to relativistic beaming. Pulse widths are thus slightly wider than for the limiting case $\gamma\to+\infty$. Therefore, increasing the Lorentz factor of the particles will sharpen the sky maps diagrams if the thickness of the layer is less than a typical length scale of $\rlight/\gamma$.

Previous works assumed that photons are emitted exactly in the direction of motion of the emitting particle as observed in the inertial frame. In our case, this would correspond to an infinite Lorentz factor $\gamma=+\infty$. Notice however that an exact comparison with previous results is prohibited by the fact that we assume constant spatial emissivity whereas when considering single photon emission, emissivity is usually assumed constant along field lines that diverge due to their dipolar nature. This could be corrected for by introducing a complicated correction factor in our emissivity function. Nevertheless such careful comparisons are useless because none of the current model can account for the plethora of light-curve diversity.

\begin{figure*}
\centering
\resizebox{0.95\textwidth}{!}{\input{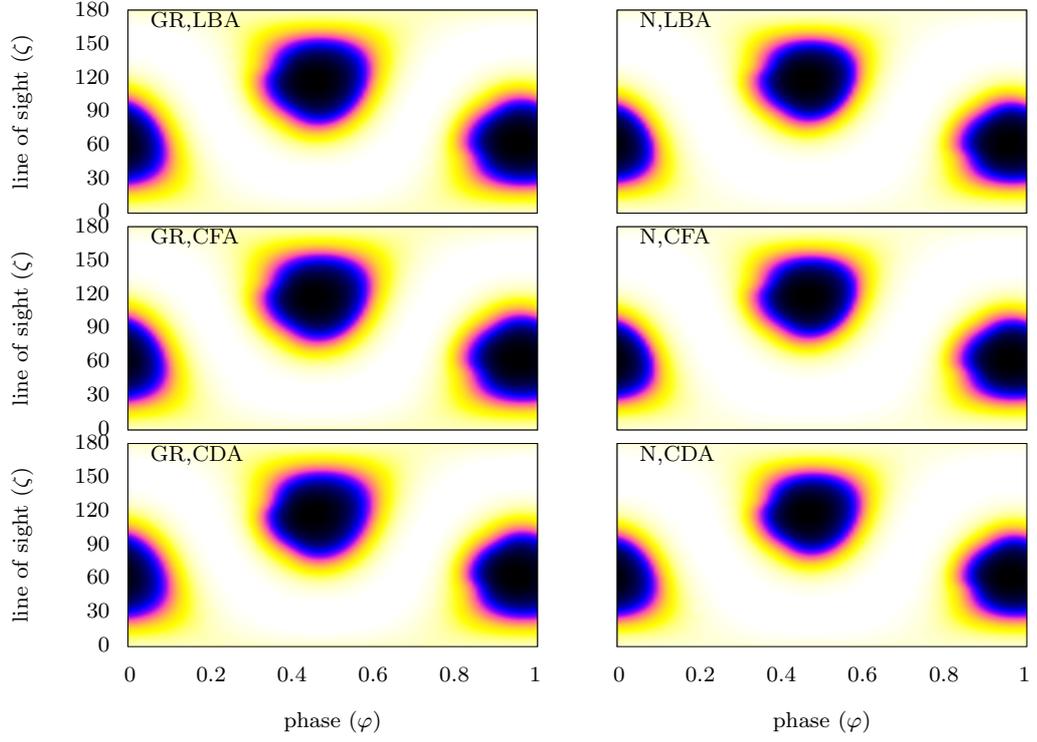}} 
\caption{Sky maps for the polar caps with obliquity~$\chi=60\degr$, power law index~$p=0$ and Lorentz factor~$\gamma=10$. The reminder is the same as in fig.~\ref{fig:phase_plot_m1_c60_g10_p2_q0_qe0}.}
\label{fig:phase_plot_m1_c60_g10_p0_q0_qe0}
\end{figure*}

Other power law indices~$p$ for the particle distribution directly impact on the beaming factor due to eq.~(\ref{eq:EmissiviteCourbure}) and eq.~(\ref{eq:EmissiviteSynchrotron}). Increasing the index~$p$ reinforces the relativistic beaming. Somehow it then changes the sky maps diagram in the same way as an increase in the Lorentz factor that also favours beaming. This is seen by inspection of Fig.~\ref{fig:phase_plot_m1_c60_g10_p0_q0_qe0} showing sky maps for the polar cap model in the same six regimes as before. The parameters for these plots are now a smaller power law index of $p=0$ but still the same particle Lorentz factor $\gamma=10$. The sky maps are very similar to the $p=2$ case except that the contrast between off pulse and peak pulse is less pronounced owing t a weaker relativistic beaming effect.

\begin{figure*}
\centering
\resizebox{0.95\textwidth}{!}{\input{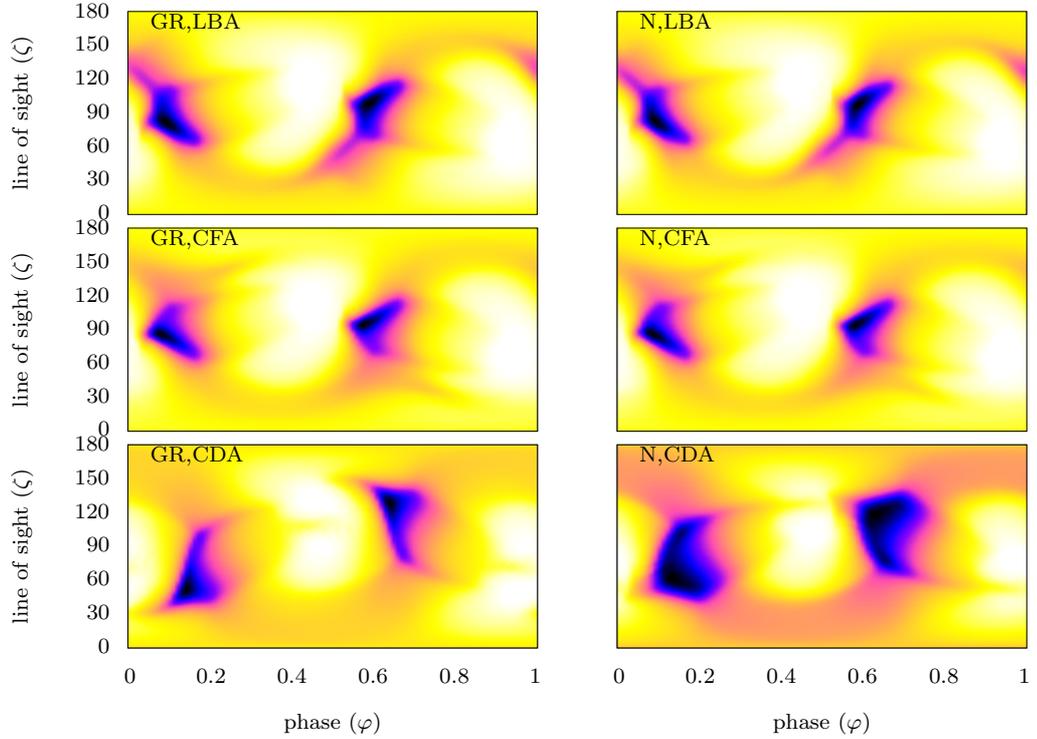}} 
\caption{Sky maps for the slot gaps with the same parameters as in fig.\ref{fig:phase_plot_m1_c60_g10_p2_q0_qe0}.}
\label{fig:phase_plot_m2_c60_g10_p0_q0_qe0}
\end{figure*}

\begin{figure*}
\centering
\resizebox{0.95\textwidth}{!}{\input{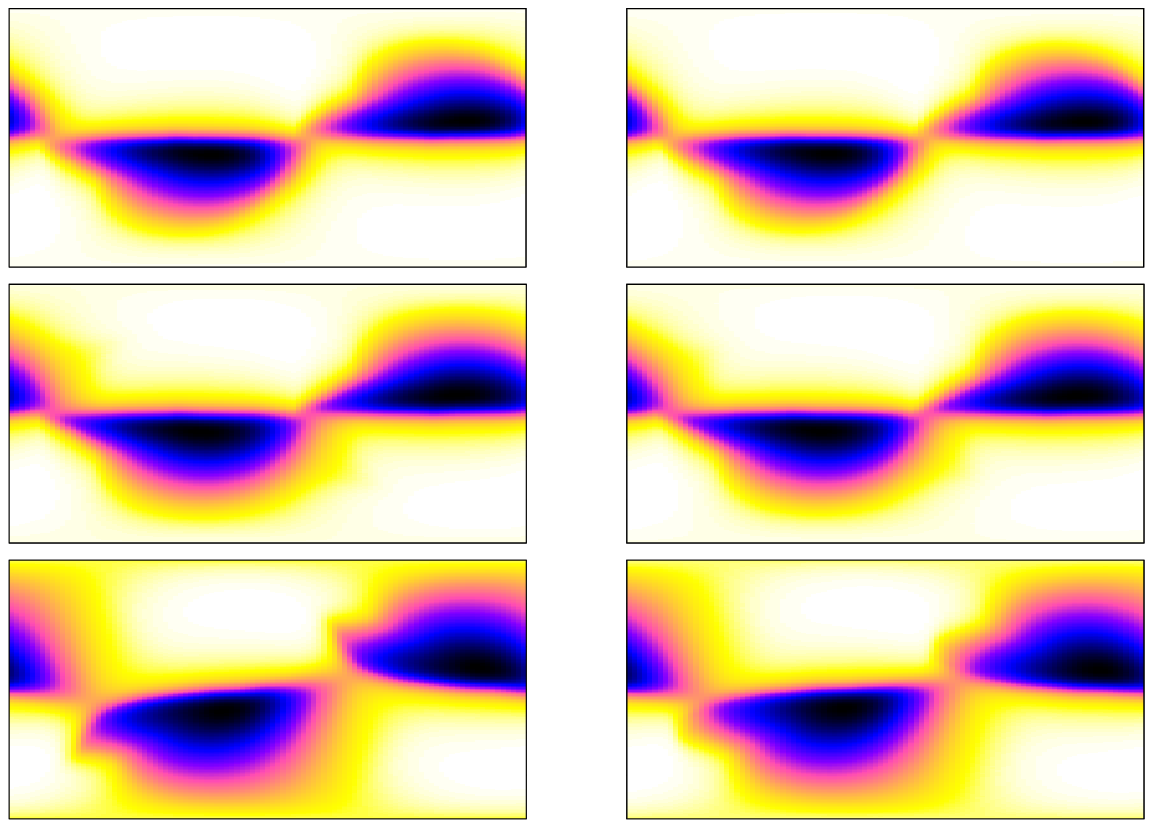}} 
\caption{Sky maps for the outer gaps with the same parameters as in fig.\ref{fig:phase_plot_m1_c60_g10_p2_q0_qe0}.}
\label{fig:phase_plot_m3_c60_g10_p0_q0_qe0}
\end{figure*}

Fig.~\ref{fig:phase_plot_m2_c60_g10_p0_q0_qe0} shows sky maps for the slot gap model and fig.~\ref{fig:phase_plot_m3_c60_g10_p0_q0_qe0} for the outer gap model with $p=0$ and $\gamma=10$. The maps are similar but the contrast between off pulse and peak pulse is fainter than for the case $p=2$ for which beaming effects are sharper.

A varying power-law index~$p$ is meant to mimic a variation in the spectral energy density (SED) of the emission. Indeed, the beaming effect strongly depends on the Doppler factor $\mathcal{D}$ to a given power fixed by the local slope of the photon spectrum. Without specifying the radiation mechanism at hand, it suffices to change the index~$p$ in order to investigate the impact of the SED onto the sky maps. This is what we did succinctly in this paragraph.

\subsection{Light-curves comparison}

\begin{figure}
\centering
\resizebox{0.95\textwidth}{!}{\input{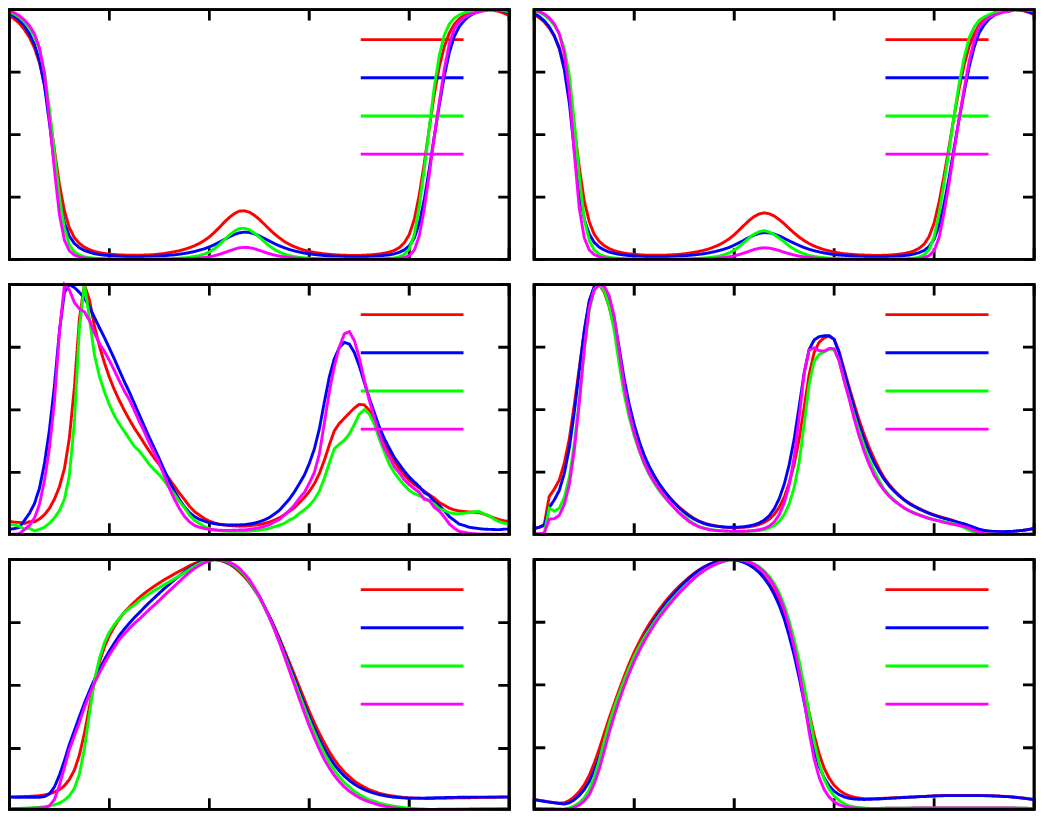}}
\caption{Light-curves for the polar cap (PC), slot gap (SG) and outer gap (OG) with $\chi = 60\degr, \zeta = 70\degr$, $\gamma = 10$ and the LBA and CDA aberration in Newtonian an GR space-times. The power law index is $p=\{0,2\}$.}
\label{fig:courbe_lumiere_mx_fx_c60_g10_px_q0_qe0}
\end{figure}

To better assess the impact of the different assumptions on emission properties, we compare some light-curves for the three aberration cases and the three emission sites in Newtonian and GR space-times.
A sample of relevant results are shown in Fig.~\ref{fig:courbe_lumiere_mx_fx_c60_g10_px_q0_qe0} for the polar cap (PC) in the upper panel, for the slot gap (SG) in the middle panel and for the outer gap (OG) in the lower panel. The left column assumes CDA whereas the right column assumes LBA. Minkowskian metric is depicted with N and general-relativistic metric with GR. The generic parameters are an obliquity $\chi=60\degr$, particle Lorentz factor $\gamma=10$ and power law index $p=\{0,2\}$ for all aberration prescriptions. The inclination of the line of sight is $\zeta=70\degr$. The different colours segregate between metric (M) and index (P) with the format (M,P) in the upper right legend.
For polar cap models, we observe a discrepancy between Newtonian and GR light-curve mainly due to the strong light-bending effect that is maximal at the stellar surface. As a corollary, the light curves (shape and peak intensity) also depend on the power law index. Increasing $p$ augments the ratio between the bright and the faint pulse.

For the polar cap, because the corotation speed is much less than~$c$, the difference between LBA and CFA (not shown in the figures) is irrelevant. Both light curves overlap and are not distinguished by eye. However, for CDA, we notice a slight delay in the pulse profile, although it looks very similar to the two previous one. For slot gaps, the pulse profile, amplitude and width, depends on the power law index $p$ in Newtonian and GR metric. We also observe a phase shift in the peak intensity of the leading pulse. This is clearly noticed in CDA although less obvious in LBA. For outer gap models, all prescriptions and metric assumptions give only slightly different profiles. GR and power law index variations do not significantly alter the profiles except for a slight spreading in CDA with respect to LBA.

The difference between LBA/CFA and CDA is most prominent in the slot gap models where the light-curve shapes depend on the space-time approximation and on the power law index~$p$. In GR, the high-energy pulses are sharper compared to Newtonian metric. This conclusion is valid irrespective of the index~$p$. For outer gaps, differences are visible but less pronounced than in slot gaps.

\subsection{Combined radio and high-energy light-curves}

To close this study, having in mind a joint radio and high-energy fitting of gamma-ray pulsar light curves, we investigated the multi-wavelength phase-resolved pulse profiles. Multi-wavelength observations will help to constrain the location of radio and high-energy emission via the time lag between their respective pulses. As an example, a small sample of combined radio and high-energy pulse profiles is summarized, showing the variety of possible configurations. The pulsar obliquity, the observer line of sight and the emission sites could be unveiled by such analysis.

\begin{figure}
\centering
\input{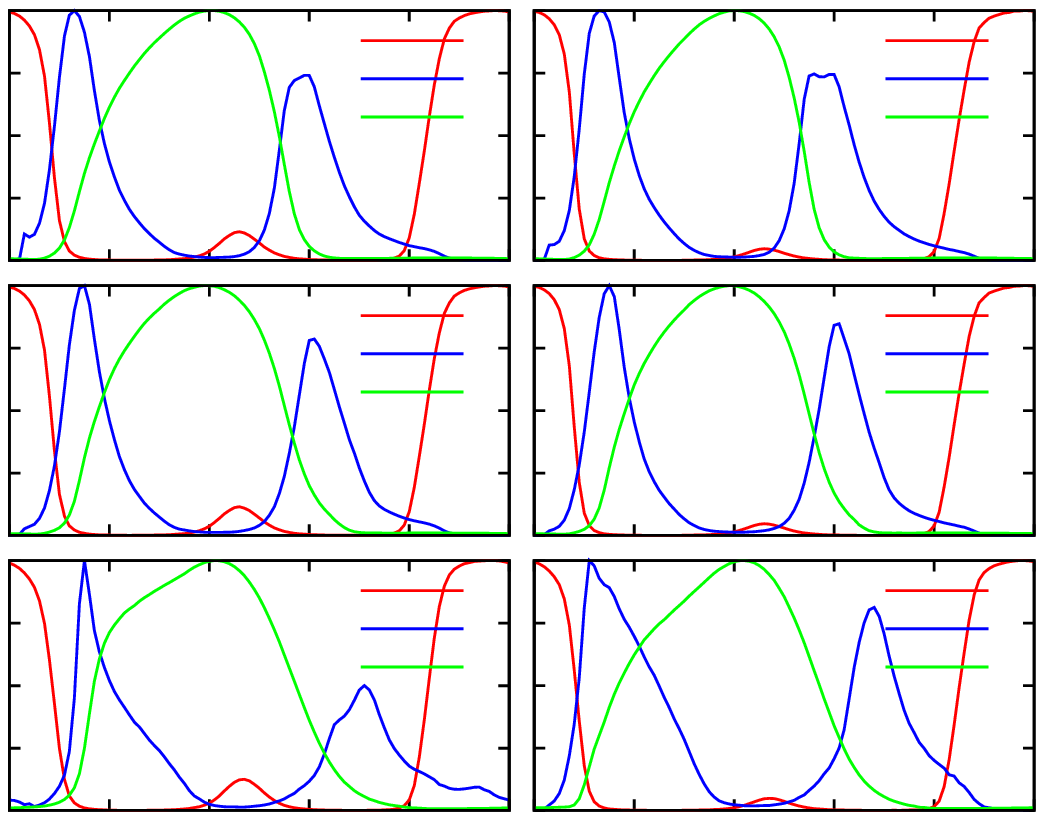}
\caption{Multiwavelength light-curves for the polar cap, the slot gap and the outer gap with $\chi=60\degr$, $\zeta=70\degr$, $\gamma=10$ and $p=2$.}
\label{fig:courbe_lumiere_multi_lambda_mx_fx_c60_g10_p2_q0_qe0}
\end{figure}

Fig.~\ref{fig:courbe_lumiere_multi_lambda_mx_fx_c60_g10_p2_q0_qe0} shows a sample of light-curves extracted for $\chi=60\degr, \zeta=70\degr$, $\gamma=10$ and $p=2$ for radio photons coming from the polar cap, in red, the high-energy photons coming from the slot gap, in green, and from the outer gap, in blue. The CFA is used. The radio pulse profiles almost anti-align with the unique outer gap pulse, both intensities reaching maximum value around a phase lag of $0.4$. Looking at the polar cap and outer gap sky-maps respectively in fig.~\ref{fig:phase_plot_m1_c60_g10_p0_q0_qe0} and in fig.~\ref{fig:phase_plot_m3_c60_g10_p0_q0_qe0}, such large misalignment persists for any inclination of the line of sight. This conclusion seems contradictory to observations made by Fermi/LAT for radio-loud gamma-ray pulsars showing a phase-lag frequently much less than $0.4$, usually about $0.2$. So the pure outer gap model must be rejected to a high confidence level. Unlike outer gaps, the slot gap behaviour with respect to the radio pulse seems more consistent with Fermi/LAT catalogue. Phase lag less than $0.2$ are easily pick out irrespective of the aberration prescription and metric.

\section{Discussions}
\label{sec:Discussions}

Because the pulsar produces and expels positive and negative charges according to the magnetospheric electromagnetic field, the net flux at large distances, for instance at the light cylinder is not necessarily fully compensated on a timescale equal to its period. On average, it is conceivable that the mean electric charge flux vanishes and that the total electric current flux also vanishes but on a timescale much longer than the rotation period. In other words, electric charging and discharging of pulsar magnetospheres is allowed erratically on short times, comparable to its period. An accurate picture requires deep studies of pair creation rate and particle dynamics, outside the scope of this paper. Nevertheless, let us give an estimate of the timescale to fully charge or discharge the magnetosphere from/to vacuum state. The neutron star magnetosphere is filled with an electron/positron pair plasma and contains a total mass that can be compared to the stellar mass. Taking typically the Goldreich-Julian charge density \citep{goldreich_pulsar_1969} with a multiplicity factor $\kappa$ and the $B_z$ component of a static oblique dipole
\begin{equation}
 B_z = \frac{B\,R^3}{r^3} \, [ (3\,\cos^2\vartheta - 1) \, \cos\chi + 3\,\cos\vartheta\,\sin\vartheta\,\cos\varphi\,\sin\chi ]
\end{equation}
the total charge within the magnetosphere extending up to a radius~$R_\infty$ is (integration is done within a spherical shell comprised between an inner radius $R$ and an outer radius $R_\infty$)
\begin{subequations}
\begin{align}
 Q_{\rm mag} & = \iiint \rho_{\rm GJ} \, d^3\mathbf r = -2\,\varepsilon_0 \, \kappa \, \Omega \iiint B_z \, d^3\mathbf r = 0.
\end{align}
\end{subequations}
This total charge is therefore null. In contrast, the particle number in the magnetosphere is
\begin{subequations}
\begin{align}
 N_{\rm mag} & = \iiint \frac{|\rho_{\rm GJ}|}{e} \, d^3\mathbf r \\
  &= 16 \, \kappa \, \frac{\varepsilon_0 \, \Omega \, B \, R^3}{e} \, \log \left(\frac{R_\infty}{R} \right) \, \left[ \frac{2\,\pi}{3\sqrt{3}} \, \cos \chi + \sin\chi \right] .
\end{align}
\end{subequations}
Knowing the particle content of this magnetosphere, we can estimate the time required to fill it from the vacuum. The charge flux generated by the polar caps each of radius $r_{\rm cp} \approx R \, \vartheta_{\rm pc}$ and given by the Goldreich-Julian current modulo the pair multiplicity~$\kappa$, is
\begin{equation}
 \dot N_\pm = \pi\,r_{\rm cp}^2 \, n_{\rm GJ} \, c = 2\,\pi \,\varepsilon_0\,\Omega\,B_z\, \kappa \frac{R^3}{\rlight} \, \frac{c}{e} .
\end{equation}
In the polar cap, the field being equal to $B_z = 2\,B\,\cos\chi$, the refilling time is of the order 
\begin{equation}
 \frac{N_\pm}{\dot{N}_\pm} = \frac{8}{\pi} \, \frac{\ln (R_\infty/R)}{\Omega} \, \left[ \frac{2\,\pi}{3\sqrt{3}} + \tan\chi \right] \approx \frac{4}{\pi^2} \, P \, \ln (R_\infty/R) \, \left[ \frac{2\,\pi}{3\sqrt{3}} + \tan\chi \right] \approx P \, \log \left(\frac{R_\infty}{R} \right) .
\end{equation}
In the last expression, we drop a factor of order unity and the $\tan\chi$ term. Therefore, within one period, the magnetosphere is completely drained or refilled, independently of the multiplicity factor~$\kappa$. However, for rotators close to the orthogonal case, the particle flux is tiny and charging or discharging time becomes longer.

In addition, the variation of the total charge of the neutron star in one period, induced by this particle flux, is of the order of
\begin{equation}
 \frac{\dot{N}_\pm \, e \, P}{Q_c} \approx \frac{3\,\pi}{2} \, \kappa \, \cos \chi.
\end{equation}
The charge of the star can thus change about a characteristic quantity~$Q_c=\frac{8\,\upi}{3}\,\varepsilon_0\,\Omega\,B\,R^3$ in an extremely short time of the order~$T\approx P/(\kappa\,\cos\chi)$, thus much lower than the pulsar period for high multiplicity~$\kappa\gg1$. The associated fluctuations in the magnetospheric electric field due to this loading are therefore drastic in amplitude and in timescales.

\begin{figure*}
\centering
\resizebox{0.95\textwidth}{!}{\input{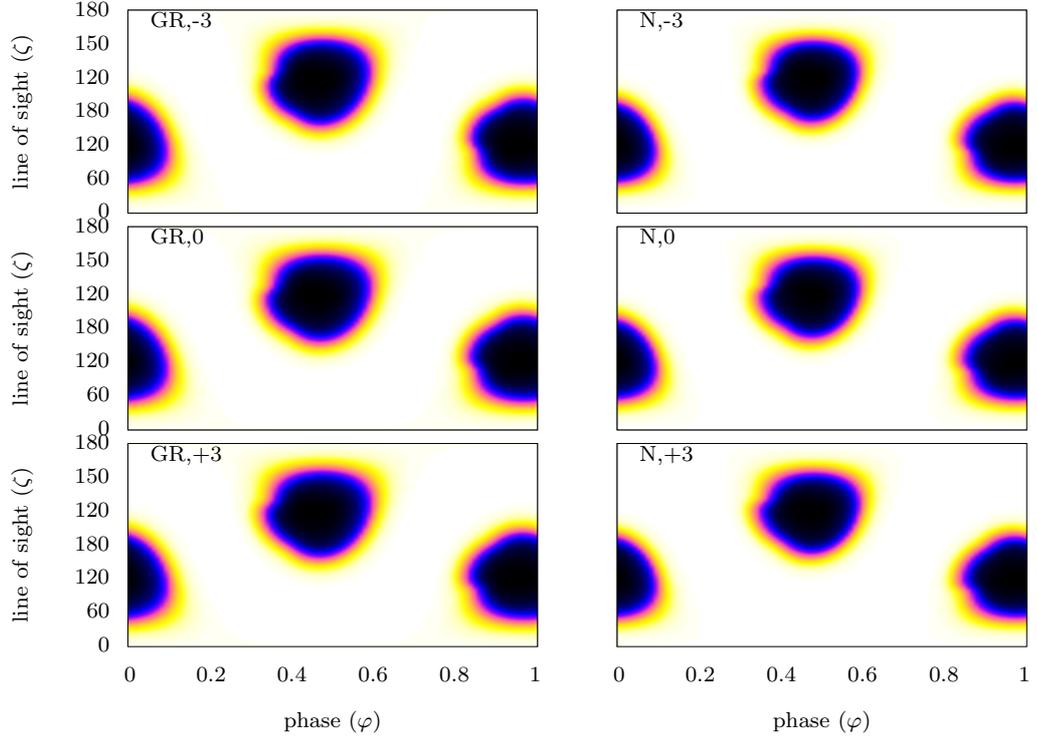}} 
\caption{Sky maps for the polar caps with $\chi=60\degr$, $p=2$ and $\gamma=10$. In each panel, N stands for Newtonian and GR for general-relativistic cases. The charge is indicated on the top right legend with $-3$, $0$ or $+3$, in units of $Q_c$.}
\label{fig:phase_plot_m1_c60_g10_p2_q0_qex}
\end{figure*}

\begin{figure*}
\centering
\resizebox{0.95\textwidth}{!}{\input{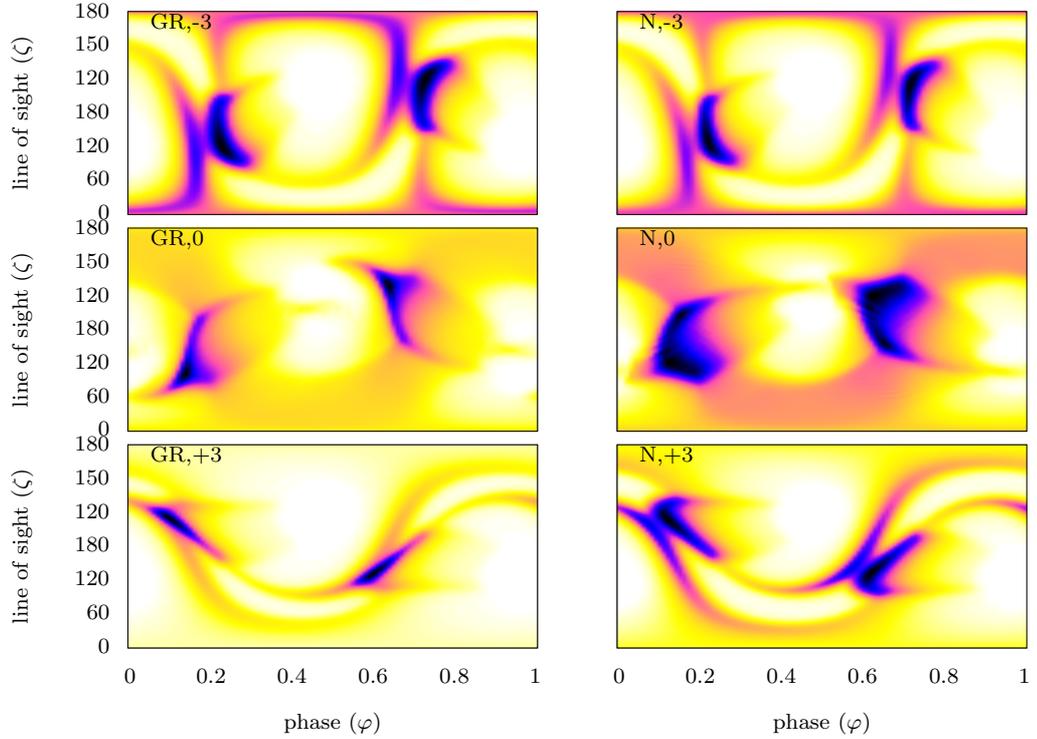}} 
\caption{Same as fig.~\ref{fig:phase_plot_m1_c60_g10_p2_q0_qex} but for the slot gap.}
\label{fig:phase_plot_m2_c60_g10_p2_q0_qex}
\end{figure*}

\begin{figure*}
\centering
\resizebox{0.95\textwidth}{!}{\input{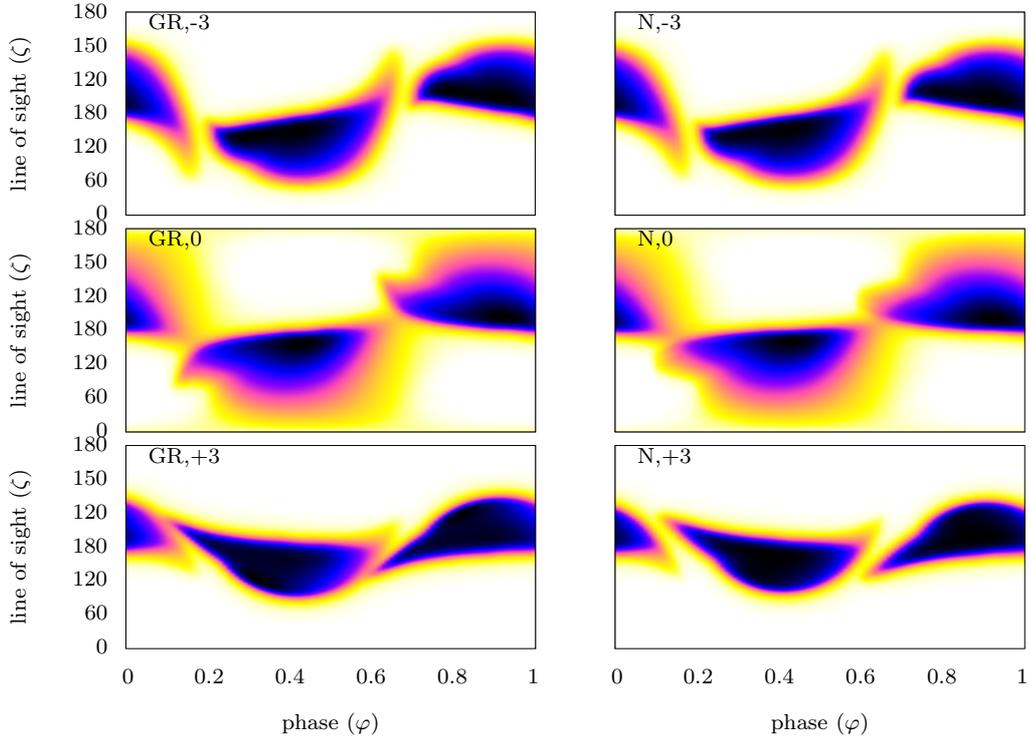}} 
\caption{Same as fig.~\ref{fig:phase_plot_m1_c60_g10_p2_q0_qex} but for the outer gap.}
\label{fig:phase_plot_m3_c60_g10_p2_q0_qex}
\end{figure*}

Because CDA is impacted not only by the magnetic field but also by the electric field, it is easily understood that such electric charge fluctuations will affect the light-curves and sky maps presented in the previous paragraphs. In order to quantify these modifications, we replot the sky maps for CDA in Newtonian and GR metric for three value of the electric charge: $Q_{\rm tot}/Q_c = \{-3,0,+3\}$. Fig.~\ref{fig:phase_plot_m1_c60_g10_p2_q0_qex} shows the sky maps generated by the polar caps. For Newtonian gravity, all three sky maps look very similar. In GR space-time, increasing the net total charge leads to a saturation of the pulse profile with a plateau seen as a large black spot in the middle of each polar cap.
Results for charge variation in the slot gap model are shown in fig.~\ref{fig:phase_plot_m2_c60_g10_p2_q0_qex}. Here the changes are drastic. The sky maps for null charge are very different from sky maps for non vanishing charges. For non neutral magnetospheres, a piece of double S-shaped curves are drawn in the $(\varphi,\zeta)$ plane. Two dominant spots are visible and interpreted as the two main pulses for favourable inclination angle $\zeta$. Some other significant interpulse emission emerges at fainter intensity level with respect to the two main pulses. Consequently, high-energy pulsar emission modelling is strongly influenced by another free parameter being the charge of the system neutron star + magnetosphere. Surprisingly this charge freedom is often neglected both in force-free, MHD, PIC simulations of the plasma dynamics and in radiation mechanisms predictions. We showed in this discussion that this approach is not sustainable on physical ground.
Lastly, for the outer gap model, the charge has a little impact on the sky maps as shown in  fig.~\ref{fig:phase_plot_m3_c60_g10_p2_q0_qex}. The two large spots are always visible, but their orientation with respect to the $\zeta$ axe reverses when switching from negative to positive charges. Nevertheless, the ensuing light-curves remain marginally modified compared to those extracted from the slot gaps.

\begin{figure*}
\centering
\resizebox{0.95\textwidth}{!}{\input{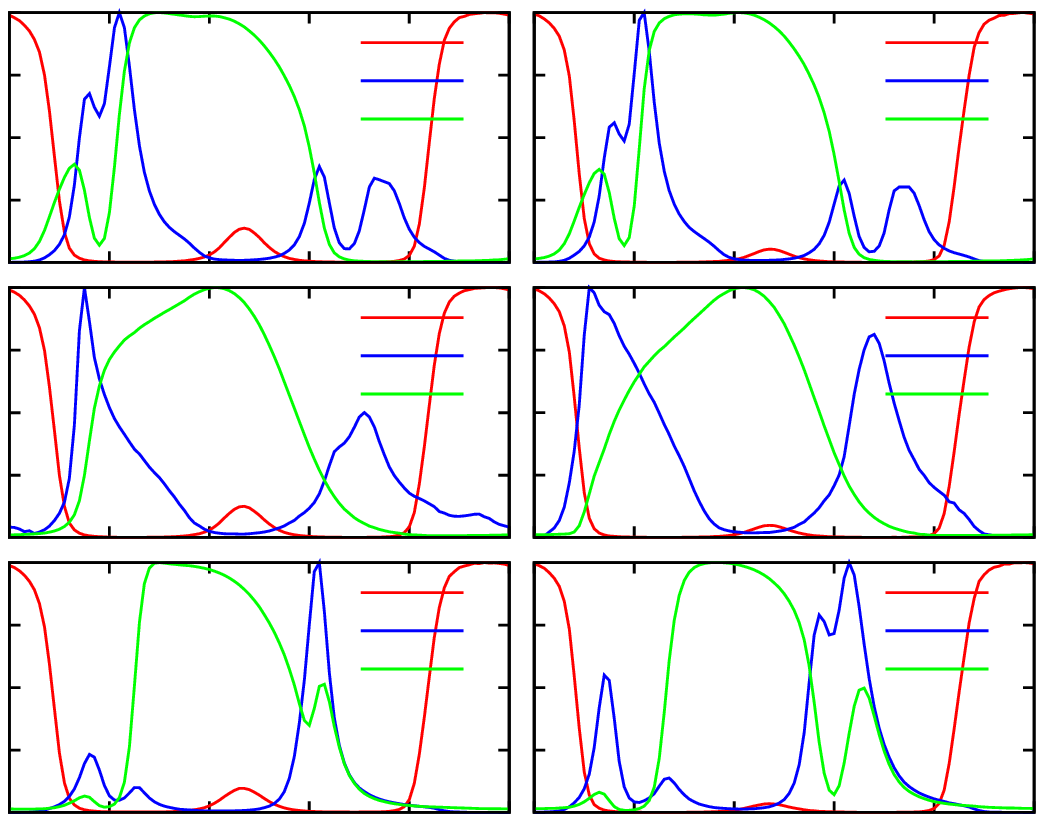}} 
\caption{Multiwavelength light-curves for the polar cap, the slot gap and the outer gap with $\chi=60\degr$, $\zeta=70\degr$, $\gamma=10$ and $p=2$.}
\label{fig:phase_plot_mx_c60_g10_p2_q0_qex}
\end{figure*}

Eventually, a multi-wavelength light-curve sample is presented in fig.~\ref{fig:phase_plot_mx_c60_g10_p2_q0_qex} to better catch the outcome of the electric charge on pulse profile fitting. The discrepancies between the three cases is clearly visible, showing a change in the pulse profile, width and shape, sharply pronounced for slot gaps and less for outer gaps.

\section{Conclusions}
\label{sec:Conclusions}

Predictions and fitting of pulsar light-curves offer a great insight into the electrodynamics of its magnetosphere: electromagnetic field topology, particle distribution function, radiation mechanisms and their emission sites. We developed a general purpose numerical algorithm to compute any kind of pulsed emission profiles and spectra by integrating the emissivity in full three-dimensional space in spherical polar coordinates. Our new approach contrasts with previous attempts to computed light-curves in that it includes general-relativistic electromagnetic field for a rotating dipole in Schwarzschild space-time and emphasized the non negligible role of the total electric charge of the system. Our algorithm is also able to compute pulsed radiation to very large distances well outside the light-cylinder and smoothly joining the inner magnetosphere to a possible striped wind well beyond the light-cylinder. We showed that the best and most promising aberration prescription corresponds to a combination between electric and magnetic field leading to a frame where both are parallel to each other. In a sense, it is a generalization of several radiation reaction limit prescriptions for particle velocity, relaxing the constrain about the speed of light. Our approach applies to vacuum fields but also to any force-free or MHD fields. There is no explicit reference to motion along field line, as this is ill defined and misleading in a general electromagnetic field.

This introductory work shows the large possibilities permitted by our new method of computing pulsar light-curves. Fitting real data from radio and gamma-ray pulsars requires good knowledge about the particle distribution function as well as the field topology and emission sites. Such variations in the model can easily be accommodated in our algorithm by relaxing some restrictive assumptions like a constant density and a pure power law distribution function. But this happens at the expense of adding several other parameters to the emission model. Attempts to fit radio-loud gamma-ray pulsars will be shown in a forthcoming work. We also plan to deduce the multi-wavelength phase-resolved polarization properties by adding the computation of the Stokes parameters $(I,Q,U,V)$. This should put even more stringent constraints to current competing high-energy emission models especially in the coming exciting era of space-based X-ray polarimeters such as IXPE \citep{weisskopf_imaging_2016}, XIPE \citep{soffitta_xipe:_2016} and eXTP \citep{zhang_extp:_2017}.

\section*{Acknowledgements}

I am grateful to the anonymous referee for its useful insight and comments that improved the paper quality. This work has been supported by the French National Research Agency (ANR) through the grant No. ANR-13-JS05-0003-01 (project EMPERE). It also benefited from the computational facilities available at Equip@Meso (Universit\'e de Strasbourg).








\appendix

\section{Polar cap shapes in flat spacetime}
\label{app:PolarCapShape}

In this first part of the appendix, we remember the shape of polar caps in flat spacetime for an arbitrary obliquity~$\chi$ in the static and rotating magnetic dipole regime. It is intended to give a brief summary of well known discrepancies between both approximations of the magnetic field topology.

\subsection{Static dipole}

For a static dipole, exact analytical solutions for the polar cap shape have been found. A simple example for the orthogonal rotator is presented in the main text. Such solutions are useful benchmarks to test the code accuracy and correctness.

\begin{figure}
\centering
\input{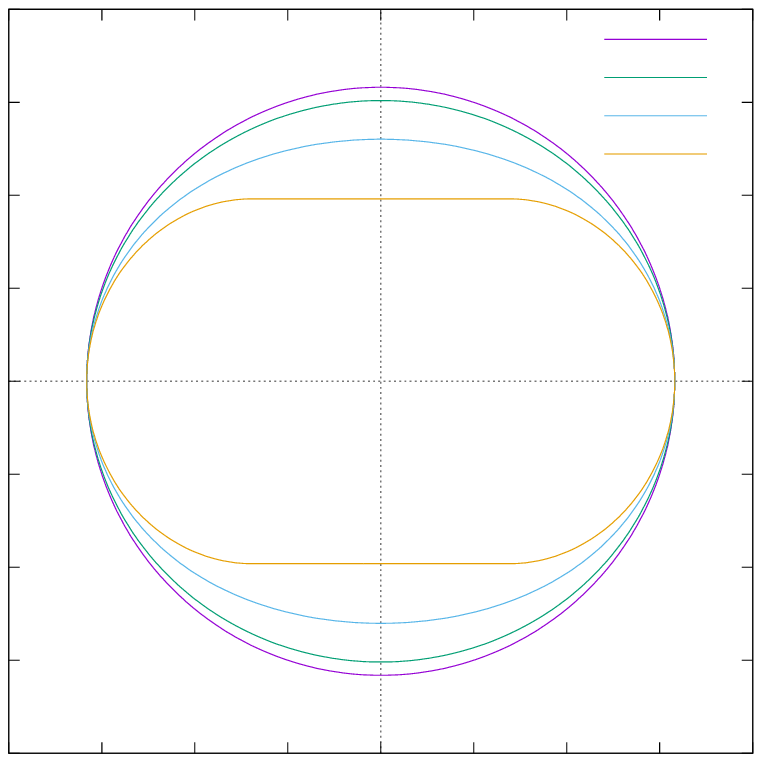}
\caption{Numerical shape of polar caps for the oblique static dipole with $R/\rlight=0.1$ and different inclination angles $\chi$ as shown in the legend.}
\label{fig:PolarCap}
\end{figure}

Numerical solutions of these polar caps are shown in fig.~\ref{fig:PolarCap} for $R/\rlight=0.1$. To better compare between several inclination of the magnetic axis, the polar cap rim has been centred with respect to the magnetic axis. The size of the cap in the~$x$ direction does not depend in the inclination~$\chi$ but the transversal size in the $y$ direction diminishes by almost a factor 2 when moving from an aligned $\chi=0\degr$ to an orthogonal $\chi=90\degr$ rotator. Note that distances are normalized with respect to the neutron star radius~$R$.
The polar cap is symmetric with respect to the $x$ and $y$ axis. There is no asymmetry induces by the stellar rotation. The asymmetric shape is only induced by a rotating dipole as shown in the next paragraph.

\subsection{Rotating dipole}

\begin{figure}
\centering
\input{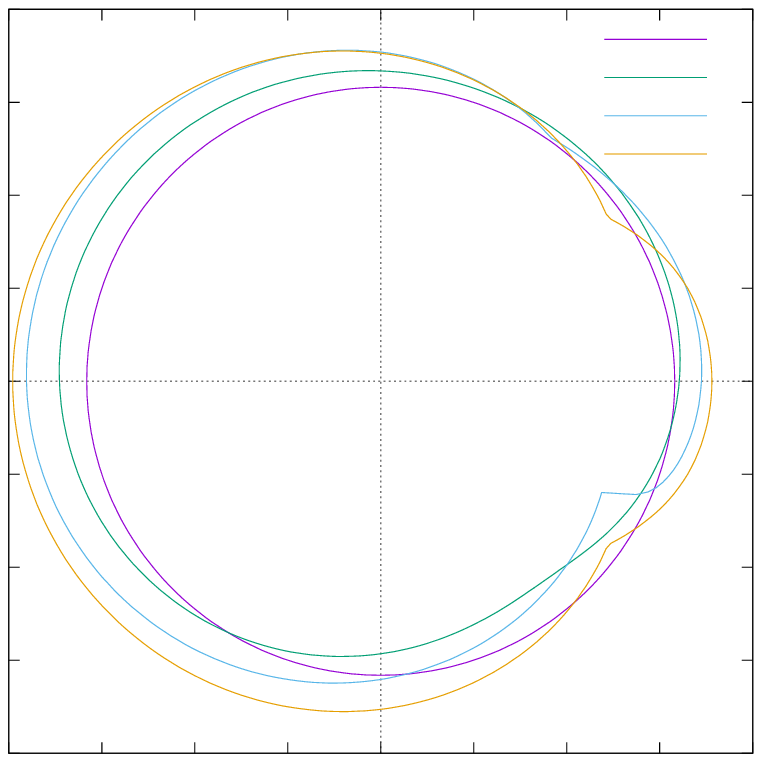}
\caption{Numerical shape of one polar cap for the oblique rotating dipole with $R/\rlight=0.1$ and different inclination angles $\chi$ as shown in the legend.}
\label{fig:PolarCapRotating}
\end{figure}

For a rotating dipole, there exist a privileged direction given by the rotation axis, breaking the above symmetry. These swept back magnetic field lines are produced when the displacement current is taken into account as in \citet{deutsch_electromagnetic_1955} solution. A bump appears on the leading side of the polar cap for increasing obliquity as shown in Fig.~\ref{fig:PolarCapRotating}.
Comparing the size of static and rotating polar caps we note a monotonic increase in the surface area.

\begin{figure}
\centering
\resizebox{0.5\textwidth}{!}{\input{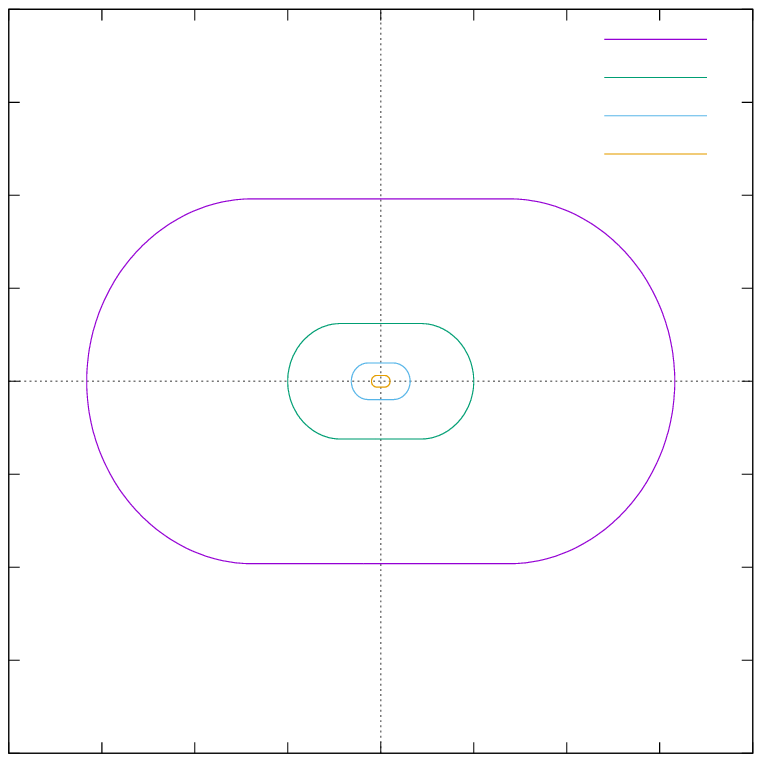}}
\caption{Analytical shape of polar caps for the orthogonal static dipole with $R/\rlight=\{0.1,0.01,0.001,0.0001\}$ as shown in the legend with index~$s$ such that $R/\rlight=10^{-s}$.}
\label{fig:PolarCapCHI90}
\end{figure}

\begin{figure}
\centering
\resizebox{0.5\textwidth}{!}{\input{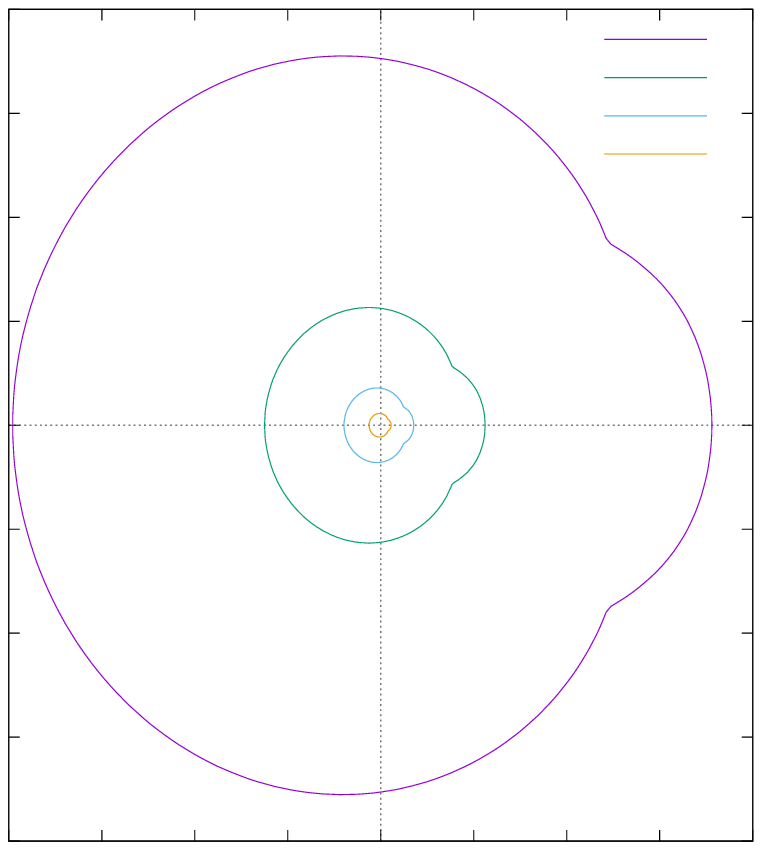}}
\caption{Polar cap shapes for the orthogonal dipole with $R/\rlight=\{0.1,0.01,0.001,0.0001\}$ as shown in the legend with index~$s$ such that $R/\rlight=10^{-s}$.}
\label{fig:PolarCapCHI90Deutsch}
\end{figure}

Obviously, the size of the polar cap depends on the period of the pulsar, decreasing with increasing period with a scaling roughly as $\sqrt{R/\rlight}$. In Fig.~\ref{fig:PolarCapCHI90}, the size of four polar caps in the static dipole limit are shown for $R/\rlight=\{0.1,0.01,0.001,0.0001\}$. The decrease in dimension according to the square root law is clearly recognized. 
The same study has been done for the more realistic rotating dipole model and the results for $R/\rlight=\{0.1,0.01,0.001,0.0001\}$ are presented in Fig.~\ref{fig:PolarCapCHI90Deutsch} showing the same scaling.

The polar cap rim defines the base of the separatrix, that is the surface separating the closed corotating volume from the open region where plasma is assumed to flow outwards towards the nebula and/or interstellar medium.

\subsection{Separatrix}

\begin{figure}
\centering
\resizebox{0.5\textwidth}{!}{\input{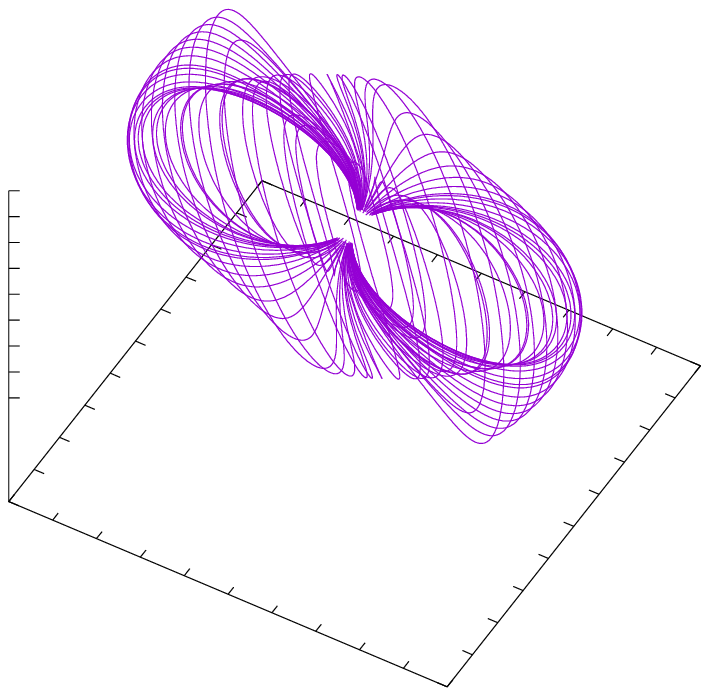}}
\caption{The separatrix surface for $\chi=60\degr$ and $R/\rlight=0.1$. The separatrix does not graze the light-cylinder for each azimuth~$\varphi$.}
\label{fig:Separatrice}
\end{figure}

The separatrix~$\mathcal{S}$ is defined as the full set of last closed field lines leaving the star on one pole and returning to it through the other pole. It draws a warped closed surface. Inside it, plasma is assumed to corotate without radiating, whereas outside a relativistic plasma flows to ultra-relativistic speeds, feeding the wind with electron/positron pairs and copiously radiating synchrotron and curvature photons and maybe inverse Compton light. The separatrix grazes the light-cylinder at different altitudes~$z$ and azimuths~$\varphi$ in a cylindrical coordinate system. A three-dimensional view of $\mathcal{S}$ is shown in Fig.~\ref{fig:Separatrice} for $\chi=60\degr$ and $R/\rlight=0.1$.
The grazing intersection points between the separatrix and the light-cylinder accumulate at preferred azimuths, tending to two very distinctive curves in the $\varphi-z$ plane as we now detail. These locus are a fortiori artefact introduced by the prescribed separation between closed and open field lines in vacuum that do not reflect confidently reality.

\subsection{Base of the striped wind}

Although the striped wind requires loading by a plasma flowing out from the magnetosphere we can guess the shape of this current sheet by inspection of the last open/closed field lines. The last open field lines inside the magnetosphere sustain an electric current, along the separatrix, that flows through the light-cylinder to form the wobbling current sheet characteristic of the striped pulsar wind. The transition between the closed corotating magnetosphere and the wind occurs right at the light cylinder. The precise shape of this current sheet is determined from the intersection of the last closed field lines with the light-cylinder. In the case of the vacuum Deutsch solution, we represent this locus in a $\varphi-z$ plane where $\varphi$ denotes the azimuth in cylindrical coordinates and $z$ the altitude related to spherical coordinates by $z=r\,\cos\vartheta$. Once this region known, the curve can be propagated radially outwards at high Lorentz factor and meanwhile, rotating it at the neutron star angular velocity~$\Omega$. This is the general procedure to construct a current sheet starting from the base of the wind at exactly the light-cylinder. The same technique has been applied to the split monopole solution except that in this latter case the wind starts immediately at the stellar surface for a monopole \citep{bogovalov_physics_1999}.

\begin{figure}
\centering
\resizebox{0.45\textwidth}{!}{\input{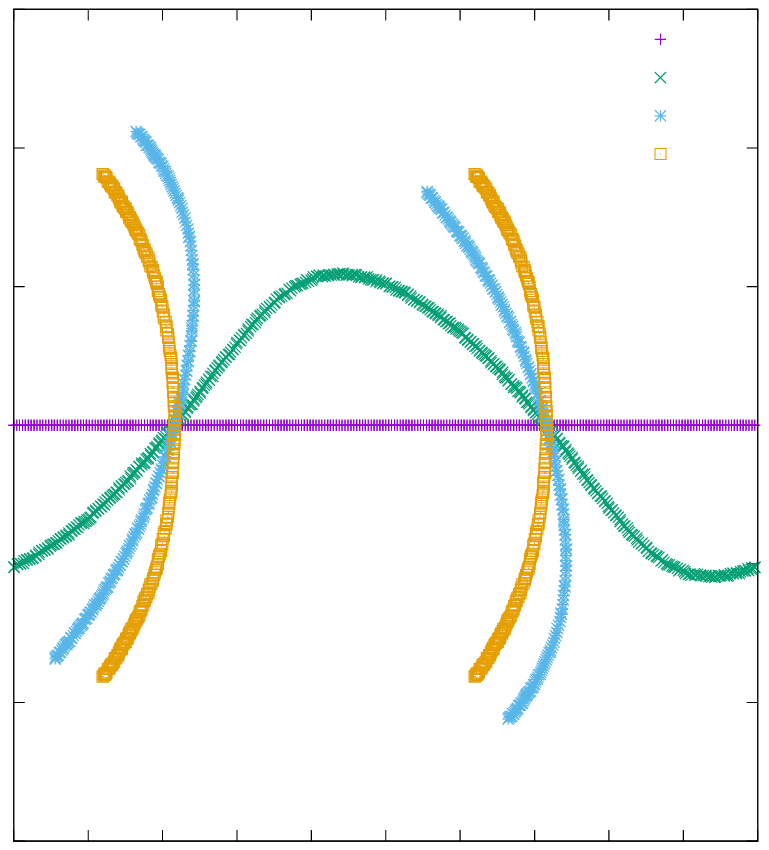}}
\caption{Base of the striped wind for various obliquities~$\chi$ as shown in the legend.}
\label{fig:off_z_phi}
\end{figure}

\begin{figure*}
\centering
\begin{tabular}{cc}
\resizebox{0.5\textwidth}{!}{\input{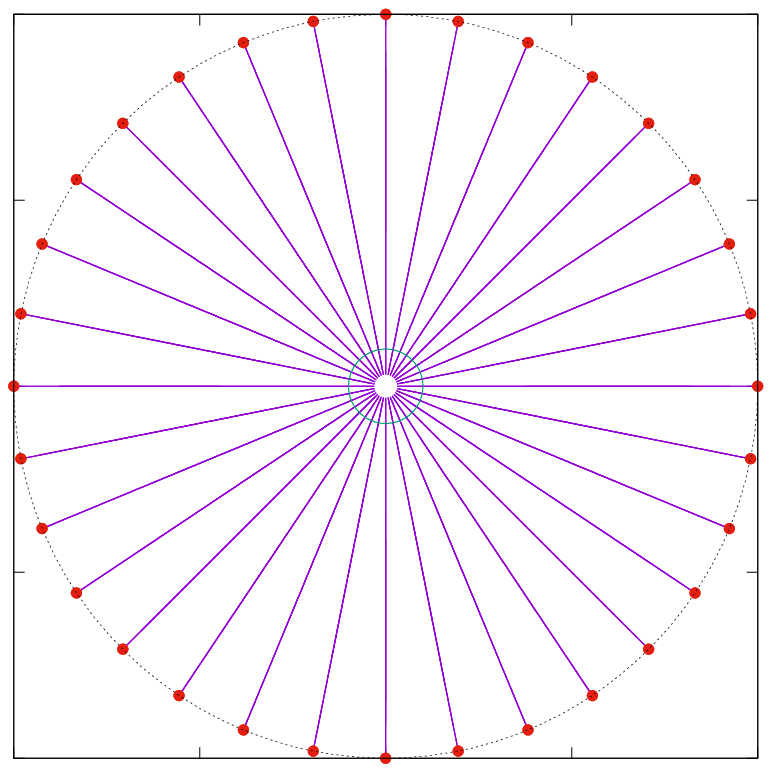}} &
\resizebox{0.5\textwidth}{!}{\input{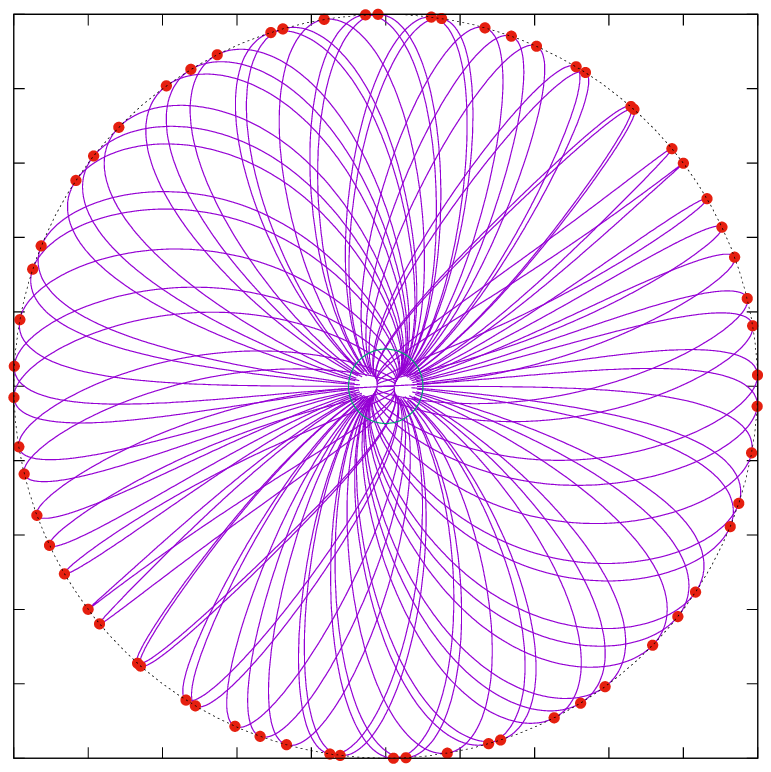}} \\
\resizebox{0.5\textwidth}{!}{\input{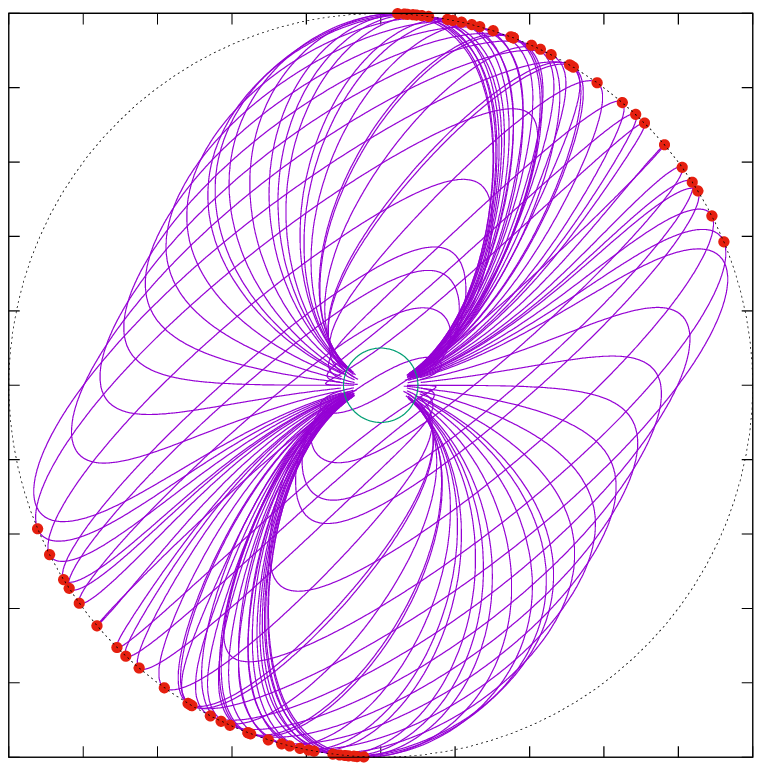}} &
\resizebox{0.5\textwidth}{!}{\input{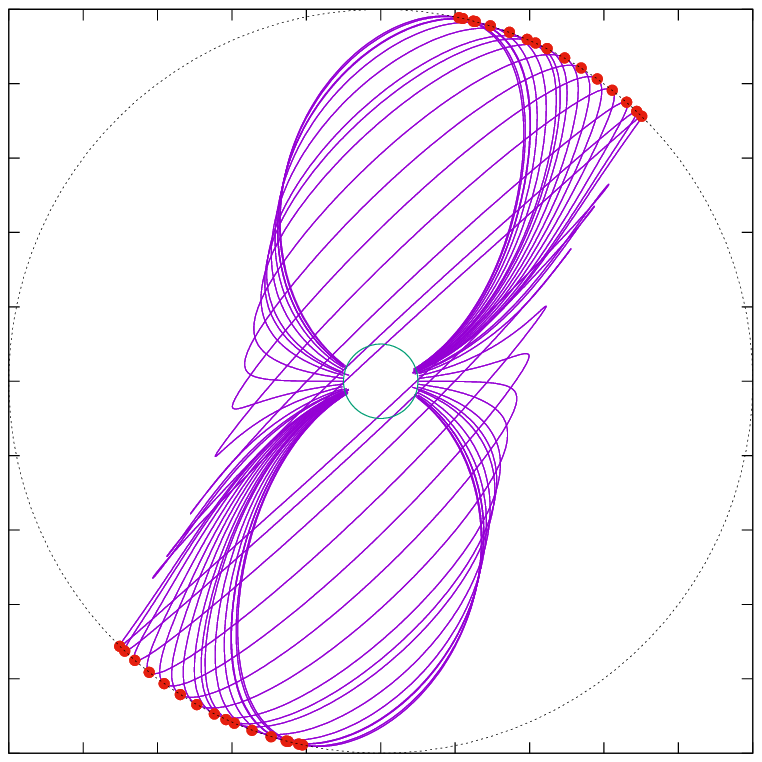}}
\end{tabular}
\caption{Locus of the separatrix surface in blue solid lines and its intersection with the light-cylinder in red points for obliquities $\chi = \{ 0\degr, 30\degr, 60\degr, 90\degr \}$.}
\label{fig:SeparatriceComparaison}
\end{figure*}

Here we take into account the dipolar topology inside the light-cylinder. Some special cases of curves in the $\varphi-z$ plane are shown in Fig.~\ref{fig:off_z_phi}. For small inclination angles $\chi$, the curve is continuous with respect to $\varphi$, there is bijection between $z$ and $\varphi$. However for large inclination angles $\chi$, at some phases there are no solutions for all $\varphi$. We notice that around phase $0\degr$ and phase $180\degr$, no solutions for $z$ were found. This fact is clearly pointed out in fig.\ref{fig:SeparatriceComparaison}. For such high inclinations, it becomes impossible to extrapolate the current sheet in the separatrix to the region outside the light-cylinder just by assuming grazing of field lines. In other words, the vacuum solution with last closed field lines prescribed in the usual manner is unable to be properly extrapolated to large distances. This picture is inconsistent with a striped topology existing for any phase $\varphi$. The only way to circumvent this drawback is to take into account the backreaction of the plasma flow leading to a self-consistent current sheet flowing inside the magnetosphere and smoothly joining the striped wind. Another possibility is to segregate between closed and open field lines by another criterion different from grazing the light-cylinder (for instance including the electric field topology).

This short discussion shows that the vacuum solution with the usual determination of the closed part of the magnetosphere with the light-cylinder is inconsistent. In such a way it is impossible to get a complete description of the magnetosphere and wind. This also reflects in the shape of the polar caps that will be distorted compared to this vacuum approximation. In a more realistic regime, the plasma sweeps the magnetic field lines to retrieve an unique solution to $z$ for each $\varphi$. Another possibility, as explored in this paper, assumed that particles do not strictly follow field lines because the definition of field line is misleading in vacuum, not offering a useful insight about particle motion in an arbitrary electromagnetic field.

\subsection{Field line projection}

The topology of the electromagnetic field is set by the solution of Maxwell equations. However, due to light travel time and aberration effects, the distant observer will see an electromagnetic structure distorted with respect to its actual geometry at a fixed coordinate time. Actually, the reception (or retarded) time depends on the location where photons are emitted and on their direction of propagation in the observer frame. Photons radiated behind the neutron star with respect to the observer require more time to reach him thus they should be emitted at early coordinate times, provided they do not hit the stellar surface. This introduces a strong distortion of the electromagnetic field structure as seen by the observer. We emphasize that these are purely propagation or kinematic effects not related to any rotational deformation of the fields. These distortions are already seen for a static dipole. It is an instantaneous picture of the field lines taken by the distant observer, including aberration, retardation and possible general-relativistic effects (light bending and Shapiro delay).

In flat spacetime, we identify three main deformations of field lines as measured at large distances
\begin{itemize}
\item magnetic field sweep back, a true deformation including displacement currents.
\item retardation, due to travel time from receiver to observer, and finite speed of light.
\item aberration of light, due to relative speed between the emission frame (not necessarily inertial) and the observer frame (inertial).
\end{itemize}
It is useful to quantify the merit of each of these corrections with respect to a static dipole to get full insight into each contribution. We study separately the time lag induced by field line sweep back, retardation and aberration according to our three prescriptions for aberration.

\begin{figure*}
\centering
\begin{tabular}{cc}
\resizebox{0.5\textwidth}{!}{\input{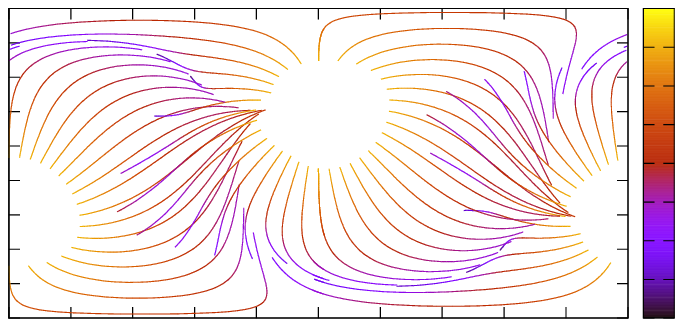}} &
\resizebox{0.5\textwidth}{!}{\input{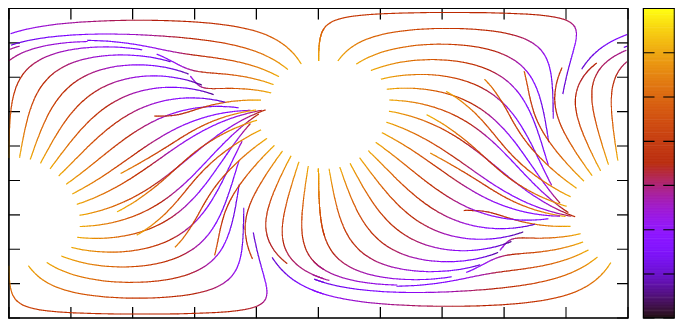}} \\
\resizebox{0.5\textwidth}{!}{\input{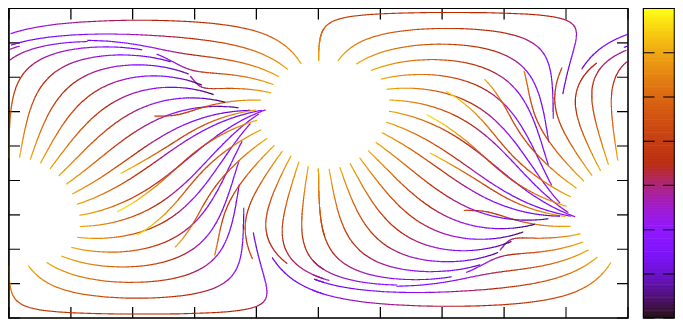}} &
\resizebox{0.5\textwidth}{!}{\input{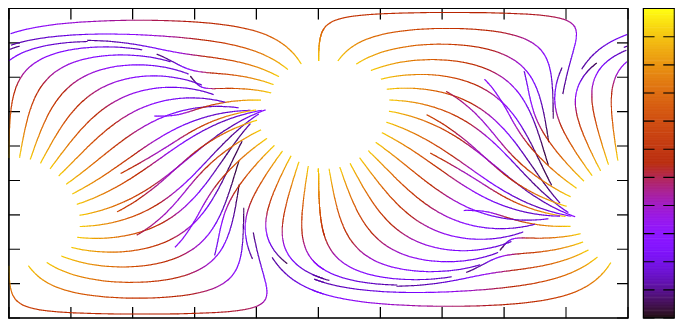}}
\end{tabular}
\caption{Field line projection onto the sky without retardation and without aberration for $\chi=60\degr$ showing the retardation time with and without aberration. In panel a) for RAD, in panel b) for LBA, in panel c) for CFA, and in panel d) for CDA.}
\label{fig:ligne_carte_00_ret}
\end{figure*}

Fig.~\ref{fig:ligne_carte_00_ret} shows the propagation time $\Delta t = -\frac{\mathbf n_{\rm obs} \cdot \mathbf r}{c}$ with and without aberration for a rotating dipole in Minkowski spacetime. Each line corresponds to the projection of a field line onto the plane of the sky. For ease of comparison, the projection does not take into account neither retardation nor aberration. Thus the field line projection remains the same for all plots. The retardation time, in units of the spin period is depicted by the colour coded legend. It is always negative because it requires less time to reach the observer compared to a photon that would be emitted right at the stellar centre. The time retardation is of the order
\begin{equation}
\frac{\Delta t}{P} \approx - \frac{r}{2\,\upi\,\rlight} .
\end{equation}
At the neutron star surface for $R/\rlight=0.1$ it is approximately $\Delta t/P \approx -0.015$ whereas at the light cylinder it approaches $\Delta t/P \approx 1/(2\,\upi) \approx -0.16$. The precise expression of the aberration formula has only minor impact on this retardation as seen from comparison of the upper and lower panels a), b), c) and d) in Fig.~\ref{fig:ligne_carte_00_ret}. Panel a) includes propagation along the radial direction (RAD) thus no true aberration effect, panel b) uses a Lorentz boost for aberration (LBA), panel c) uses the instantaneous corotating frame for aberration (CFA) and panel d) starts from the common direction of $\mathbf{E}$ and $\mathbf{B}$ to look for aberration (CDA). However, although LBA and CFA look similar, for CDA, the influence of the electric field becomes notable close to the light-cylinder. Aberration effects in CDA differ then from the other two prescriptions. As a consequence, retardation becomes important only very close to the end of each field line, that is, close to the light-cylinder and reach only up to 15\% of the pulsar period.

\begin{figure*}
\centering
\begin{tabular}{cc}
\resizebox{0.5\textwidth}{!}{\input{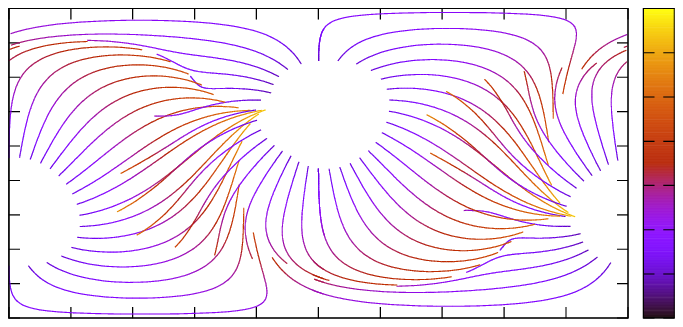}} & 
\resizebox{0.5\textwidth}{!}{\input{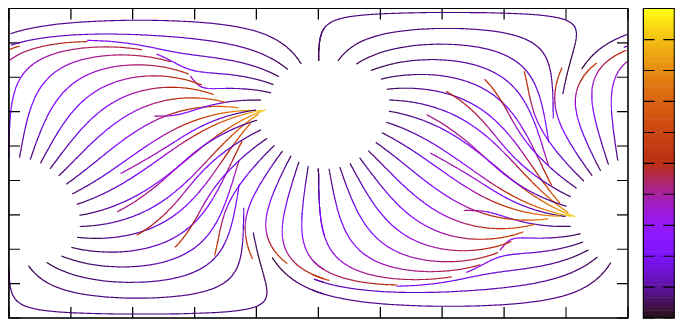}} \\
\resizebox{0.5\textwidth}{!}{\input{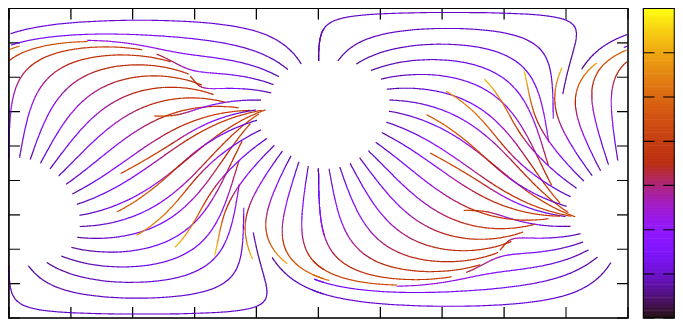}} &
\resizebox{0.5\textwidth}{!}{\input{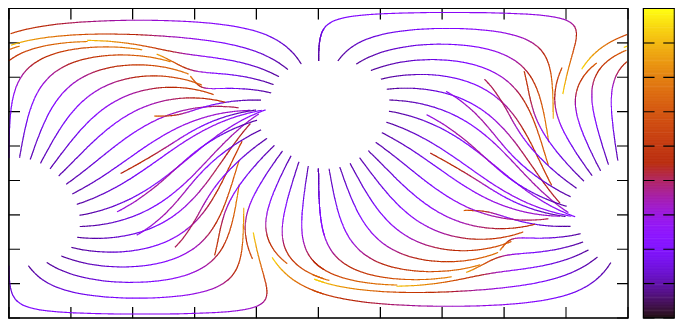}}
\end{tabular}
\caption{Field line projection onto the sky without retardation and without aberration for $\chi=60\degr$ showing the correction in the aberration angle~$\Delta\vartheta$ in degrees when the usual formula is used eq.~(\ref{eq:AberrationLorentz}) in panel b), or corotation aberration is used in eq.~(\ref{eq:AberrationCorot}) in panel c), or common direction aberration in panel d). Panel a) show the angle between the normal to the surface and field line direction.}
\label{fig:ligne_carte_cor_aber}
\end{figure*}

We also show the correction~$\Delta \vartheta$ due to aberration by computing the angle between the photon direction before $\mathbf{n}$ and after $\mathbf{n}'$ aberration has been applied according to the variation
\begin{equation}
\Delta \vartheta = \arccos(\mathbf{n}' \cdot \mathbf{n}).
\end{equation}
Fig.~\ref{fig:ligne_carte_cor_aber} shows this correction for several prescriptions of the aberration formula. Angles are given in degrees.

\begin{figure*}
\centering
\begin{tabular}{cc}
\resizebox{0.5\textwidth}{!}{\input{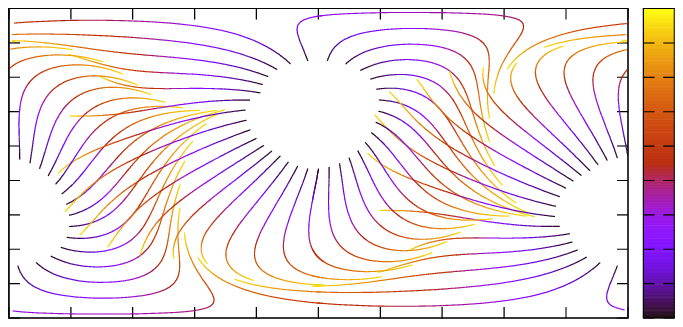}} &
\resizebox{0.5\textwidth}{!}{\input{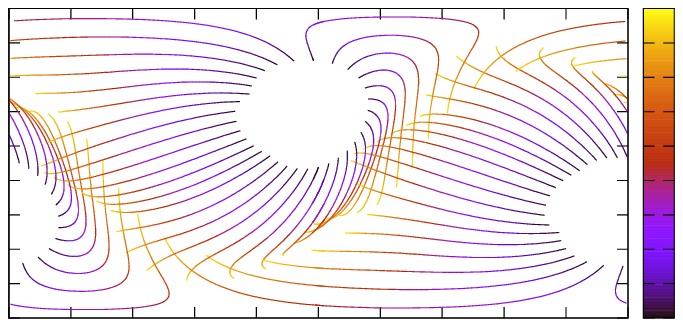}} \\
\resizebox{0.5\textwidth}{!}{\input{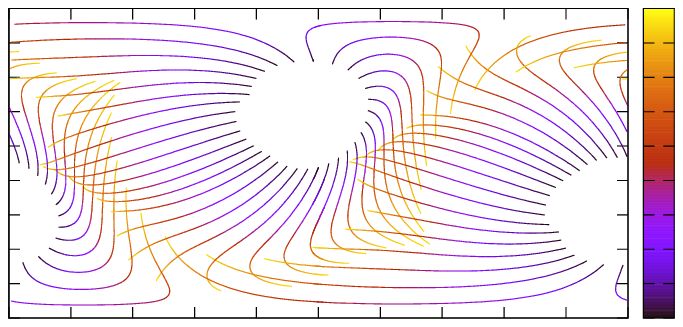}} &
\resizebox{0.5\textwidth}{!}{\input{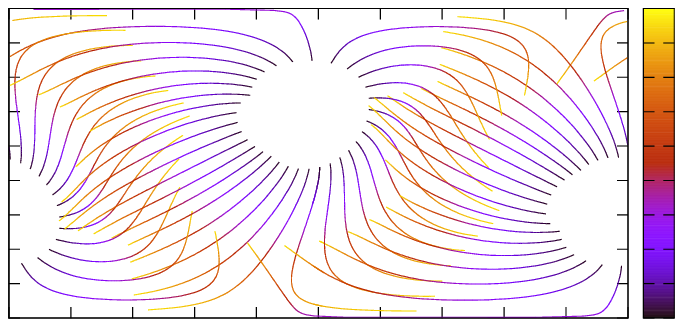}}
\end{tabular}
\caption{Field line projection onto the sky with retardation and with or without aberration for $\chi=60\degr$ showing the altitude~$r$ where photons have been emitted. In panel a) for RAD, in panel b) for LBA, in panel c) for CFA, and in panel d) for CDA.}
\label{fig:ligne_carte_alt}
\end{figure*}

Knowing the altitude of emission is a strong indicator of the field line curvature and therefore about curvature photon energies. Thus in Fig.~\ref{fig:ligne_carte_alt} we show the altitude of emission along field lines depending on the aberration formula. Emission starts at the base of the field lines which is here $r=0.1\,\rlight$ for $R/\rlight=0.1$. It reflects the shape of the polar caps. When leaving the star, the emission height obviously increases up to a spherical distance close to $\rlight$ imposed by the gap models. Projection of the field lines onto the plane of the sky are significantly distorted by aberration effects, especially close to the light cylinder where the corotation speed becomes important with respect to the speed of light and where the electric field is comparable to the magnetic field strength, $E\lesssim c\,B$.

\begin{figure*}
\centering
\begin{tabular}{cc}
\resizebox{0.5\textwidth}{!}{\input{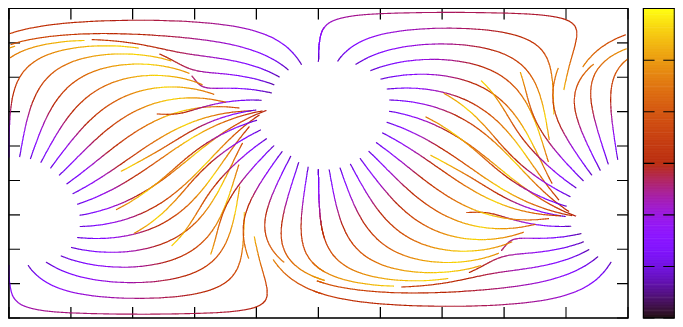}} &
\resizebox{0.5\textwidth}{!}{\input{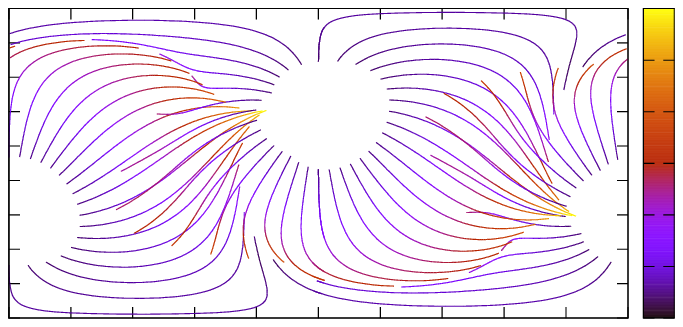}} \\
\resizebox{0.5\textwidth}{!}{\input{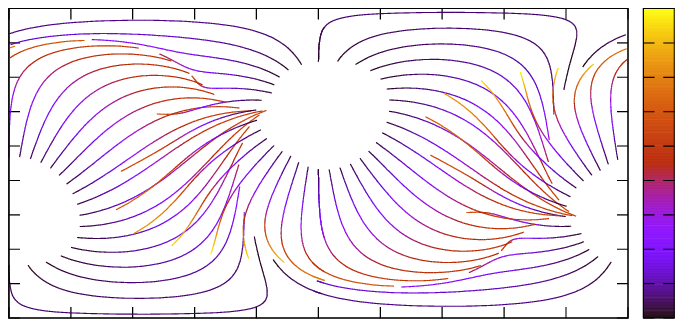}} &
\resizebox{0.5\textwidth}{!}{\input{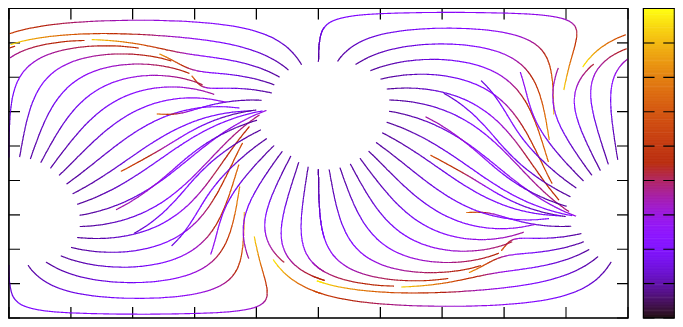}}
\end{tabular}
\caption{GR corrections due to light-bending showing the correction in the aberration angle. Angle showing deviation from the flat spacetime aberration direction for $\chi=60\degr$ in degrees when the usual formula is used eq.~(\ref{eq:AberrationLorentz}) in panel b), or corotation aberration is used in eq.~(\ref{eq:AberrationCorot}) in panel c), or common direction aberration in panel d). Panel a) show the angle between the normal to the surface and field line direction.}
\label{fig:ligne_carte_light_bending}
\end{figure*}

The same plots can be computed for general-relativistic dipolar magnetic field. The two new contributions with no Newtonian equivalent are
\begin{enumerate}
\item light bending in curved spacetime.
\item Shapiro delay because of the longer path to travel to the distant observer.
\end{enumerate}
In fig.~\ref{fig:ligne_carte_light_bending}, we show the additional change in the direction of propagation of the photon as seen by a distant observer when light bending is taken into account. Panel a) shows the angle between the normal to the surface and the field line direction without aberration or light bending effects. Panel b) shows light deflection in the LBA regime. Maximum deflection is about $30\degr$ and happens at largest distances because there the photon travels in a direction far from being radial. The same conclusion applies for CFA but with maximum up to $20\degr$ deflection and in CDA only up to $18\degr$ deflection.

\begin{figure*}
\centering
\begin{tabular}{cc}
\resizebox{0.5\textwidth}{!}{\input{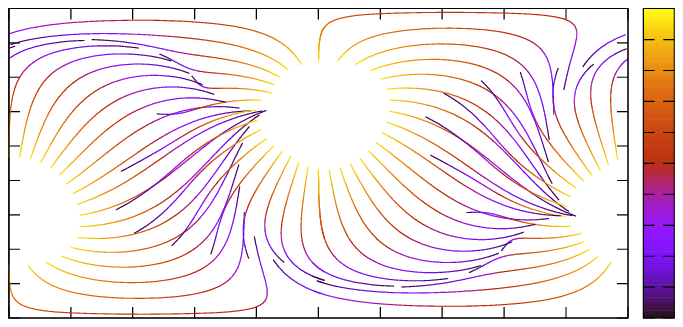}} &
\resizebox{0.5\textwidth}{!}{\input{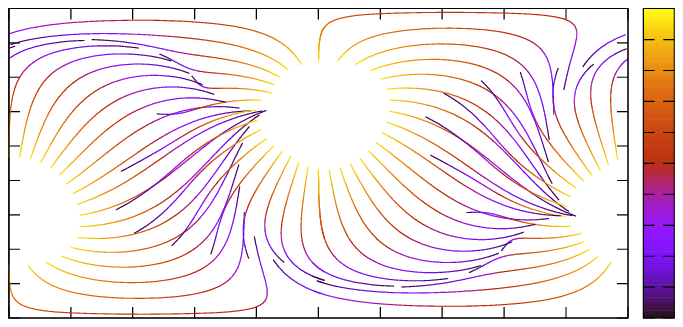}} \\
\resizebox{0.5\textwidth}{!}{\input{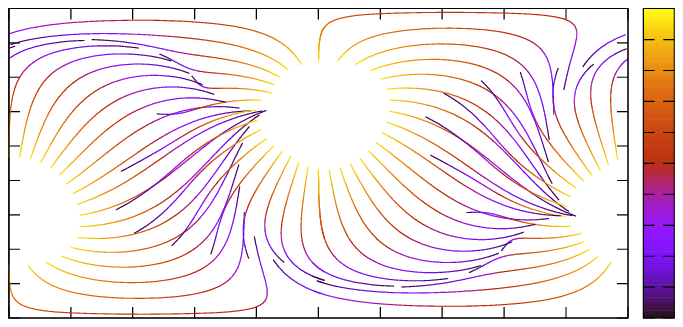}} &
\resizebox{0.5\textwidth}{!}{\input{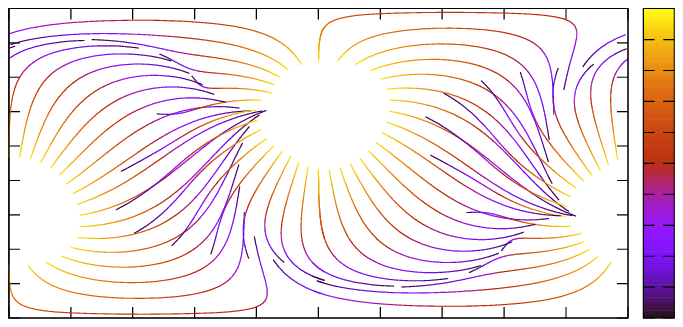}}
\end{tabular}
\caption{Photon travel time due to Shapiro delay showing the correction in propagation time. The normalized time required to reach the observer is given for non aberrated photons in panel a),  for LBA in panel b), for CFA in panel c), and for CDA in panel d).}
\label{fig:ligne_carte_delai_shapiro_shapiro}
\end{figure*}

In fig.~\ref{fig:ligne_carte_delai_shapiro_shapiro}, we show the associated travel time of photon to reach a spherical distance of $D=1000\,\rlight$. The full Shapiro delay is computed according to the integral of $dt/dr$. The average propagation time is about $D/c$. Close to the poles, the delay is perceptible with a small increase in arrival time whereas for photons emitted close to the light cylinder, they arrive earlier than $D/c$ with little GR effects. All for panels show very similar behaviours, the Shapiro delay is insensitive to the aberration expression used.

\begin{figure*}
\centering
\begin{tabular}{cc}
\resizebox{0.5\textwidth}{!}{\input{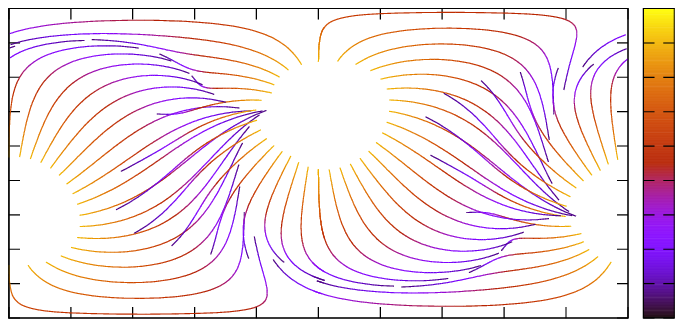}} &
\resizebox{0.5\textwidth}{!}{\input{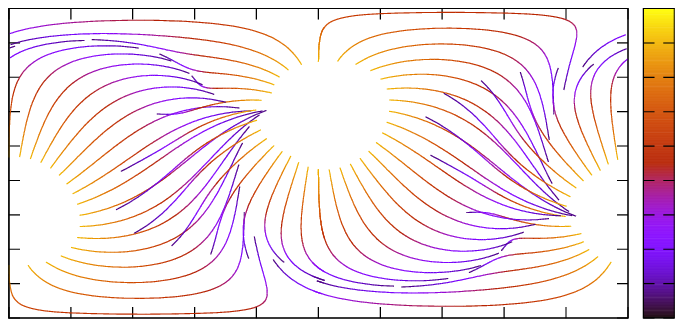}} \\
\resizebox{0.5\textwidth}{!}{\input{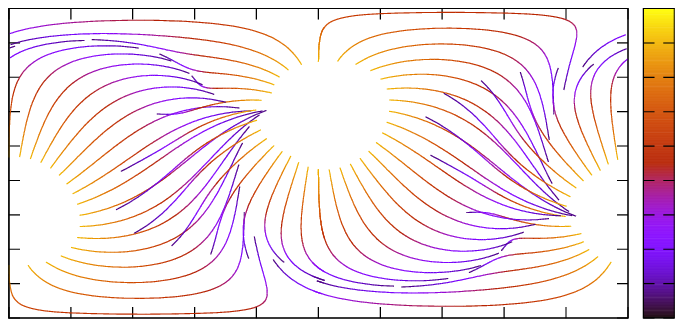}} &
\resizebox{0.5\textwidth}{!}{\input{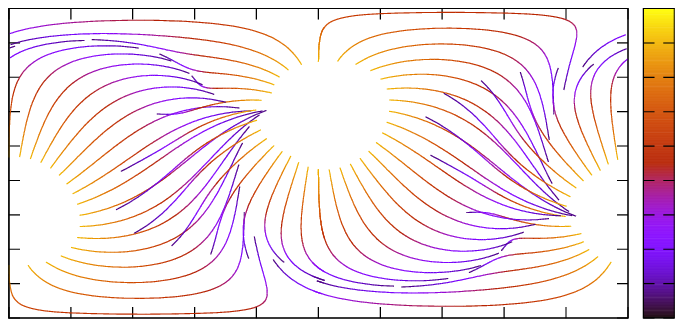}}
\end{tabular}
\caption{Relative difference between retarded time~$t_{\rm r}$ in flat spacetime compared to Shapiro delay $t_{\rm s}$ given by $t_{\rm s}/t_{\rm r}-1$.}
\label{fig:ligne_carte_delai_shapiro_tret}
\end{figure*}

\begin{figure*}
\centering
\begin{tabular}{cc}
\resizebox{0.5\textwidth}{!}{\input{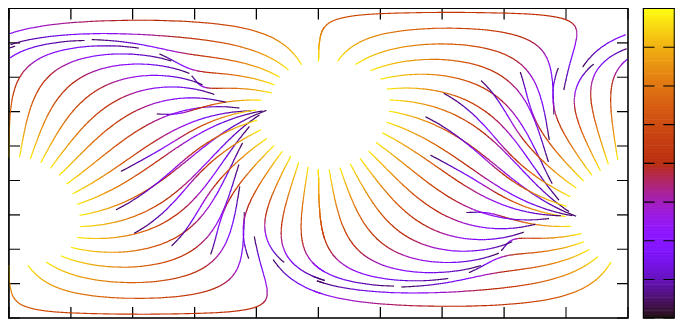}} &
\resizebox{0.5\textwidth}{!}{\input{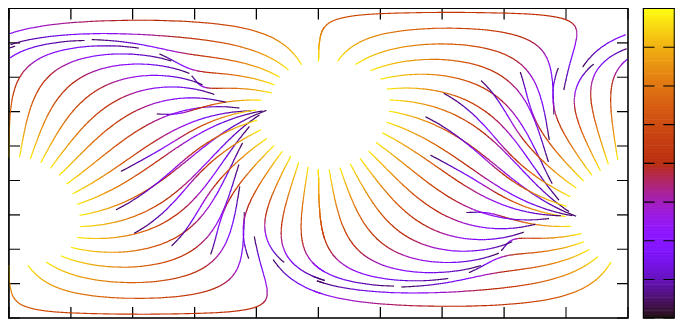}} \\
\resizebox{0.5\textwidth}{!}{\input{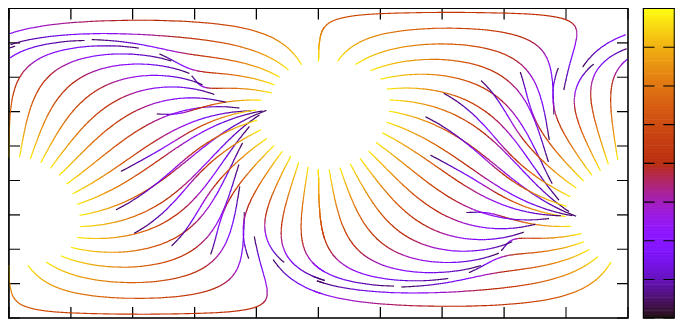}} &
\resizebox{0.5\textwidth}{!}{\input{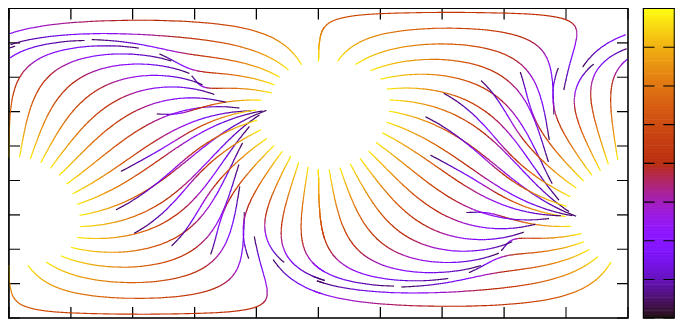}}
\end{tabular}
\caption{True Shapiro delay compared to a reference photon path in general relativity. All for aberration approximation give very similar results.}
\label{fig:ligne_carte_delai_shapiro_true}
\end{figure*}

In fig.~\ref{fig:ligne_carte_delai_shapiro_tret}, we show the associated travel time of photon to reach a spherical distance of $D=1000\,\rlight$ if spacetime would be flat. The average propagation time is again about $D/c$. We compare both times by computing the relative difference between retarded time~$t_{\rm r}$ in flat spacetime compared to Shapiro delay $t_{\rm s}$ given by $t_{\rm s}/t_{\rm r}-1$. Close to the poles where gravity is important, the difference is largest as expected, reaching about 22\% at most. For photons emitted close to the light cylinder, the difference tends to zero as gravity decreases sensitively. All for panels show very similar behaviours, the relative difference is insensitive to the aberration expression used. We emphasize that this 22\% difference should not be interpreted as a time lag observable by a distant observer. What can really be measured at large distant is the time lag between photons emitted at different time and/or location, taking gravity into account or not. Therefore, to get a relevant physical insight into the effect of GR on pulse profiles, we have to compute the difference in photon arrival time between a reference photon trajectory and any other photon. This has been done in fig.~\ref{fig:ligne_carte_delai_shapiro_true} where we compute the time lag between photons emitted on a given field line and a photon going straight from the surface. Time is normalized with respect to one period of the pulsar. Photons emitted from the poles are almost in phase with the reference path as expected because bending is weak (almost radial trajectories). Photons emanating from high altitude arrive early to the observer because they are closer. The maximum time lag between polar photons and light-cylinder photons is about 16\% of the pulsar period. This has to be compared with the usual retarded time expression $(\rlight-R)/2\,\pi\,\rlight \approx 1/2\,\pi \approx 0.16$. Consequently, we conclude that time propagation effects in curved spacetime has little impact on pulsar light-curves and sky maps, the retarded time can be account for only by time of flight in flat spacetime. This explain why we did not implement this effect in a first stage.

\section{Photon orbits and Shapiro delay}

\begin{figure}
 \centering
 \resizebox{0.5\textwidth}{!}{\input{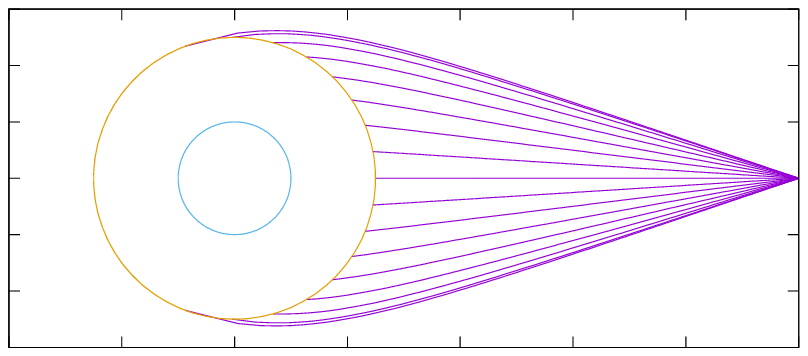}}
 \caption{A sample of photon trajectories arriving at the same location where the observer is located. The compacity is $\Rs/R=0.4$, with $D=10\,\Rs$ and $\Rs=1$. The neutron star surface are depicted as an orange circle and its Schwarzschild radius is shown by a blue circle.}
 \label{fig:PhotonTrajectoires}
\end{figure}

\begin{figure}
 \centering
 \resizebox{0.5\textwidth}{!}{\input{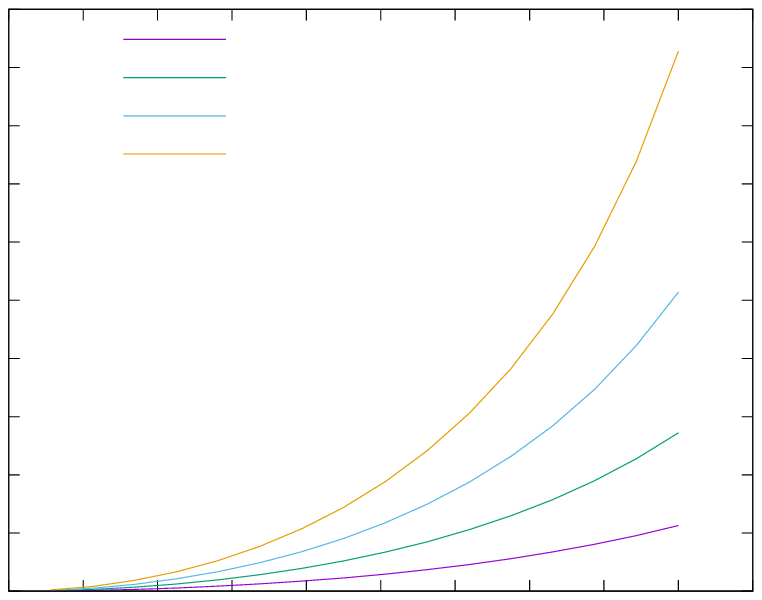}}
 \caption{A sample of time retardation effects corresponding to trajectories similar to those shown in fig.~\ref{fig:PhotonTrajectoires}. Several compacities and initial photon angle~$\xi$ with respect to the radial direction are shown.}
 \label{fig:PhotonTemps}
\end{figure}

In this section we show some photon path integration curves with the associated time of flight including curved space-time. Photons are emitted from the neutron star surface and travel to a distance of~$D=1$~kpc, typical for pulsars. If photons travel in flat spacetime their trajectories would be straight lines starting from an emission point of coordinates $(x_e,y_e,z_e)$ up to the observation point at $(x_r,y_r,z_r)$. The time of flight would then simply be $c\,\Delta t_N = \sqrt{(x_r-x_e)^2 + (y_r-y_e)^2 + (z_r-z_e)^2}$. In Schwarzschild spacetime we have to compute the time integrals along the curved photon path. A relevant sample is shown in fig.~\ref{fig:PhotonTrajectoires} with a low distance of $D=10\,\Rs$ for visualization purposes.
For the 1~kpc distance, the associated time delay is represented in fig.~\ref{fig:PhotonTemps}.
As expected for photons going almost in the radial direction at the emission point, the discrepancy between $\Delta t_N$ and $\Delta t_{GR}$ is negligible. The discrepancy becomes relevant only for high compacities and for photon initial angles $\xi \gtrsim 50\degr$. For magnetospheric emission models, the angle with respect to the radial direction is always $\xi \lesssim 50\degr$ so there is indeed no need to worry about additional Shapiro delay. This has been shown explicitly in the previous section for emission along curved magnetic field lines. In this paper, we do not include Shapiro time delays into the sky map computations. Refined and very accurate models could precisely quantify this discrepancy that we not explore further in this paper.

\section{General relativistic aberration in rotating frames}
\label{app:Aberration}

Relativistic aberration is well described in inertial frames according to the Lorentz transformation between uniformly moving observers. When accelerated frames are considering, especially rotating frames, aberration formulas are less well known. The situation is even worth in general relativity. In this paper, we show how to derive the change in the direction of photon propagation by introducing transfer matrices between inertial frames and general-relativistic frames. Such aberration effects are crucial in our understanding of neutron star radio emission where curved spacetime is important at the surface, but also for high energy emission where the corotation speed becomes large at the light-cylinder. We give explicit and univoque expression for aberration in accelerated frames and rotating frames in Schwarzschild spacetime. We emphasize the discrepancy between the Lorentz aberration in the instantaneous rest frame and our results.

We first remind the usual aberration formula for special relativistic motion, that is two observers moving at constant relativistic velocity. Then we focus on an accelerated frame which should mimic locally the effect of gravitation. Follows then the important discussion about rotating frames. We explicitly add gravity by considering the Schwarzschild background metric.

\subsection{Inertial frame}

Let us consider an observer~$\mathcal{O}'$ moving in the $x$ direction at the speed $v$ with respect to another observer~$\mathcal{O}$ at rest in the Cartesian coordinate system $(t,x,y,z)$. The observer~$\mathcal{O}$ drags an orthonormal basis $\mathcal{R} = \{\mathbf{e}_i\}_{i=0,3}$ and the observer~$\mathcal{O}'$ another orthonormal basis $\mathcal{R}' = \{\mathbf{e}_i'\}_{i=0,3}$. Introducing the Lorentz factor $\gamma=(1-\beta^2)^{-1/2}$ and the rapidity $\tanh \alpha = \beta$ with $\beta = v/c$ where $c$ is the speed of light, the transfer matrix $A_i^k$ going from $\mathcal{O}$ to $\mathcal{O}'$ with $\mathbf{e}_i' = A_i^k\,\mathbf{e}_k$ is
\begin{equation}
\label{eq:MatricePassageInertiel}
 A_i^k = 
 \begin{pmatrix}
  \gamma & \gamma \, \beta & 0 & 0 \\
  \gamma \, \beta & \gamma & 0 & 0 \\
  0 & 0 & 1 & 0 \\
  0 & 0 & 0 & 1
 \end{pmatrix} =
 \begin{pmatrix}
 \cosh \alpha & \sinh \alpha & 0 & 0 \\
 \sinh \alpha & \cosh \alpha & 0 & 0 \\
 0 & 0 & 1 & 0 \\
 0 & 0 & 0 & 1
 \end{pmatrix} .
\end{equation}
Because the coordinate systems are orthonormal, in their respective frames, both observers will measure a wave vector $K^i=\frac{\omega}{c}\,(1,\mathbf n)$ and ${K'}^i=\frac{\omega'}{c}\,(1,\mathbf n')$ where $\omega$ and $\omega'$ are the photon frequency and $\mathbf n$ and $\mathbf n'$ the direction of propagation in their own frame thus normalized spacelike vectors by construction. Transforming the vector from $\mathcal{R}'$ to $\mathcal{R}$ then gives
\begin{subequations}
	\label{eq:AberrationInertiel}
	\begin{align}
	\omega & = \gamma \, \omega' \, (1+\beta\,n_x') \\
	n_x & = \frac{\beta+n_x'}{1+\beta\,n_x'} \\
	n_y & = \frac{n_y'}{\gamma \,(1+\beta\,n_x')} \\
	n_z & = \frac{n_z'}{\gamma \,(1+\beta\,n_x')} .
	\end{align}
\end{subequations}
We retrieve the usual textbook results for relativistic aberration in inertial frames if we set $n_z=n_z'=0$, $n_x=\cos\vartheta$ and $n_x'=\cos\vartheta'$. When the Lorentz factor tends to infinity, $\gamma \to \infty$, beaming is sharply directed into the direction of motion given by the velocity of $\mathcal{R}'$ with respect to $\mathcal{R}$. Indeed, we find in that limit that emission is exactly in the direction of relative motion between the two frames, $\mathbf{n}=(1,0,0)$.

\subsection{Uniformly accelerated frame}

Consider now an accelerated observer~$\mathcal{O}'$ with constant acceleration~$a$ along $x$. The transformation matrix $A_i^k$ going from $\mathcal{O}$ to $\mathcal{O}'$ with $\mathbf{e}_i' = A_i^k\,\mathbf{e}_k$ with $\alpha = \frac{a\,\tau}{c}$ where $\tau$ is $\mathcal{O}'$ proper time is now
\begin{equation}
 A_i^k = 
 \begin{pmatrix}
  \cosh \alpha & \sinh \alpha & 0 & 0 \\
  \sinh \alpha & \cosh \alpha & 0 & 0 \\
  0 & 0 & 1 & 0 \\
  0 & 0 & 0 & 1
 \end{pmatrix}.
\end{equation}
This matrix is deduced from the world line expression of an accelerated observer \citep{gourgoulhon_relativite_2010}. These are exactly the same transformations as between two inertial observers, eq.~(\ref{eq:MatricePassageInertiel}). The aberration formulas are therefore identical to eq.~(\ref{eq:AberrationInertiel}) except that the rapidity varies with proper time~$\tau$. The aberration effect will now depend on the relative position between both observers which was not the case for two inertial observers.

\subsection{Uniformly rotating frame in Schwarzschild metric}

In order to study light aberration around a Schwarzschild black hole, we consider three distinct bases
\begin{itemize}
\item a distant observer with its Cartesian coordinate system $(ct,x,y,z)$ and Minkowskian metric.
\item an observer in uniform rotation around the black hole.
\item a spherical coordinate system non necessarily orthonormal.
\end{itemize}
The rotating observer is located at the point of spherical coordinates $(r,\vartheta,\varphi=\Omega\,t)$ where we introduced $\Omega=\frac{d\varphi}{dt}$. Using the Schwarzschild metric and normalizing the azimuthal velocity to $\beta = \frac{r\,\Omega}{c} \,\sin\vartheta$, its proper time as measured by a comoving clock is
\begin{equation}
 d\tau = \sqrt{\alpha^2 - \beta^2 } \, dt = \frac{dt}{\gamma} .
\end{equation}
with the Lorentz factor $\gamma = \left(\alpha^2 - \beta^2 \right)^{-1/2}$ (a Lorentz factor associated to the coordinate system, not the one measured locally by an observer, see below in the text). Expressed in the Boyer-Lindquist spherical coordinate system, the associated 4-velocity is
\begin{equation}
u^i = \gamma \, (c,0,0,\Omega) .
\end{equation}
The unit vector along the time direction is
\begin{equation}
 \mathbf e_\tau = \gamma \, (1,0,0,\Omega/c) .
\end{equation}
The unit vector orthogonal to $\mathbf e_\tau$ is then defined by 
\begin{equation}
 \mathbf e_{\hat \varphi} = \gamma \, \left( \frac{\beta}{\alpha}, 0, 0, \frac{\alpha}{r\,\sin\vartheta} \right) .
\end{equation}
The other two unit vectors are
\begin{subequations}
\begin{align}
 \mathbf e_{\hat r} & = \alpha \, \mathbf e_r \\
 \mathbf e_{\hat \vartheta} & = \frac{\mathbf e_\vartheta}{r} .
\end{align}
\end{subequations}
The set $(\mathbf e_\tau, \mathbf e_{\hat r}, \mathbf e_{\hat \vartheta}, \mathbf e_{\hat \varphi})$ forms an orthonormal basis for the rotating observer.

To compare with a distant observer, we must project quantities on an orthonormal basis in spherical Boyer-Lindquist coordinates. This is equivalent to switching from one orthonormal frame to another. Note however that the $(t,r,\vartheta,\varphi)$ coordinates system is not an orthonormal basis. The new coordinates are transformed according to
\begin{subequations}
\begin{align}
\mathbf e_\tau & = \gamma \, (\alpha,0,0,\beta) \\
\mathbf e_{\hat r} & = (0,1,0,0) \\
\mathbf e_{\hat \vartheta} & = (0,0,1,0) \\
\mathbf e_{\hat \varphi} & = \gamma \, \left( \beta, 0, 0, \alpha \right) .
\end{align}
\end{subequations}
We remind that in the Schwarzschild metric
\begin{subequations}
\begin{align}
\mathbf e_{\hat t} & = \frac{1}{\alpha} \, \mathbf e_t \\
\mathbf e_{\hat r} & = \alpha \, \mathbf e_r \\
\mathbf e_{\hat \vartheta} & = \frac{1}{r} \, \mathbf e_\vartheta \\
\mathbf e_{\hat \varphi} & = \frac{1}{r\,\sin\vartheta} \, \mathbf e_\varphi .
\end{align}
\end{subequations}
The matrix transformation from the Minkowskian frame to the rotating frame is written as follows
\begin{equation}
A_i^k = 
 \begin{pmatrix}
  \gamma \, \alpha & 0 & 0 & \gamma \, \beta \\
  0 & 1 & 0 & 0 \\
  0 & 0 & 1 & 0 \\
  \gamma \, \beta & 0 & 0 & \gamma \, \alpha
 \end{pmatrix} .
\end{equation}
The frame speed measured locally by an observer at rest (with $dr = d\vartheta = d\varphi = 0$) but located at the photon emission point is
\begin{equation}
 v^{\hat \varphi} = \frac{r\,\sin\vartheta\,d\varphi}{\alpha\,dt} = \frac{\beta}{\alpha} \, c
\end{equation}
because the proper time of the observer is $d\tau_{\rm obs} = \alpha \, dt$. The matrix $A_i^k$ is therefore the same as in special relativity. More precisely its Lorentz factor is $\gamma_{\rm obs} = (1-\beta_{\rm obs}^2)^{-1/2} = \alpha\,\gamma$ and $\beta_{\rm obs} = v^{\hat \varphi}/c = \beta/\alpha$ thus $\gamma_{\rm obs} \, \beta_{\rm obs} = \gamma \, \beta$. This matrix is summarized as
\begin{equation}
A_i^k = 
 \begin{pmatrix}
  \gamma_{\rm obs} & 0 & 0 & \gamma_{\rm obs} \, \beta_{\rm obs} \\
  0 & 1 & 0 & 0 \\
  0 & 0 & 1 & 0 \\
  \gamma_{\rm obs} \, \beta_{\rm obs} & 0 & 0 & \gamma_{\rm obs}
 \end{pmatrix} .
\end{equation}
The aberration of light is then expressed by
\begin{subequations}
\begin{align}
 \omega & = \gamma \, \omega' \, ( \alpha + \beta \, n_\varphi' ) \\
 n_r & = \frac{n_r'}{\gamma \, ( \alpha + \beta \, n_\varphi' )} \\
 n_\vartheta & = \frac{n_\vartheta'}{\gamma \, ( \alpha + \beta \, n_\varphi' )} \\
 n_\varphi & = \frac{\beta + \alpha \, n_\varphi'}{\alpha + \beta \, n_\varphi'} .
\end{align}
\end{subequations}
The vectors $\mathbf n$ et $\mathbf n\,'$ are unit vectors by construction. It can be verified by performing the transformation for the different frames.

Replacing the coordinate Lorentz factor~$\gamma$ and the coordinate speed~$\beta$ by their counterparts (physically measurable) for a stationary observer located at $(r,\vartheta,\varphi)$, we find the special relativistic formulae such that
\begin{subequations}
\begin{align}
 \omega & = \gamma_{\rm obs} \, \omega' \, ( 1 + \beta_{\rm obs} \, n_\varphi' ) \\
 n_r & = \frac{n_r'}{\gamma_{\rm obs} \, ( 1 + \beta_{\rm obs} \, n_\varphi' )} \\
 n_\vartheta & = \frac{n_\vartheta'}{\gamma_{\rm obs} \, ( 1 + \beta_{\rm obs} \, n_\varphi' )} \\
 n_\varphi & = \frac{\beta_{\rm obs} + n_\varphi'}{1 + \beta_{\rm obs} \, n_\varphi'} .
\end{align}
\end{subequations}
The Lorentz factor tends to infinity when $\alpha\rightarrow\beta$. Under these conditions, the direction of propagation of the photon in the inertial frame becomes $\mathbf n = \mathbf e_\varphi$ thus in the direction of motion as it should be.

We can never emphasize too much the fact that the special-relativistic aberration formulae are valid in any arbitrary frame, accelerated or not, with gravitation field or not. The important point to notice is that these expressions remains valid as long as they are performed locally between two orthogonal frame located at the same point in spacetime (a sufficient but not necessary condition).


\bsp	
\label{lastpage}
\end{document}